\documentclass[letterpaper]{aa}
\usepackage{epsfig}
\usepackage{graphicx}
\usepackage{psfrag}
\usepackage{natbib}
\usepackage{amsmath}
\newcommand{\be}{\begin{equation}}
\newcommand{\ee}{\end{equation}}
\newcommand{\bea}{\begin{eqnarray}}
\newcommand{\eea}{\end{eqnarray}}
\newcommand{\bc}{\begin{center}}
\newcommand{\ec}{\end{center}}

\renewcommand{\vec}[1]{ {\bf #1} }

\newcommand{\e}{{\rm e}}

\newcommand{\msun}{{\rm M}_{\odot}}
\newcommand{\Msol}{\,\msun}
\newcommand{\mhalo}[1]{{$10^{#1}\,h^{-1}{\rm M}_\odot$}}

\newcommand{\kpc}{\,\unit{kpc}}
\newcommand{\Gyr}{\,\unit{Gyr}}

\newcommand{\gadget}{{\small GADGET-2}}

\newcommand{\B}{{\mathcal B}}

\newcommand{\bra}{\left< }
\newcommand{\ket}{\right> }

\newcommand{\unit}[1]{\text{#1}}
\newcommand{\diff}[1]{\unit{d}#1}

\newcommand{\nablavec}{\vec{\nabla}}

\begin{document}

\title{Cosmic ray feedback in hydrodynamical simulations of galaxy
  formation}

\titlerunning{Cosmic rays in hydrodynamical simulations}

\author{Martin Jubelgas\inst{1},
Volker Springel\inst{1}, 
Torsten En\ss{}lin\inst{1} and
Christoph Pfrommer\inst{1,2}}

\institute{Max-Planck-Institut
  f\"{u}r Astrophysik, Karl-Schwarzschild-Stra\ss{}e 1, 85740 Garching
  bei M\"{u}nchen, Germany\\
\email{jubelgas@mpa-garching.mpg.de; vspringel@mpa-garching.mpg.de; ensslin@mpa-garching.mpg.de} \and
Canadian Institute for Theoretical Astrophysics, University of Toronto,
60 St. George Street, Toronto, Ontario, M5S 3H8, Canada\\
\email{pfrommer@cita.utoronto.ca}
}

\authorrunning{M.~Jubelgas, V.~Springel, T.~En\ss{}lin, C.~Pfrommer}

\abstract{It is well known that cosmic rays contribute significantly
  to the pressure of the interstellar medium in our own Galaxy,
  suggesting that they may play an important role in regulating star
  formation during the formation and evolution of galaxies. We here
  discuss a novel numerical treatment of the physics of cosmic rays
  and its implementation in the parallel smoothed particle
  hydrodynamics code {\small GADGET-2}.  In our methodology, the
  non-thermal cosmic ray population of each gaseous fluid element is
  approximated by a simple power law spectrum in particle momentum,
  characterized by an amplitude, a cut-off, and a fixed
  slope. Adiabatic compression, and a number of physical source and
  sink terms are modelled which modify the cosmic ray pressure of each
  particle. The most important sources considered are injection by
  supernovae and diffusive shock acceleration, while the primary sinks
  are thermalization by Coulomb interactions, and catastrophic losses
  by hadronic interactions. We also include diffusion of cosmic rays.
  Using a number of test problems, we show that our scheme is
  numerically robust and efficient, allowing us to carry out the first
  cosmological structure formation simulations that self-consistently
  account for cosmic ray physics. In simulations of isolated galaxies,
  we find that cosmic rays can significantly reduce the star formation
  efficiencies of small galaxies, with virial velocities below $\sim
  80\,{\rm km\,s^{-1}}$, an effect that becomes progressively stronger
  towards low mass scales. In cosmological simulations of the
  formation of dwarf galaxies at high redshift, we find that the total
  mass-to-light ratio of small halos and the faint-end of the
  luminosity function are strongly affected. The latter becomes
  flatter. When cosmic ray acceleration in shock waves is followed as
  well, we find that up to $40\%$ of the energy dissipated at
  structure formation shocks can appear as cosmic ray pressure at
  redshifts around $z\sim 3-6$, but this fraction drops to $\sim 10\%$
  at low redshifts when the shock distribution becomes increasingly
  dominated by lower Mach numbers.  Despite this large cosmic ray
  energy content in the high-redshift intergalactic medium, the flux
  power spectrum of the Lyman-$\alpha$ forest is only affected at very
  small scales of $k>0.1\,{\rm km^{-1}s}$, and at a weak level of
  $5-15\%$. Within virialized objects, we find lower contributions of
  CR-pressure, due to the increased efficiency of loss processes at
  higher densities, the lower Mach numbers of shocks inside halos, and
  the softer adiabatic index of CRs, which disfavours them when a
  composite of thermal gas and cosmic rays is adiabatically
  compressed. The total energy in cosmic rays relative to the thermal
  energy within the virial radius drops from 20\% for \mhalo{12} halos
  to 5\% for rich galaxy clusters of mass \mhalo{15} in non-radiative
  simulations.  Interestingly, the lower effective adiabatic index
  also increases the compressibility of the intrahalo medium, an
  effect that slightly increases the central concentration of the gas
  and the baryon fraction within the virial radius. We find that this
  can enhance the cooling rate onto central cluster galaxies, even
  though the galaxies in the cluster periphery become slightly less
  luminous as a result of cosmic ray feedback.    }

\maketitle

\section{Introduction}

In recent years, the $\Lambda$CDM model has emerged as a highly
successful `concordance' model for cosmological structure
formation. It conjectures that the dominant mass component in the
universe consists of cold dark matter, and that a cosmological
constant or dark energy field add sufficient energy density to yield a
spatially flat spacetime. This model is impressively successful in
matching observational data on a large range of scales and epochs,
including the cosmic microwave background fluctuations
\citep[e.g.][]{Spergel2003}, galaxy clustering
\citep[e.g.][]{Tegmark2004,Cole2005}, cosmic flows in the present
universe \citep[e.g.][]{Willick1997,Hudson2004}, or the observational
data on distant supernovae \citep{Riess1998,Perlmutter1999}.

While the dynamics of the dark matter component in the $\Lambda$CDM model is
now quite well understood and can be followed with high accuracy in numerical
simulations \citep{Power2003,Navarro2004,Heitmann2005}, the baryonic processes
that regulate the formation of the luminous components of galaxies are much
less well understood. Direct hydrodynamical simulations that follow the
baryonic gas as well as the dark matter, face a number of `small-scale'
problems. For example, they tend to produce too many stars as a result of a
`cooling catastrophe', unless effects like galactic outflows are included in a
phenomenological way \cite[e.g.][]{SpringelSFRHistory2003}. They lead to too
concentrated disk galaxies \citep[e.g.][]{Abadi2003} and fail to reproduce the
observed shape of the luminosity function of galaxies in detail
\citep{Murali2002,Nagamine2004}.

By invoking strong feedback processes, semi-analytic models of galaxy
formation are able to overcome these problems and to explain a wide array of
galaxy properties
\citep{White1991,Kauffmann1993,Baugh1998,Sommerville1999,Cole2000,Croton2006}.
While this supports the notion that feedback is crucial for the regulation of
galaxy formation, it is unclear whether the physical nature of the feedback
processes is correctly identified in the present semi-analytic models, or
whether they merely give a more or less correct account of the consequences of
this feedback.  Direct hydrodynamic simulations can in principle be used to
lift this ambiguity and to more directly constrain the physical processes at
work.

In most current models of galaxy formation, feedback effects due to supernovae
explosions and due to a photoionizing background are usually included, and
more recently, some studies have considered quasar and radio activity by AGN
as well \citep[e.g.][]{DiMatteo2005,Sijacki2006}.  However, perhaps
surprisingly, magnetic fields and non-thermal pressure components from cosmic
rays have received comparatively little attention thus far \citep[with notable
exceptions,
including][]{Kang1996,Miniati2001COSMOCR,Miniati2001,Miniati2002,RyuKang2003,RyuKang2004},
despite the fact that cosmic rays are known to contribute substantially to the
pressure in the ISM of our own Galaxy. This is probably at least in part due
to the complexity of the cosmic ray dynamics, which when coupled to the galaxy
formation process is very hard to describe analytically.  Even when numerical
methods are invoked, the cosmic ray physics is so involved that a number of
simplifying approximations are required to make it tractable in a cosmological
simulation setting, as we discuss here.

In this study, our goal is to introduce the first cosmological code of
galaxy formation that treats cosmic rays self-consistently during the
structure formation process. Our principal approach for capturing the
cosmic ray physics has been laid out in a companion paper
\citep{Ensslin2005}, where we introduced a number of approximations to
reduce the complexity of the problem.  Fundamentally, we model the
cosmic ray population of each fluid element with a power law spectrum
in particle momentum, characterized by an amplitude, a cut-off, and a
fixed slope. Our model then accounts for adiabatic advection of cosmic
rays, and for injection and loss terms due to a variety of physical
sources. Finally, we also include cosmic ray diffusion. The primary
injection mechanisms we consider are supernova shocks and diffusive
shock acceleration at structure formation shock waves. Since the
efficiency of the latter is a sensitive function of the Mach number of
the shock, we have also developed an on-the-fly shock finder for SPH
calculations, which is described in a second companion paper
\citep{Pfrommer2005}.

In this paper, we use the theoretical model of \citet{Ensslin2005} and cast it
into a numerical formulation of cosmic ray physics that we implement in the
cosmological TreeSPH code {\small GADGET-2} \citep{Springel2001,Springel2005}.
We discuss our numerical approach in detail, including also various
optimisations needed to keep the scheme efficient and robust. We then move on
to show first results from applications of the model, ranging from isolated
galaxies of different sizes, to cosmological simulations of galaxy cluster
formation, and of homogeneously sampled boxes.  Interestingly, cosmic rays can
have a substantial effect on dwarf galaxies, suppressing their star formation
considerably. We show that this should leave a noticeable imprint in the
luminosity function of galaxies, leading to a shallower faint-end slope.

This paper is laid out as follows. In Section~2, we describe the details of
our implementation of cosmic ray physics, and in Section~3 we discuss our
treatment of cosmic ray diffusion.  Section~4 presents a number of test
problems, which we used to verify the validity of results obtained by the
code.  We then describe in Section~5 a first set of simulations of isolated
galaxies carried out with the new code. This establishes a number of principal
effects found for the model. In Section~6, we then extend our analysis to more
sophisticated, fully cosmological simulations of structure formation. We
consider both galaxy clusters and dwarf galaxy formation at high redshift.
Finally, Section~7 summarizes our conclusions and gives an outlook for future
studies of cosmic rays in a cosmological context.

\section{Modelling cosmic ray physics}

In \citet{Ensslin2005}, we have introduce a new theoretical formalism for
a simplified treatment of cosmic ray physics during cosmological
structure formation. We also gave a detailed discussion of the
physical background and the relative importance of various physical
source and sink processes, and how they can be incorporated within the
simplified framework. In this section of the present study, we
describe the practical implementation of this model within the
Lagrangian TreeSPH code \gadget{}, including also a concise summary of
the those parts of the framework of \citet{Ensslin2005} that we
included in the code thus far.

\subsection{The cosmic ray spectrum and its adiabatic evolution}

\begin{figure}
\begin{center}
\includegraphics[width=7cm]{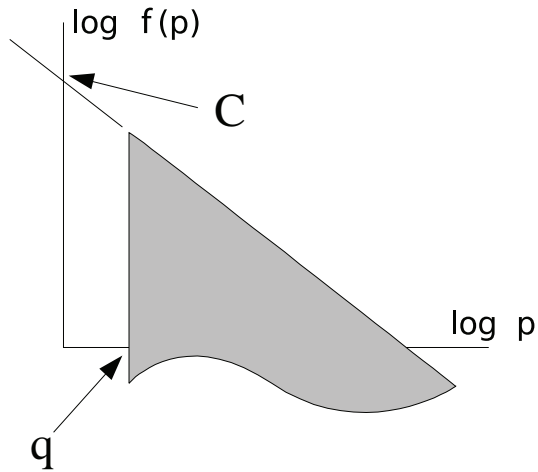}
\end{center}
  \caption{Schematic illustration of the cosmic ray momentum spectrum in our
    two parameter model. We adopt a simple power-law description, where the
    slope of the cosmic ray spectrum is given by a spectral index $\alpha$,
    kept constant throughout our simulation. The normalization of the spectrum
    is given by the variable $C$, and the low-momentum cut-off is expressed in
    terms of a dimensionless variable $q$, in units of $m_p c$ where $m_p$ is
    the proton rest mass. }
\end{figure}

\begin{figure}
\begin{center}
  \resizebox{8cm}{!}{\includegraphics{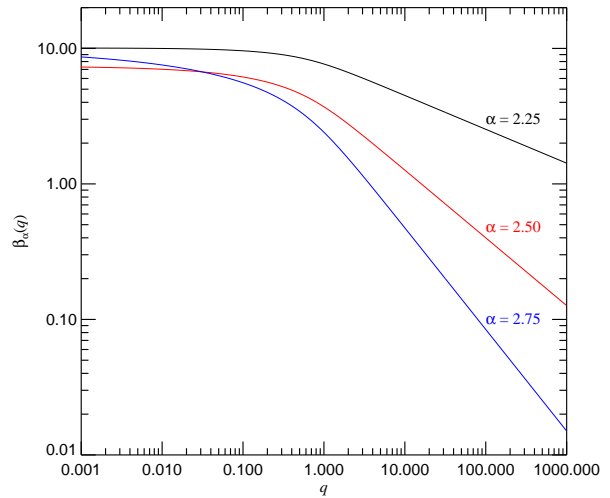}}
\end{center}
  \caption{The function $\beta_\alpha(q)$ introduced in equation
    (\ref{eqnbeta}), for several different values of the spectral slope
    $\alpha$.
 \label{FigBeta}}
\end{figure}

As discussed in full detail in \citet{Ensslin2005}, we assume that the cosmic
ray population of each fluid element is made up of relativistic protons with
an isotropic momentum distribution function of the form \be f(p)=\frac{{\rm
    d}N}{{\rm d}p\,{\rm d}V} = C p^{-\alpha}\,\theta(p-q), \ee where $C$ gives
the normalization, $q$ is a low momentum cut-off, and $\alpha$ is the power
law slope. The momenta are expressed in dimensionless form in units of $m_{\rm
  p}c$, where $m_{\rm p}$ is the proton mass.  For the purposes of this paper,
we will generally take $\alpha$ to be constant ($\alpha\sim 2.5-2.8$), which
should be a reasonable first order approximation in most cases relevant to galactic structure
formation.  We note however that $\alpha$ can in principle be made to vary in
our formalism, at the price of a substantially increased computational cost
and complexity \citep{Ensslin2005}. The pressure of this cosmic ray population
is given by \be P_{\rm CR} = \frac{C\, m_{\rm p} c^2\,}{6} \,
\B_{\frac{1}{1+q^2}} \left( \frac{\alpha-2}{2},\frac{3-\alpha}{2} \right), \ee
while the number density is simply $ n_{\rm CR} = {C\, q^{1-\alpha}}/
({\alpha-1})$.  Here \be \B_n(a,b)\equiv \int_0^n x^{a-1} (1-x)^{b-1}\,{\rm
  d}x \ee denotes incomplete Beta functions.  To describe the kinetic energy
per cosmic ray particle for such a power-law population we define the function
\be T_{\rm CR}(\alpha,q) \equiv \left[\frac{1}{2}q^{\alpha-1} \beta_\alpha(q)
  + \sqrt{1+q^2} -1\right] m_{\rm p}c^2 ,
\label{eqnTCR}
\ee which will be of later use.
The quantity
\begin{equation}
\beta_\alpha(q) \equiv {\mathcal B}_{\frac{1}{1+q^2}} \left( \frac{\alpha-2}{2},
  \frac{3-\alpha}{2} \right ) \text{.}
\label{eqnbeta}
\end{equation}
is here introduced as a convenient abbreviation for the incomplete Beta
function. We show $\beta_\alpha(q)$ as a function of $q$ for a few values of
$\alpha$ in Figure~\ref{FigBeta}.

We here implement the cosmic ray model in a Lagrangian simulation
code, where the advection of the cosmic ray population can be
conveniently described simply in terms of the motion of gas particles. In this
approach, the normalization of the spectrum should be expressed
in terms of a quantity normalized to mass, instead of the
volume-normalized quantity $C$, with the translation between the two
being simply given by the local gas density $\rho$.
In our case, it is convenient to absorb the proton mass into a
redefinition of the amplitude, so we define
\begin{equation}
\tilde{C} = C \frac{m_{\rm p}}{\rho}
\end{equation}
as Lagrangian amplitude of the spectrum.  Upon adiabatic changes of the gas density,
the normalization of the spectrum changes according to
\begin{equation}
\tilde{C}(\rho) = \left( \frac{\rho}{\rho_0}
\right)^{\frac{\alpha-1}{3}} \tilde{C}_0 \text{,}
 \label{eqadiab}
\end{equation}
while the momentum cut-off shifts as
\begin{equation}
q(\rho) = \left( \frac{\rho}{\rho_0}
\right)^{\frac{1}{3}} q_0 .
\label{eqadiab2}
\end{equation}
Here, we introduced a reference density $\rho_0$ (for example set
equal to the mean cosmic density) and a corresponding normalization
$\tilde{C}_0$ and cut-off $q_0$ at this density. In our numerical
implementation, we only have to follow the evolution of the adiabatic invariants
$\tilde{C}_0$ and $q_0$ due to non-trivial physical source and sink
processes, releasing us from the task to compute adiabatic changes of
the normalization and cut-off explicitly. Instead, they are simply
accounted for by equations (\ref{eqadiab}) and (\ref{eqadiab2}).

The number density $n_{\rm CR}$ of relativistic CR protons can also be
conveniently expressed in terms of the total proton density\footnote{We here
  loosly call $\rho/m_{\rm p}$ the proton density. In our cosmological
  applications, we of course use the mean particle mass where appropriate to
  account for the presence of heavier elements and the ionization state of the
  gas.}, yielding
\begin{equation}
\tilde{n} = n_{\rm CR} \frac{m_p}{\rho}  = \tilde{C}
\frac{q^{1-\alpha}}{\alpha - 1} = \tilde{C}_0
\frac{q_0^{1-\alpha}}{\alpha - 1}\text{.}
\label{eqn:baryonfraction}
\end{equation}
We can thus interpret $\tilde{n}$ as something like a ``cosmic ray to baryon
ratio''.  This quantity is an adiabatic invariant which can be
followed accurately in dynamical simulations with our Lagrangian approach.

Note that in our model we do not explicitely remove baryons from the reservoir of
ordinary thermal matter when they are accelerated to become relativistic
cosmic ray particles.  This is a valid approximation, provided we have
$\tilde{n} \ll 1$, which is always expected in our applications.  In
the calculations we performed so far, the fraction of protons contained in the
relativistic phase typically remained far below a maximum value of $\tilde{n}
\approx 10^{-4}$. The latter is already an exceptionally large value which we
encountered only in our most extreme tests, but it is still small enough that
the reduction of the number density of thermal particles can be safely
neglected. We note that cosmic ray confinement by magnetic fields holds
in ideal magneto-hydrodynamics (MHD) when the mass fraction in relativistic
particles is small.

\begin{figure}
\begin{center}
  \resizebox{8cm}{!}{\includegraphics{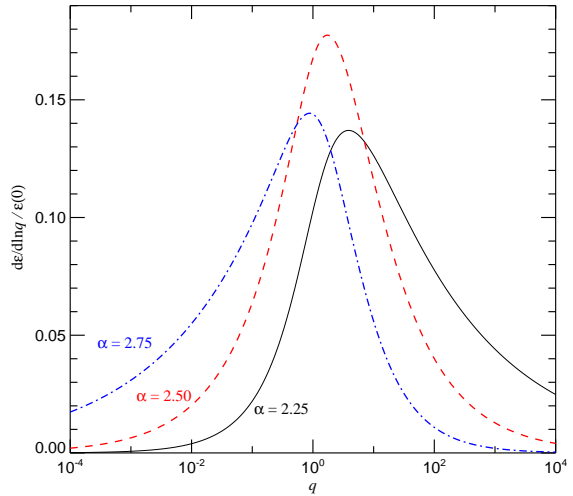}}
\end{center}
  \caption{The distribution of cosmic ray energy per unit logarithmic
    interval of proton momentum, for several different values of the
    spectral slope $\alpha$. The distributions have been normalized to
    $\tilde{\varepsilon}(0)$ in each case.
 \label{FigEnergySpec}}
\end{figure}

In a Lagrangian code, it is natural to express the cosmic ray energy content
in terms of energy per unit gas mass, $\tilde{\varepsilon}$, which is given 
 by
\begin{equation}
\tilde{\varepsilon}
= c^2 \frac{\tilde{C}}{\alpha - 1} \left[ \frac{1}{2} \beta_\alpha(q) + q^{1-\alpha}\left(\sqrt{1+q^2}-1\right)\right],
 \label{eqenergy}
\end{equation}
Note that $\tilde{\varepsilon}$ refers to the energy normalized by the
total gas mass, not by the mass of the cosmic ray particles alone. The specific
energy content can also be expressed as
\begin{equation}
\tilde \varepsilon = \frac{T_{\rm CR}\,n_{\rm CR}}{\rho} = \frac{T_{\rm CR}\,\tilde n}{m_{\rm p}} .
\label{eqnegyspec}
\end{equation}
In Figure~\ref{FigEnergySpec}, we show the distribution ${\rm d}\tilde
\varepsilon / {\rm d}\ln q$ of energy per logarithmic momentum
interval, normalized to a spectrum with vanishingly small cut-off. For
spectral indices in the range $2 < \alpha <3$, most of the energy is
typically contained around $q \simeq 1$, unless the cut-off of the
actual spectrum lies higher than that, in which case the particles
just above the cut-off will dominate the total energy. Due to our
assumption that the momentum distribution extends as a power-law to
infinity, the spectral index $\alpha$ is restricted to $\alpha > 2$,
otherwise the energy would diverge. For $\alpha <3$, the energy stays
finite also for an arbitrarily low spectral cut-off.

In our numerical scheme, every baryonic SPH particle carries the adiabatic
invariants $q_0$ and $\tilde{C}_0$ as internal degrees of freedom for the
description of the cosmic ray physics.  These variables are then used to
compute all physical properties of the cosmic ray sector, as required for the
force evaluations.  For the gas dynamics, we are primarily interested in the
effective pressure term due to the relativistic particles, which are confined
by the ambient magnetic field. In our set of variables, this is compactly
given as
\begin{equation}
P_{\rm CR} = \frac{\tilde{C} \rho\, c^2}{6} \beta_\alpha(q) \text{.} \label{eqpressure}
\end{equation}
In the calculation of the hydrodynamic accelerations with the Euler equation,
we can simply add this partial pressure due to CRs to the ordinary thermal
pressure \citep[see][for further discussion]{Ensslin2005}. This makes the
interface with the ordinary hydrodynamical code conveniently small, requiring
only a small number of changes at well defined places.

\begin{figure}
\begin{center}
  \resizebox{8cm}{!}{\includegraphics{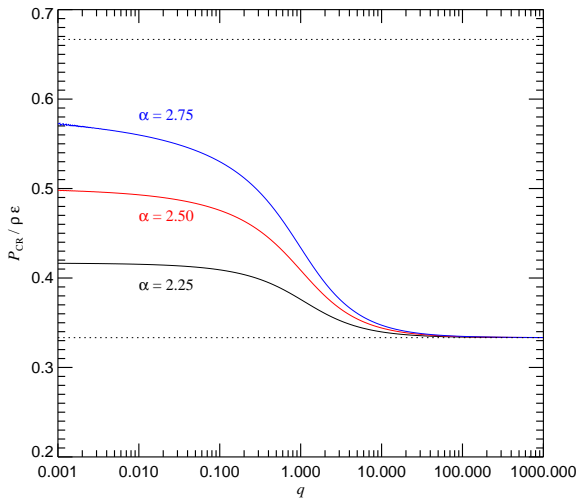}}
\end{center}
  \caption{Cosmic ray pressure in units of the cosmic ray energy density, as a
    function of the spectral cut-off $q$. Except in the transition region from
    non-relativistic to relativistic behaviour, the cosmic ray pressure
    depends only weakly on $q$. In the ultra-relativistic regime, the ratio
    approaches $P_{\rm CR}/({\rho \, \tilde\varepsilon}) \simeq (4/3-1)$,
    which is shown as the lower dotted line. The upper dotted line gives the
    expected value of $(5/3-1)$ for an ideal gas. For the same energy content,
    cosmic rays always contribute less pressure than thermal gas.
 \label{FigPressure}}
\end{figure}

The effective adiabatic index of the cosmic ray pressure component
upon local isentropic density changes is
\begin{equation}
\gamma_{{\rm CR}} \equiv \frac{\partial \log P_{\rm CR}}{\partial
\log  \rho} =\frac{\alpha + 2}{3} - \frac{2}{3}\frac{q^{3-\alpha}}{\beta_{\alpha}(q)\sqrt{1+q^2}}.
\end{equation}
On the other hand, when the pressure is
expressed in terms of the cosmic ray energy density, we obtain 
\begin{equation}
\frac{P_{\rm CR}}{\rho \, \tilde\varepsilon} = \frac{(\alpha-1)\,
  \beta_\alpha(q)}{3 \beta_\alpha(q) + 6\, q^{1-\alpha} ( \sqrt{1+q^2}
  -1)} \label{EqnPress3} .
\end{equation}
In Figure~\ref{FigPressure}, we show the dependence of the
 right-hand-side of equation (\ref{EqnPress3}) on the spectral cut-off
 $q$, for different values of the slope $\alpha$. For large values of
 $q$ we obtain $P_{\rm CR}/({\rho \, \tilde\varepsilon}) \simeq
 (4/3-1)$, as expected for particles in the ultra-relativistic regime,
 while for low values of $q$ the ratio is still significantly below
 the $(5/3-1)$ expected for an ideal gas. However, it is clear that
 for {\em given} cosmic ray energy density, the pressure depends only
 weakly on the spectral cut-off; the value of $\tilde\varepsilon$ is
 hence much more important for the dynamics than the value of
 $\alpha$.

\subsection{Including non-adiabatic CR processes}

The adiabatic behaviour of cosmic rays that are locally locked into the fluid
by magnetic fields can be well traced with the above prescriptions. However,
there are also a multitude of physical processes that affect the CR spectrum
of a gaseous mass-element in a non-adiabatic fashion. For instance, particles
can be accelerated in strong shock waves to relativistic momenta and become
cosmic rays. This process of diffusive shock acceleration should be
particularly effective in high Mach number accretion shocks during
cosmological structure formation, which can be traced by the hydrodynamical
solver of our simulation code. On sub-resolution scales, violent shocks due to
supernova explosions associated with stellar evolution may inject cosmic rays
as well. Other astrophysical sources include the ejection of high-energy
particles in a jet from an accreting black hole.

On the other hand, the cosmic ray population suffers a number of loss
processes which will diminish the abundance over time if there is no new
supply of freshly injected or accelerated protons. We shall here consider only
the most prominent loss processes in the form of Coloumb losses that thermalize
the cosmic ray energy, catastrophic losses that let the energy escape as
radiation, and diffusion which washes out cosmic ray pressure gradients.

As discussed above and in \citet{Ensslin2005}, our cosmic ray model
requires three parameters to describe the state of the relativistic particle
component of the gas.  One parameter is the spectral index $\alpha$, which we
set to a constant value throughout the simulation volume, specified at the
start of the simulation with a value motivated by the typically observed index
in galactic systems.  However, we do not restrict the range of non-adiabatic
processes we consider to those with a similar injection index. Rather, we
translate the injected cosmic ray properties into changes of amplitude and
momentum cut-off within the framework of our simplified, fixed-slope model for
the cosmic ray spectrum.  This translation is based on basic principles of
mass and energy conservation. Despite this considerable simplification, it is
clear that the thermodynamic state of CR gas is considerably more complex than
that of an ideal gas, where essentially everything is determined by the
specific entropy alone.

Given some changes within a simulation time step of the cosmic ray
specific energy ($\diff{\tilde{\varepsilon}}$) and relative CR density
($\diff{\tilde{n}}$) associated with a fluid element, these changes
can be cast into variations of the adiabatic invariants $q_0$ and
$\tilde{C}_0$ of the cosmic ray population using a Jacobian matrix. We
then obtain
\begin{equation}
\diff{\tilde{C}_0} = \left( \frac{\rho}{\rho_0} \right)^{-\frac{\alpha-1}{3}}
\diff{\tilde{C}} = 
\tilde{C}_0 \frac{m_{\rm p}\, \diff{\tilde{\varepsilon}}  - T_{\rm p}(q)
  \,\diff{\tilde{n}}}{m_{\rm p} \tilde{\varepsilon}  - T_{\rm p}(q)\, \tilde{n}}
\label{eqnJacobi1}
\end{equation}
and
\begin{equation}
\diff{q}_0 = \left( \frac{\rho}{\rho_0} \right)^{-\frac{1}{3}} \diff{q}
= \frac{q_0}{\alpha-1}\frac{ m_{\rm p} \,\diff{\tilde{\varepsilon}} - T_{\rm CR}\,
  \diff{\tilde{n}}}{m_{\rm p}\tilde{\varepsilon}  - T_{\rm p}(q)\,
  \tilde{n}} \text{,}
\label{eqnJacobi2}
\end{equation}
where we used the mean kinetic energy per cosmic ray particle, viz.
  $T_{\rm CR} = {\tilde{\varepsilon}\, m_{\rm p}}/{\tilde{n}}$, as
  given by equation (\ref{eqnTCR}), and defined 
\begin{equation}
T_{\rm p}(q) =
  (\sqrt{1+q^2} -1)m_{\rm p} c^2
\end{equation}
 as the kinetic energy of a proton
  with normalized momentum $q$. Recall that $T_{\rm CR}$ only depends
  on $q$ and $\alpha$, but does not depend on the normalization of the
  spectrum.

However, this simple and fast translation scheme will only work for
sufficiently small changes of the cosmic ray population.  In our
implementation, we therefore apply equations (\ref{eqnJacobi1}) and
(\ref{eqnJacobi2}) only if the relative changes in cosmic ray energy
and number density are less than a few percent. Otherwise, new cosmic
ray spectral parameters in terms of $\tilde{C}_0$ and $q_0$ are
computed by explicitly solving equations (\ref{eqn:baryonfraction})
and (\ref{eqenergy}), after applying the principles of energy and
particle conservation.  While (\ref{eqn:baryonfraction}) can be easily
solved for either $q_0$ or $\tilde C_0$, equation (\ref{eqenergy}) for
the specific energy needs to be inverted analytically, but this can be
done efficiently numerically \citep[see also the discussion in
section \ref{SecNumDetails}, and in][]{Ensslin2005}.

Still, a naive application of energy and particle conservation when
adding a newly injected CR component to the current spectrum can cause
unphysical results if the spectral cut-offs involved are very
different. The reason lies in the strong dependence of the cosmic ray
loss processes on particle momentum, together with our simplified
spectral representation. As we will see, the life-time of cosmic ray
particles grows monotonically with particle momentum. This dependence
is particularly steep in the non-relativistic regime ($\tau_{\rm losses}(p)
\sim p^3$), but becomes much shallower and eventually nearly flat in
the mildly relativistic and strongly relativistic regimes. Simply
injecting a CR component with very low cut-off to one with high
cut-off while enforcing total energy and CR particle number
conservation will then result in a new composite spectrum where many
of the original CR particles are represented as lower momentum
particles. Consequently, their cooling times would be artificially
reduced. Ultimately, this problem arises because the number density of
the injected particles is dominated by low momenta, and these
have cooling times much shorter than any relevant dynamical
timescale.  If this is the case, then it would make more sense to
never inject this population to begin with, and to rather thermalize
it instantly, thereby avoiding an unphysical distortion of the
composite spectrum.

The two injection processes we consider in this paper (shocks and
supernova) both supply power-law distributions of cosmic ray particles
which start at very low thermal momenta.  For them, we define an
injection cut-off $q_{\rm inj}$ such that only the particles and the
energy above the cut-off are injected, while the rest of the energy is
instantly thermalized and added to the thermal
reservoir. The criterion we choose for defining $q_{\rm inj}$ is 
\be 
\tau_{\rm losses}(q_{\rm inj}) = \tau_{\rm inj}  (\tilde\varepsilon ,
q_{\rm inj})
 \label{EqInjectCrit}\ee
where $\tau_{\rm losses}(q)$ is the cooling timescale \be \tau_{\rm
losses}(q) \equiv \frac{\tilde{\varepsilon}} { |{\rm d}
\tilde{\varepsilon} / {\rm d}t |_{\rm losses}}\ee due to cosmic ray
loss processes for a spectrum with cut-off at $q$ and spectral slope
$\alpha$.  As we will see, this timescale is independent of the
normalization of the spectrum, and inversely proportional to the
density, provided the  weak 
density-dependence of the Coulomb logarithm  in the Coulomb loss-rate is
neglected.  Given the present cosmic ray
energy content $\tilde{\varepsilon}$, the timescale \be \tau_{\rm
inj}(\tilde{\varepsilon}, q_{\rm inj}) \equiv
\frac{\tilde{\varepsilon}}{ f_{\alpha_{\rm inj}}(q_{\rm 0}, q_{\rm
inj})\, ({\rm d} \tilde{\varepsilon} / {\rm d}t)_{\rm inj} }\ee
defines the current heating time due to the injection source, assuming
that only the fraction $f(q_0, q_{\rm inj})$ above $q_{\rm inj}$ of
the raw energy input rate $({\rm d} \tilde{\varepsilon} / {\rm
d}t)_{\rm inj}$ contributes efficiently to the build up of the cosmic
ray population. Here $q_0$ is the intrinsic injection cut-off of the
source, which typically lies at very small thermal momenta, and
$\alpha_{\rm inj}$ is the slope of the source process.  The factor
$f_{\alpha_{\rm inj}}(q_{\rm 0}, q_{\rm inj})$ is given by \be
f_{\alpha_{\rm inj}}(q_{\rm 0}, q_{\rm inj}) = \left(\frac{q_{\rm
inj}}{q_{\rm 0}}\right)^{1-\alpha_{\rm inj}} \frac{T_{\rm
CR}(\alpha_{\rm inj}, q_{\rm inj})}{T_{\rm CR}(\alpha_{\rm inj},
q_{\rm 0})} \ee for $q_{\rm inj} \ge q_0$, otherwise by unity.

Equation~(\ref{EqInjectCrit}) simply says that we only inject cosmic
ray particles at momenta where a spectral component {\em could} grow,
given the rate of energy injection. It would be unphysical to assume
that a spectrum develops that extends as a power-law to momenta lower
than $q_{\rm inj}$. The addition of this injection rule is hence
necessary to make our simple model with a fixed spectral shape
physically well-behaved.

We note that equation~(\ref{EqInjectCrit}) will typically have two solutions,
or may not have a solution at all if the
current energy $\tilde{\varepsilon}$ in cosmic rays is high enough. In the
former case, the
physical solution is the smaller of the two, which lies at $q_{\rm inj} \le
q_{\rm max}$, where $q_{\rm max}$  is the place where the expression 
$\tau_{\rm losses}(q_{\rm inj}) f(q_0, q_{\rm inj})$ attains its maximum, i.e.
\be
\left. \frac{{\rm d}}{{\rm d}q}\left[
{ \tau_{\rm losses}(q) f(q_0, q)    }\right]\right|_{q=q_{\rm max}} = 0 \,.
\label{InjMax}\ee
In the latter case, we choose $q_{\rm inj}=q_{\rm max}$, which comes closest to
solving equation~(\ref{EqInjectCrit}) and naturally corresponds to the point
where one expects the largest amount of cosmic ray energy that can be present
as a power-law with a balance between loss and source processes.

\subsection{Shock acceleration}

In our present model, we consider two primary sources for cosmic rays,
diffusive shock acceleration and supernovae. In the former, a small fraction
of the particles streaming through a shock front is assumed to diffuse back
and forth over the shock interface, experiencing multiple accelerations This
can only happen to particles in the high-energy tail of the energy
distribution, and eventually results in a power-law momentum distribution
function for the accelerated particles.

In the linear regime of CR acceleration, particles above a threshold momentum
$q_{\rm inj}$ can be accelerated. They are redistributed into a power-law
distribution in momentum that smoothly joins the Maxwell-Boltzmann
distribution of the thermal post-shock gas. The slope of the injected CR
spectrum is given by \be \alpha_{\rm inj} = \frac{r+2}{r-1}, \ee where
$r=\rho_2/\rho_1$ is the density compression ratio at the shock
\citep{Bell1978a,Bell1978b}.  If the shock is dominated by the thermal
pressure, the spectral index can also be expressed through the Mach number $M$
as \be \alpha_{\rm inj} = \frac{4-M^2+3\gamma M^2}{2(M^2-1)}. \ee The stronger
the shock becomes, the flatter the spectrum of the accelerated CR particles,
and hence the more high-energy particles are produced.  Weak shocks on the
other hand produce only rather steep spectra where most of the particles
thermalize quickly.

Due to the continuity between the power law and the thermal spectrum, the
injected cosmic ray spectrum is completely specified by the injection
threshold $q_{\rm 0}$, provided the shock strength is known.  We will assume
that $q_{\rm 0}$ is at a fixed multitude $x_{\rm inj}$ of the average thermal
post-shock momentum, $p_{\rm th}=\sqrt{2 k T_2/(m_{\rm p}c^2)}$, i.e. $q_{\rm
  0} = x_{\rm inj} p_{\rm th}$.  In this case, the fraction of particles
that experience shock acceleration does not depend on the post-shock
temperature $T_2$, and is given by
\begin{equation}
\Delta \tilde n_{\rm lin}  = \frac{4}{\sqrt{\pi}} \frac{x_{\rm inj}^3}{\alpha_{\rm inj} - 1}
\e^{-x_{\rm inj}^2}.
\end{equation}
We will typically adopt a fixed value of $x_{\rm inj}\simeq 3.5$, motivated by
theoretical studies of shocks in galactic supernova remnants
\citep[e.g.][]{Drury1989}. The fraction of injected supernovae particles in
strong shocks is then a few times $10^{-4}$
\citep{Drury1989,Jones1993,Kang1995}.

In the linear regime of CR acceleration, the specific energy per unit gas mass in the injected
cosmic ray population is given by 
\be 
\Delta \tilde \varepsilon_{\rm lin}  =  \frac{T_{\rm CR}(\alpha_{\rm
    inj}, q_{\rm inj})\, \Delta \tilde n_{\rm lin}}{m_{\rm p}} \, .
 \ee We can use this value to define a shock injection efficiency
for CRs by relating the injected energy to the dissipated energy per unit mass
at the shock front. The latter appears as extra thermal energy above the
adiabatic compression at the shock and is given by $\Delta u_{\rm diss} = u_2
- u_1 r^{\gamma -1}$, where $u_1$ and $u_2$ are the thermal energies per unit
mass before and after the shock, respectively. The injection efficiency of
linear theory is then given by
\begin{equation}
\zeta_{\rm lin} \equiv \frac{\Delta \tilde \varepsilon_{\rm lin}} {\Delta u_{\rm diss}}.
\label{eqnlinefficiency}
\end{equation}
However, the shock acceleration effect experiences saturation when the
dynamical CR becomes comparable to the upstream ram pressure $\rho_1 v_1^2$ of the
flow. We account for this by adopting the limiter suggested by
\citet{Ensslin2005}, and define as final acceleration efficiency
\begin{equation}
\zeta_{\rm inj}=\left[ 1 - \exp \left(- \frac{\zeta_{\rm lin}}{\zeta_{\rm
      max}} \right) \right] \zeta_{\rm max}.
\end{equation}
We will adopt  $\zeta_{\rm max}=0.5$ for the results of this study.
Thus, we take the injected energy to be
\begin{equation}
\Delta \tilde \varepsilon_{\rm inj} = \zeta_{\rm inj} \Delta
u_{\rm diss},
\label{eqndissinj}
\end{equation}
and correspondingly, 
the injected particle number is given by
\be
\Delta \tilde n_{\rm inj} = \frac{m_{\rm p}\, \Delta \tilde
  \varepsilon_{\rm inj}}{T_{\rm CR}(\alpha_{\rm inj}, q_{\rm 0})},
\ee
where $T_{\rm CR}(\alpha_{\rm inj}, q_{\rm 0})$ is the mean kinetic
energy of the accelerated cosmic ray particles. In practice, both
$\Delta \tilde \varepsilon_{\rm inj}$ and $\Delta \tilde n_{\rm inj}$ will be
lowered when we shift the actual injection point from $q_0$ to $q_{\rm inj}$,
as determined by equation~(\ref{EqInjectCrit}), with the difference of the energies 
fed to the thermal reservoir directly.

It is clear that the efficiency of CR particle acceleration depends strongly
on the compression ratio, or equivalently on the Mach number of shocks.
Interestingly, accretion shocks during cosmological structure formation can be
particularly strong. Here we hence expect potentially interesting effects both
for the forming intragroup and intracluster media, as well as for the
intergalactic medium. However, in order to accurately account for the cosmic
ray injection by structure formation shocks, we somehow need to be able to
estimate the strength of shocks in SPH simulations. As SPH captures shocks
with an artificial viscosity instead of an explicit shock detection scheme,
this is a non-trivial problem.

In \citet{Pfrommer2005}, our second companion paper to this study, we have
proposed a practical solution to this problem and developed a novel method for
measuring the Mach number of shocks on the fly during cosmological SPH
calculation. The method relies on the entropy formulation for SPH by
\citet{SprHe01}, and uses the current rate of entropy injection due to
viscosity, together with an approximation for the numerical shock transit
time, to estimate the shock Mach number. The scheme works better than one may
have expected, and it is in fact capable of producing quite accurate Mach
number estimates, even for the case of composite gases with a thermal and a
cosmic ray pressure component.  In cosmological simulations, the method
delivers Mach number statistics which agree well with results obtained with
hydrodynamical mesh codes that use explicit Riemann solvers
\citep{Ryu2003,Pfrommer2005}.

Having a reliable Mach number estimator solves an important problem when
trying to account for cosmic ray injection by shocks in SPH. Another
complication is posed by the shock broadening inherent in SPH, which implies a
finite shock transit time for particles, i.e.~a SPH particle may require
several timesteps before it has passed through a shock and received the full
dissipative heating.  Note that the number of these steps depends on the
timestep criterion employed, and can in principle be made very large for a
sufficiently conservative choice of the Courant coefficient. Unlike assumed in
the above treatment of diffusive shock acceleration, we hence are not dealing
with a discrete injection event, but rather need to formulate the cosmic ray
acceleration in a `continuous fashion', in parallel to the thermal
dissipation, such that the final result does not depend on how many timesteps
are taken to resolve a broadened shock front.

Fortunately, the above treatment can be easily adjusted to these conditions.
We can simply insert for $\Delta u_{\rm diss}$ in equation~(\ref{eqndissinj})
the dissipated energy in the {\em current timestep}. This quantity is computed
in the SPH formalism anyway, and in fact, we know that SPH will integrate
$\Delta u_{\rm diss}$ correctly through the shock profile, independent of the
number of steps taken. This is because the correct pre- and post-shock state
of the gas are enforced by the conservation laws, which are fulfilled by the
conservative SPH code. For the same reason, the Rankine-Hugoniot jump
conditions are reproduced across the broadened shock. Note however that for
computing the linear shock injection efficiency according to
equation~(\ref{eqnlinefficiency}), we need to continue to use an estimate for
the total energy dissipated across the shock, based on the Mach number finder.

For simplified test calculation with the code, we have also implemented an
option where we assume a constant injection efficiency of the shock
acceleration process, along with a constant spectral index and momentum
cut-off parameter.  Values for these parameters can then be chosen to
represent the energetically most important types of shocks in the environment
to be simulated. Such a simplified injection mechanism can then also be used
get an idea about the importance of the Mach-number dependence of the shock
acceleration for different environments.

\subsection{Injection of cosmic rays by supernovae}

Strong shock waves associated with supernovae explosions are believed to be
one of the most important cosmic ray injection mechanisms in the interstellar
medium. However, similar to star formation itself, individual supernovae are
far below our resolution limit in cosmological simulations where we need to
represent whole galaxies, or more challenging still, sizable parts of the
observable universe. We therefore resort to a subresolution treatment for star
formation and its regulation by supernovae, as proposed by
\citet{Springel2003}. In this model, the interstellar medium is pictured as a
multiphase medium composed of dense, cold clouds, embedded in a tenuous hot
phase. The clouds form by thermal instability out of the diffuse medium, and
are the sites of star formation.  The massive stars of each formed stellar
population are assumed to explode instantly, heating the hot phase, and
evaporating some of the cold clouds. In this way, a tight self-regulation
cycle for star formation in the ISM is established.

To model the generation of cosmic rays, we assume that a certain
fraction $\zeta_{\rm SN}\simeq 0.1-0.3$ of the supernova energy
appears as a cosmic ray population \citep{Aharonian2006,Kang2006}. The total
rate of energy injection by supernovae for a given star formation rate
$\dot \rho_\star$ depends on the IMF. Assuming a Salpeter IMF and that
stars above a mass of $8\, {\rm M}_\odot$ explode as supernova with a
canonical energy release of $10^{51}\,{\rm ergs}$, we obtain roughly
one supernova per $250\, {\rm M}_\odot$ of stellar mass formed,
translating to an energy injection rate per unit volume of
$\epsilon_{\rm SN} \dot \rho_\star$, with $\epsilon_{\rm SN}= 4 \times
10^{48}\, {\rm ergs}\,{\rm M}_\odot^{-1}$. We then model the CR energy
injection per timestep of a gas particle as \be \Delta
\tilde\varepsilon_{\rm SN} = \zeta_{\rm SN} \epsilon_{\rm SN} \,\dot
m_\star \, \Delta t, \label{EqnSNinput}\ee where $\dot { m}_\star =
\dot \rho_\star / \rho $ is the particle's star formation rate per
unit mass. Note that uncertainties in the IMF are not really important
here as we have introduced a free parameter, $\zeta_{\rm SN}$, to
control the amount of energy that is fed into cosmic rays.

For the slope of the injected cosmic ray power-law we assume a
plausible value of $\alpha_{\rm SN}= 2.4$
\citep{Aharonian2004,Aharonian2006}, and for the cut-off $q_{\rm SN}$,
we can adopt the thermal momentum $q_{\rm SN}=\sqrt{kT_{\rm
SN}/(m_{\rm p}c^2)}$ for a fiducial supernova temperature
characteristic for the involved shock acceleration. Our choice of
$\alpha_{\rm SN}= 2.4$ for the injection slope is motivated by the
observed slope of $\sim 2.75$ in the ISM, and the realization that
momentum dependent diffusion in a turbulent magnetic field with a
Kolmogorov-type spectrum on small scales should steepen the injected
spectrum by $p^{-1/3}$ in equilibrium.  Our results do not depend on
the particular choice for $T_{\rm SN}$, provided $q_{\rm SN}\ll
1$. The change of the particle number density can then be computed
with the mean kinetic energy $T_{\rm CR}(\alpha_{\rm SN}, q_{\rm SN})$
of the injected power law. Using equations (\ref{eqnTCR}) and
(\ref{eqnegyspec}), this results in \be \Delta \tilde n = m_{\rm p}
\frac{\zeta_{\rm SN} \epsilon_{\rm SN} \dot { m}_\star}{T_{\rm
CR}(\alpha_{\rm SN}, q_{\rm SN})} \, \Delta t. \ee We note that in the
formalism of \citet{Springel2003}, we need to reduce the supernovae
energy injected into thermal feedback (and to an optional wind model
if used) by the fraction $\zeta_{\rm SN}$ that we assume powers cosmic
ray acceleration.

\begin{figure}
\begin{center}
  \resizebox{7.7cm}{!}{\includegraphics{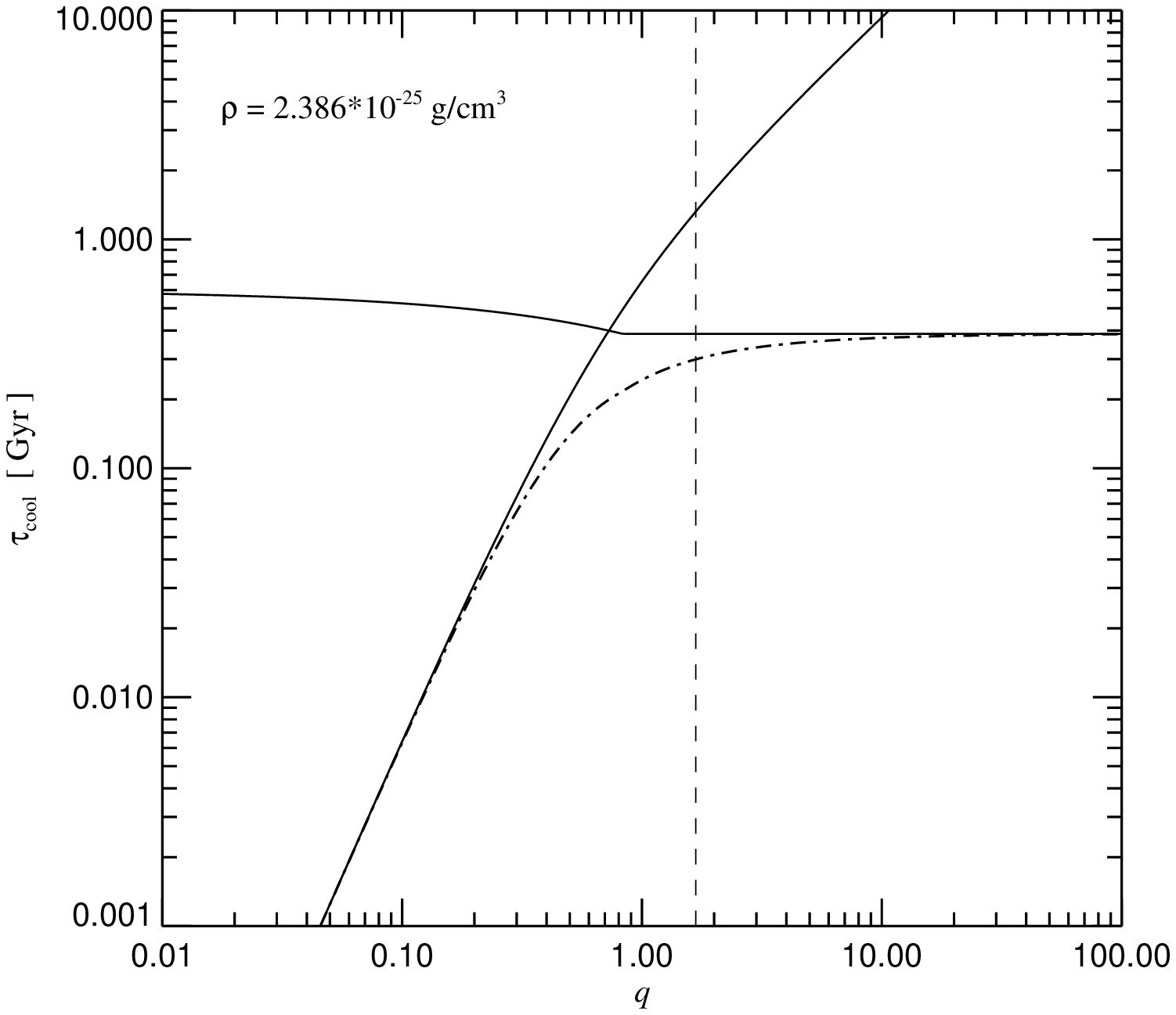}}
  \resizebox{7.7cm}{!}{\includegraphics{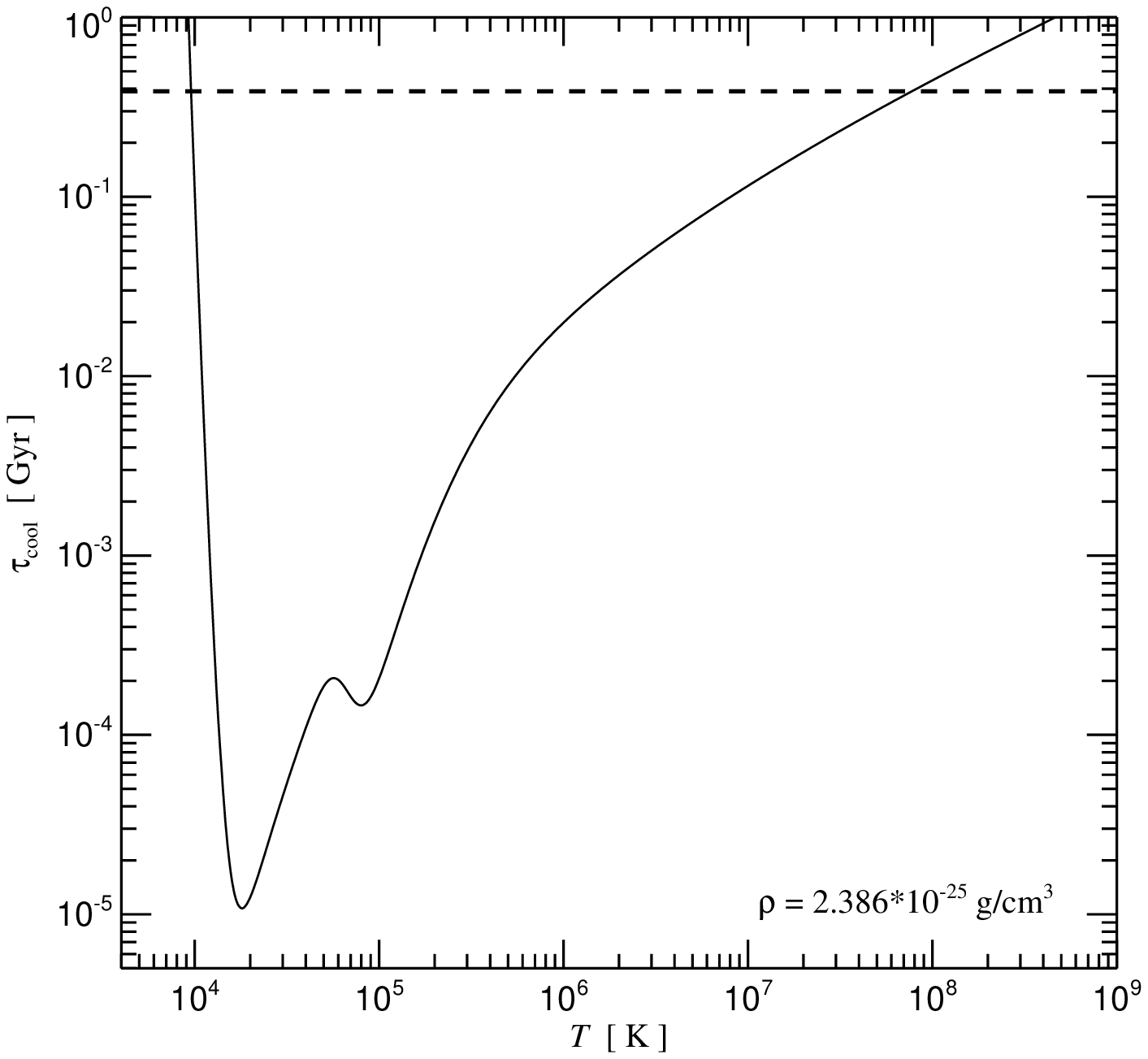}}\vspace*{-.3cm}
\end{center}
  \caption{The top panel shows the cooling times due to Coulomb losses (rising
    solid line) and hadronic dissipation (nearly horizontal line) as a
    function of the spectral cut-off. The dot-dashed line gives the total
    cooling time, while the vertical dashed lines marks the asymptotic
    equilibrium cut-off reached by the CR spectrum when no sources are present.
    The bottom panel shows the cooling time of ordinary thermal gas due to
    radiative cooling (for primordial metallicity), as a function of
    temperature. The horizontal dashed line marks the cooling time of CRs with
    a high momentum cut-off ($q\gg 1$), for comparison. In both panels, the
    times have been computed for a gas density of $\rho = 2.386\times
    10^{-25}\,{\rm g\,cm^{-3}}$, which corresponds to the density threshold
    for star formation that we usually adopt. Note however that the cooling
    times all scale as $\tau\propto 1/\rho$, i.e.~for different densities,
    only the vertical scale would change but the relative position of the
    lines would remain unaltered.
 \label{FigCoolTimes}}
\end{figure}

\subsection{Coulomb cooling and catastrophic losses}

Charged particles moving through a plasma will gradually dissipate their
kinetic energy and transfer it to the surrounding ions and electrons by
Coulomb interactions. The rate of this energy loss depends both on the
physical properties of the surrounding medium and on the detailed momentum
spectrum of the cosmic ray population. The latter in particular complicates an
accurate determination of the Coulomb loss rate.

Since particles with low momenta are most strongly affected by the Coulomb
interactions, a qualitative consequence of this effect is that it induces a
flattening of the spectrum; the high-momentum tail remains unchanged while the
low-momentum cosmic ray particles dissipate their energy effectively to the
thermal gas, and eventually drop out of the cosmic ray population altogether.

In our model, we have deliberately abandoned a detailed representation of the
cosmic ray spectrum of each fluid element, in favour of the high computational
speed and low memory consumption allowed by our simplified spectral model. We
even opted to use a globally fixed spectral index, which means that we cannot
represent a spectral flattening in detail.  However, we can account for the
effect of thermalization of the low-momentum particles in a simple and
efficient way.  To this end, we compute the energy loss by Coulomb-cooling
over the entire spectrum, and then shift the low-momentum cutoff of our
spectral model such that just the right amount of energy is removed in
low-momentum particles, keeping the high-momentum part unaffected.

More specifically, we follow \citet{Ensslin2005} and estimate the Coulomb loss
rate of the CR population as \bea \left(\frac{{\rm d}
    \tilde{\varepsilon}}{{\rm d} t}\right)_{\rm C} & = & - \frac{2\pi
  \tilde{C} e^4\, n_{\rm e}}{m_{\rm e} m_{\rm p} c} \left[ \ln\left( \frac{2
      m_{\rm e} c^2 \left<\beta p\right>}{\hbar
      \omega} \right) \times \right. \\
& & \left.
  \hspace*{-1.7cm}\B_{\frac{1}{1+q^2}}\left(\frac{\alpha-1}{2},-\frac{\alpha}{2}\right)
  -\frac{1}{2}\B_{\frac{1}{1+q^2}}\left(\frac{\alpha-1}{2},-\frac{\alpha-2}{2}\right)
\right].  \nonumber \eea Here $n_{\rm e}$ is the electron abundance, $\omega =
\sqrt{4\pi e^2 n_{\rm e}/m_{\rm e}}$ the plasma frequency, and $\left<\beta
  p\right> = 3 P_{\rm CR}/ (\rho \tilde n c^2)$ is a mean value for our
assumed spectrum.  We also define a cooling timescale due to Coulomb cooling
as \be \tau_{\rm C}(q) = \frac{\tilde{\varepsilon}} {\left|{{\rm
        d}\tilde{\varepsilon}}/{{\rm d} t}\right|_{\rm C}}, \ee which depends
only on the spectral cut-off $q$, and is inversely proportional to density,
modulo a very weak additional logarithmic density-dependence though the plasma
frequency.  For a given energy loss $\Delta \tilde{\varepsilon}_{\rm C}$ in a
timestep based on this rate, we can then estimate the corresponding change in
cosmic ray number density as
\begin{equation}
\Delta \tilde{n}_{\rm C} =
\Delta \tilde{\varepsilon}_{\rm C} \, \frac{m_{\rm
    p}}{T_{\rm p}(q)} \text{,}
\end{equation}
i.e.~we assume that the particles are removed at the low momentum
cut-off $q$.  From equations (\ref{eqnJacobi1}) and
(\ref{eqnJacobi2}), we can see that this will result in a gradual rise
of the spectral cutoff $q$ while the normalization will remain
unchanged. The corresponding energy loss $\Delta \tilde{\varepsilon}_{\rm C}$
is added to the thermal energy in the ordinary gas component, i.e.~the
Coulomb cooling process leaves the  total energy content of the
gas unaffected. We note that for large Coulomb cooling rates, we use
an implicit solver to compute the new position of the spectral
cut-off, leaving the amplitude parameter exactly invariant. This ensures
stability even if the cut-off increases substantially in one timestep.

Another class of loss processes for cosmic rays results occurs through
inelastic collisions with atoms of the ambient gas, resulting in the
hadronic production of pions, which subsequently decay into
$\gamma$-rays, secondary electrons, and neutrinos. In this case, the
energy is ultimately dissipated into photons which tend to escape.  So
here the net effect is a loss of energy from the system, unlike in the
case of Coulomb losses, where the dissipated cosmic ray energy heats
the thermal reservoir.

However, these `catastrophic losses' can only proceed efficiently when the
cosmic ray particles exceed the energy threshold of $q_{\rm thr} m_{\rm p} c^2
= 0.78\,{\rm GeV}$ for pion production ($q_{\rm thr}=0.83$).  The total loss
rate can then be described by \citep{Ensslin2005}
\begin{equation}
\left(\frac{\diff{\tilde{\varepsilon}}}{\diff{t}}\right)_{\rm had} = \;
- \frac{c \,\rho \overline{\sigma}_{\rm pp}\, \tilde{C}
\, q_\star^{1-\alpha}\,  T_{\rm  CR}(\alpha,q_\star)}
{2 (\alpha-1)\,m_{\rm p}^2} \,   \text{,}
\end{equation}
where $\overline{\sigma}_{\rm pp} \simeq 32 \,{\rm mbarn}$ is the
averaged pion production cross section, and $q_\star$ denotes $q_\star =
\max\left\{q,q_{\rm thr}\right\}$. The number density of cosmic ray particles stays
constant, however, due to conservation of baryon number in strong and
electroweak interactions, i.e.~we have $\Delta \tilde n_{\rm had}
=0$. This condition in turn implies that the changes of 
amplitude and cut-off of our
spectral model due to this cooling process are related by 
\be
\frac{\Delta\tilde C}{\tilde C} = (\alpha -1)\,\frac{\Delta q}{q}.
\ee
As before, we also define a  cooling timescale due to catastrophic losses, which is
given by
\be
\tau_{\rm had}(q) \equiv \frac{\tilde{\varepsilon}}
{\left|{{\rm d}\tilde{\varepsilon}}/{{\rm d} t}\right|_{\rm had}}
=
\frac{2\, m_{\rm p}}{c \rho \,\overline{\sigma}_{\rm pp}} \; \frac{T_{\rm
    CR}(\alpha, q)}{T_{\rm
    CR}(\alpha, q_\star)} \left(\frac{q}{q_\star}\right)^{1-\alpha}.
\ee
Note that $\tau_{\rm had}$ becomes constant for $q\ge q_{\rm thr}$.

In the top panel of Figure~\ref{FigCoolTimes}, we show the cooling timescales
for Coulomb and catastrophic losses as a function of the cut-off parameter
$q$, at a fiducial density corresponding to our star formation threshold,
assuming a spectral index $\alpha=2.5$. As expected, catastrophic losses
dominate for high momentum cut-offs, and therefore limit the lifetime of any
cosmic ray population to $\tau \sim 2\, m_{\rm p}/({c \rho
  \,\overline{\sigma}_{\rm pp}})$, unless an injection process provides a
resupply. For small cut-offs, Coulomb losses dominate and will rapidly
thermalize the low momentum cosmic rays. The dot-dashed line shows the total
loss timescale, defined by
\be
\frac{1}{\tau_{\rm losses}(q)} =  \frac{1}{\tau_{\rm C}(q)} + \frac{1}{\tau_{\rm had}(q)}\,.
\ee
This timescale is monotonically rising with $q$.

\begin{figure}
\begin{center}
\resizebox{8cm}{!}{\includegraphics{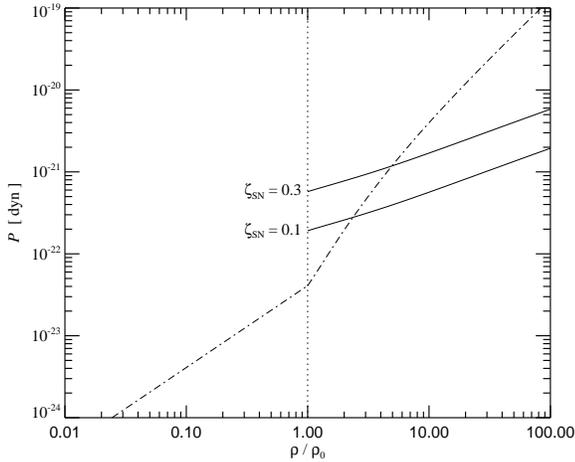}}
\end{center}
  \caption{Pressure of the cosmic ray population predicted for
    equilibrium between supernova injection on one hand, and Coulomb
    cooling and catastrophic losses on the other hand. The sold lines
    mark the pressure as a function of overdensity for two values of
    the injection efficiency $\zeta_{\rm SN}$. The assumed threshold
    density for star formation, $\rho_0 =2.4\times10^{-25}\,{\rm
    g\,cm^{-3}}$, is derived from the multiphase model of
    \cite{Springel2003}. The latter also predicts an effective
    equation of state for the star forming phase, shown as a
    dot-dashed line. The part below the threshold is an isothermal
    equation of state with temperature $10^4\,{\rm K}$.  \label{FigPessure}}
\end{figure}

In the absence of any injection, the cosmic ray population will
always evolve towards a fixed cut-off $q_{\rm fix}$, driven by the
tendency of Coulomb cooling to increase the cut-off, while catastrophic losses
tend to lower it.  A balance is reached at the solution of the equation \be 1
+\frac{\tau_{\rm C}(q_{\rm fix})}{\tau_{\rm had}(q_{\rm fix})} = \frac{T_{\rm
    CR}(\alpha,q_{\rm fix})}{T_{\rm p}(q_{\rm fix})}, \label{eqnEuil} \ee which
follows from equations (\ref{eqnJacobi1}) and (\ref{eqnJacobi2}).  The
vertical dotted line in Figure~\ref{FigCoolTimes} marks this equilibrium point
at $q_{\rm fix} = 1.685$ for $\alpha =2.5$. Once this fix-point is
reached, only the spectral amplitude decays due to the cosmic ray cooling and
dissipation processes.  Finally, note that both the Coulomb cooling and the
hadronic cooling time are inversely proportional to density. This also
means that the fix-point $q_{\rm fix}$ is density independent, but whether it
can be reached in the available time at low density is a separate question.

It is also interesting to compare the cosmic ray loss timescale to the thermal
cooling timescale of primordial gas. The latter is also inversely proportional
to density, but has in addition a strong temperature dependence. In the bottom
panel of Figure~\ref{FigCoolTimes}, we show the thermal cooling timescale as a
function of temperature, at the same fiducial density used in the top panel. For
comparison, we show the asymptotic cosmic ray dissipation time scale (dashed
line), which is reached if the cosmic ray population is dominated by the
relativistic regime. In this case, cosmic rays decay much slower than the
thermal gas pressure in the intermediate temperature regime. This could be
interesting for cooling flows in halos or clusters. Even a moderate cosmic ray
pressure contribution in the diffuse gas in a halo of temperature $T_{\rm
  vir}\sim 10^5 - 10^7\,{\rm K}$ should tend to survive
longer than the thermal gas pressure, which could influence the cooling rate.
We will examine this question explicitly in our numerical simulations of
isolated halos.

\subsection{Equilibrium between source and sink terms}

The above considerations lead to an interesting question: What will the cosmic
ray spectrum look like for a fluid element at a density $\rho$ high enough to allow
star formation, such that is fed at a constant rate by supernova injection?
We expect that after some time, a balance will be established between the
supernova input on one hand and the cosmic ray losses due to Coulomb cooling
and hadronic processes on the other hand. The energy content in the cosmic
rays at this equilibrium point will then determine the CR pressure, and
comparison of this pressure with the thermal pressure of the ISM will show
whether `cosmic ray feedback' could be important in regulating star formation
in galaxies.

To derive this equilibrium point, we first note that from the
conditions $\Delta\tilde\varepsilon_{\rm SN}
+\Delta\tilde\varepsilon_{\rm C} + \Delta\tilde\varepsilon_{\rm had}
=0$ and $\Delta\tilde n_{\rm SN} +\Delta\tilde n_{\rm
C}=0$\footnote{Note that we always have $\Delta\tilde n_{\rm had}=0$.}
it directly follows that the mean energy per injected particle due to
supernovae in equilibrium must be \be T_{\rm CR}(\alpha_{\rm SN}, q_{\rm inj}) =
T_{\rm p}(q_{\rm eq}) \left[ 1+ \frac{\tau_{\rm C}(q_{\rm
eq})}{\tau_{\rm had}(q_{\rm eq})}\right]. \ee 
This is a relation between the effective injection cut-off of the supernova
feeding, and the equilibrium cut-off $q_{\rm eq}$ of the CR spectrum. 
In equilibrium, we also know that the 
supernova input will just balance the cooling losses for the
spectrum with equilibrium cut-off $q_{\rm eq}$, yielding \be
\tilde\varepsilon = \tau_{\rm losses}(q_{\rm eq})\,f(q_{\rm SN}, q_{\rm inj}) \left(\frac{{\rm d}
\tilde\varepsilon}{ {\rm d} t}\right)_{\rm SN}. \label{EgyEgyEqual} \ee 
On the other hand, our injection criterion of equation (\ref{EqInjectCrit})
tells us that 
$\tilde\varepsilon = \tau_{\rm losses}(q_{\rm inj})\,f(q_{\rm SN}, q_{\rm inj}) \left(\frac{{\rm d}
\tilde\varepsilon}{ {\rm d} t}\right)_{\rm SN}$,
provided a solution for equation~(\ref{EqInjectCrit}) actually exists when the
system is in the dynamic equilibrium.
Assuming this for the moment, it follows that the CR
loss timescales at $q_{\rm inj}$ and $q_{\rm eq}$ are equal, which in turn
implies $q_{\rm inj} = q_{\rm eq}$.
In other words, the injection
cut-off coincides with the cut-off of the equilibrium spectrum.
The location of the equilibrium cut-off itself is given as solution
of \be 1 +\frac{\tau_{\rm C}(q_{\rm eq})}{\tau_{\rm had}(q_{\rm eq})}
= \frac{T_{\rm CR}(\alpha_{\rm SN},q_{\rm eq})}{T_{\rm p}(q_{\rm
eq})}. \label{EqnEquil} \ee This is almost identical to equation~(\ref{eqnEuil})
which describes the fix-point of the cut-off if there is no
injection. The only difference is the occurrence of $\alpha_{\rm SN}$
instead of just $\alpha$ in the argument of the $T_{\rm CR}$
function. The result of this will be a slight shift of the equilibrium
position once the assumed spectral indices for supernova injection and
the general cosmic ray population differ, while for $\alpha_{\rm SN}=
\alpha$, there will be no difference. At
first sight it may seem surprising  that the equilibrium position
can be shifted away from the fix-point by an {\em arbitrarily small} supernova injection
rate. However, recall that in the injection case we have described a
truly invariant spectrum with a fixed amplitude (which may take very
long time to be established), while in the case without injection, the
amplitude keeps falling on the cooling timescale.

The above assumed that the injection condition~(\ref{EqInjectCrit}) has a
solution in equilibrium. This will be the case if $q_{\rm eq}$ determined by
equation~(\ref{EqnEquil}) is smaller or equal than $q_{\rm max}$ as given by
equation (\ref{InjMax}).  Otherwise the injection cut-off is given by $q_{\rm
  inj}= q_{\rm max}$, and the position of the equilibrium cut off is
determined by equation~(\ref{EgyEgyEqual}). A detailed comparison of the
steady-state CR spectrum with and without our approximate description is
provided by \citet{Ensslin2005}. There it is shown that dynamically important
quantities like cosmic ray pressure and energy are calculated with $\sim 10\%$
accuracy by our formalism.

Of primary importance for us is the equilibrium value of the cosmic ray energy
content, because this will directly determine the CR pressure and hence the
strength of potential feedback effects on star formation. The total cosmic ray
energy injection rate by supernovae is related to the star formation rate by
equation (\ref{EqnSNinput}). Once the equilibrium cut-off is $q_{\rm eq}$ is
known, the energy content in equilibrium will be given by
equation~(\ref{EgyEgyEqual}), so that the pressure is fully specified. For
example, in the case of $q_{\rm
  eq}\le q_{\rm max}$,  the final pressure is therefore given by \be
P_{\rm CR} = \frac{ (\alpha-1) \beta_\alpha(q_{\rm eq})\, \tau_{\rm loss}(q_{\rm
    eq })\, f_{\alpha_{\rm SN}}(q_{\rm SN}, q_{\rm eq})}{3 \beta_\alpha(q_{\rm eq}) + 6\, q_{\rm
    eq}^{1-\alpha} ( \sqrt{1+q_{\rm eq}^2} -1)} \,\, \zeta_{\rm SN}
\epsilon_{\rm SN} \dot \rho_\star.  \ee

In our simulations, we combine the cosmic ray formalism with the
subresolution multiphase model for the regulation of star formation by
\citet{Springel2003}. In the latter, the mean star formation rate is
determined by the local density, and scales approximately as
$\dot\rho_\star \propto \rho^{1.5}$ above a density threshold for the
onset of star formation.
A detailed discussion of the multiphase model together
with the precise density dependence of the star formation rate is
given in \citet{Springel2003}. 
These authors also derive an effective
equation of state for the ISM, which governs the assumed two-phase
structure of the ISM. 

In Figure~\ref{FigPessure}, we show this effective pressure as a function of
density, using the parameters for gas consumption timescale, initial mass
function, and cloud evaporation efficiency discussed by \citet{Springel2003},
which result in a star formation threshold of $\rho_{\rm th}=2.4\times
10^{-25}\,{\rm g\,cm^{-3}}$. We also plot the
expected cosmic ray pressure in the same diagram, for two different injection
efficiencies of $\zeta_{\rm SN}=0.1$ and $\zeta_{\rm SN}=0.3$. Quite
interestingly, the cosmic ray pressure exceeds the thermal pressure at the
threshold density for star formation, but due to the shallow dependence of the
equilibrium pressure on density, with approximately $P_{\rm CR}\sim
\rho^{0.5}$ (note that $\tau_{\rm losses}\propto \rho^{-1}$), we find that the
pressure of cosmic rays only plays a role for the low density part of the ISM
model, while regions with very high specific star formation rates should be at
most weakly affected.

For small efficiencies $\zeta_{\rm SN}\le 0.01$, we essentially expect no
significant influence of cosmic rays from supernova on the regulation of star
formation whatsoever. Even for the fiducial choice of $\zeta_{\rm SN} = 1$,
the influence would vanish for densities $\rho > 100 \rho_{\rm th}$. Based on
these findings, we expect that galaxies which form most of their stars at
comparatively low densities should be potentially strongly affected by the
cosmic ray feedback, while this influence should be weak or absent for
vigorously star-forming galaxies with higher typical ISM densities. Our
numerical results presented later will confirm this picture. The fact that
CR-pressure seems to become dominant around the star formation threshold may
even suggest that cosmic rays could play an active role in defining this
threshold.

\section{Cosmic ray diffusion}
\label{section:diffusion}

Our treatment is based on the notion that the cosmic ray population is
approximately locked into a gas fluid element by a locally present
magnetic field. Even a weak ambient field makes the charged particle
gyrate around the field lines, preventing them from freely travelling
over macroscopic distances with their intrinsic velocity close to to
the speed of light. The cosmic ray particles may move slowly
along the field lines, but their perpendicular transport is strongly
suppressed.

However, occasional scattering of particles on magnetic irregularities of MHD
waves can displace the gyrocentres of particles, such that a particle
effectively ``changes its field line'', allowing it to move perpendicular to
the field. Since realistic magnetic field configurations are often highly
tangled, or even chaotic, this can lead to sizable cross-field speeds of the
particles. From a macroscopic point of view, the motion of the cosmic ray
population can be described as a diffusion process, which is anisotropic with
respect to the local magnetic field configuration. The theory of the
respective diffusion coefficients is complicated, and uncertain for the case
of turbulent magnetic field configurations
\citep[e.g.][]{Rechester1978,Duffy1995,Bieber1997,Giacalone1999,Narayan2001,Ensslin2003}.

In principle, one could try to simulate the magnetic field in SPH and then
treat the diffusion in an anisotropic fashion. While recent advances in
modelling MHD with SPH are promising \citep{Dolag1999,Price2004}, these
techniques still face severe numerical challenges when applied to simulations
with radiative cooling. We therefore defer such an approach to future work and
model the diffusion isotropically. Also, since we have no direct information
about the local magnetic field strength and field configuration, we will
invoke a phenomenological approach to estimate the effective diffusion
coefficient as a function of the local conditions of the gas. It would be easy
however to refine the spatial dependence of the diffusion coefficient once a
more detailed physical scenario becomes available.

Assuming isotropicy, the diffusion in the CR distribution function $f(p,
\vec{x}, t)$ can be written as \be \frac{\partial f}{\partial t} =
\nablavec(\kappa\nablavec\,f). \label{eqndiffbase} \ee The diffusion
coefficient $\kappa$ itself depends on the particle momentum (more energetic
particles diffuse faster), and on the local magnetic field configuration. For
definiteness we assume that the underlying MHD is turbulent with a Kolmogorov
power spectrum, in which case the momentum dependence of $\kappa$ is given by
\be \kappa(p)= \tilde \kappa \,p^{\frac{1}{3}}, \ee where we assumed
relativistic particle velocities with $v\simeq c$. Integrating the diffusion
equation over particle momenta then yields \be \left. \frac{\partial n_{\rm
      CR}}{\partial t}\right|_{\rm diff} = \frac{\alpha-1}{\alpha-\frac{4}{3}}
\,\, \nablavec \tilde\kappa \nablavec (q ^{\frac{1}{3}} n_{\rm CR}), \ee where
$q$ is the spectral cut-off. Because more energetic particles diffuse faster,
we expect that the diffusion speed for the cosmic ray energy density will be a
bit higher than for the number density itself. To account for this effect, we
first multiply the diffusion equation (\ref{eqndiffbase}) with $T_{\rm p}(p)$,
and then integrate over the particle momenta. This results in \be \left.
  \frac{\partial \varepsilon_{\rm CR}}{\partial t}\right|_{\rm diff} =
\frac{\alpha-1}{\alpha-\frac{4}{3}} \,\, \nablavec \tilde\kappa \nablavec
\left(q ^{\frac{1}{3}} \frac{ T_{\rm CR} (\alpha-\frac{1}{3},q)}{T_{\rm
      CR}(\alpha, q)}\, \varepsilon_{\rm CR}\right), \ee where
$\varepsilon_{\rm CR} = \rho \tilde\varepsilon$ is the cosmic ray energy
density per unit volume.  The factor ${ T_{\rm CR} (\alpha-\frac{1}{3},q)} /
{T_{\rm CR}(\alpha, q)}$ is larger than unity and encodes the fact that the
diffusion in energy density proceeds faster than in particle number density.
In \citet{Ensslin2005}, we also give more general formulae for different
power-law dependences of the diffusivity, and provide a more accurate
treatment where the reduction of the diffusion rate at sub-relativistic energies is
accounted for.

\subsection{Modelling the diffusivity}

Due to the lack of direct local information about the magnetic field
strength in our present numerical models, we parameterize the dependence
of the diffusion coefficient on local gas properties in terms of a
fiducial power-law dependence on the local gas density and gas
temperature. In particular, we make the ansatz
\begin{equation}
\tilde\kappa = \kappa_0 \left(\frac{\rho}{\rho_0}\right)^{n_{\rho}}
\left(\frac{T}{T_0}\right)^{n_{ T}},
\end{equation}
for the diffusion constant, which is effectively a three parameter
model for the diffusivity, specified by the overall strength
$\kappa_0$ of the diffusion at a reference density and temperature,
and by the power-law slopes $n_{\rho}$ and $n_{T}$ for the density
and temperature dependence, respectively.

While clearly a schematic simplification, this parameterization is general
enough to allow an analysis of a number of interesting cases, including models
where the typical magnetic field strength has a simple density dependence,
which can be a reasonable first order approximation for some systems, for
example for the diffuse gas in cluster atmospheres \citep{Dolag2004}.

For definiteness, we now construct such a very simple model, which
will be the fiducial choice for the results on diffusion presented in
this study. In Kolmogorov-like MHD turbulence, the dominant parallel diffusivity is expected
to scale as \citep{Ensslin2003}
\begin{equation}
\tilde \kappa \propto {l_B}^{2/3}\,
B^{-1/3} \text{,}
\label{eqn:kappa_approx}
\end{equation}
where $l_B$ gives a characteristic length scale
for the magnetic field of strength $B$. We start out by assuming that
the magnetic energy density is a fixed fraction of the thermal energy
content, which corresponds to
\begin{equation}
B \propto \rho^{1/2} T^{1/2} \text{.}\label{eqnscaling1}
\end{equation}
An appropriate length scale for $l_B$ is more difficult to estimate,
as it will sensitively depend on the level of local MHD turbulence,
and on the build up of the magnetic field due to structure formation
processes. For simplicity, we here assume that the length scale is
related to the local Jeans scale, which may be appropriate for the
conditions of a multiphase interstellar medium where local density
fluctuations constantly form clouds of order a Jeans mass, which are
then in part dispersed by supernova-driven turbulence. This then gives
a scaling of the form
\begin{equation}
l_B \propto \rho^{-1/2} T^{1/2} \text{.}\label{eqnscaling2}
\end{equation}

Combining equations (\ref{eqnscaling1}) and (\ref{eqnscaling2}), we
obtain a model for the conductivity in the form $\tilde\kappa \propto
T^{1/6} \rho^{-1/2}$, i.e.~$n_T =1/6$ and $n_\rho = -1/2$. We fix the
overall strength by alluding to measurements in our own Galaxy
\citep{Berezinskii1990,Schlickeiser2002}, which estimate a diffusivity
along the magnetic field lines in the interstellar medium of
approximately
\begin{equation}
\kappa_{\rm ISM} \approx 3 \times 10^{27} \frac{\unit{cm}^2}{\unit{s}}
\approx 10\, \frac{\unit{kpc}^2}{\unit{Gyr}}
\end{equation}
Adopting our typical temperature and density values of the ISM as
reference values, we end up with the following model for the
diffusivity
\begin{equation}
\kappa = 10 \frac{\unit{kpc}^2}{\unit{Gyr}}
\left(\frac{T}{10^4\unit{K}}\right)^{1/6} \left(
  \frac{\rho}{10^{6}\Msol\unit{kpc}^{-3}} \right)^{-1/2}
\label{eqndiffusivity}
\end{equation}
We will use this parameterization in the simulations with diffusion
analysed in the this study. It is clear however that this model needs
to be interpreted with a lot of caution, as the real diffusivity is
highly uncertain, and may vary widely between different parts of the
Universe. Improving the physical understanding of the strength of the
diffusion will therefore remain an important goal for cosmic ray
physics in the future.

\subsection{Discretizing the diffusion equation}

We still have to discuss how we numerically implement diffusion in our
Lagrangian SPH code. We here follow a similar strategy as in
\citet{Jubelgas04}, where a new treatment of thermal conduction in SPH
was discussed and applied to simulations of clusters of galaxies.  In
essence, the very same techniques that can be used to solved the heat
diffusion equation can also be applied to the cosmic ray diffusion
needed here.

We first rewrite the diffusion equations into a Lagrangian form that
is matched to the variables evolved in our simulation code. This
results in
\be
\rho \frac{{\rm d}\tilde \varepsilon}{{\rm d}t} = \nablavec \tilde
\kappa\nablavec D_\varepsilon
\ee
and
\be
\rho \frac{{\rm d}\tilde n}{{\rm d}t} = \nablavec \tilde
\kappa\nablavec D_n ,
\ee
where we have defined the abbreviations
\be
D_\varepsilon = \frac{\alpha-1}{\alpha-\frac{4}{3}} \rho \, q^{1/3}
\frac{ T_{\rm CR} (\alpha-\frac{1}{3},q)}{T_{\rm CR}(\alpha, q)}
\,\tilde\varepsilon
\ee
and
\be
D_n = \frac{\alpha-1}{\alpha-\frac{4}{3}} \rho \, q^{1/3}
\, \tilde n
\ee
respectively. Our method for representing these equations in SPH
is based on a discretisation scheme for the Laplace
operator that avoids second order derivatives of the SPH kernel
\citep{Brookshaw85,Mo92}, which makes the scheme robust against
particle disorder and numerical noise. 
This gives us an evolution equation in the form of
\begin{equation}
\frac{\diff{\tilde{\varepsilon}_i}}{\diff{t}} = \sum_j
\frac{m_j}{\rho_i \rho_j} \frac{2 \overline{\kappa}_{ij} (
\overline{D}_{\varepsilon,j} - D_{\varepsilon,i} )}{|\vec{x}_{ij}|^2}
\vec{x}_{ij} \nablavec_i W_{ij},
\label{eqn:energydiff}
\end{equation}
and similarly for the cosmic ray number density. We here introduced a
symmetrization of the diffusivities according to
\begin{equation}
\overline{\kappa}_{ij} = 2 \frac{ \tilde\kappa_i \tilde\kappa_j }{ \tilde\kappa_i + \tilde\kappa_j },
\end{equation}
based on the suggestion by \citet{Cleary99}. Furthermore, we replaced 
one of the $D_{\varepsilon}$ terms in the pairwise diffusion term by
the kernel interpolant
\begin{equation}
\overline{D}_{\varepsilon,j} = \sum_k \frac{m_k D_{\varepsilon,k}}{\rho_k} W_{jk} \text{.}
\end{equation}
As \citet{Jubelgas04} show, such a mixed formulation between
intrinsic particle variables and SPH-smoothed interpolants
substantially improves the numerical stability against small-scale
particle noise. The smoothing process suppresses strong small-scale
gradients, while long-range variations and their diffusive evolution
remain unchanged.  Since we use an explicit time integration scheme,
this behaviour prevents stability problems due to the typical
`overshooting' problem that otherwise may arise due to strong local
gradients from local outliers.

Nevertheless, we still need to impose a comparatively strict timestep
criterion to ensure proper integration of the diffusion.  For the
thermal conduction problem, we employed a simple criterion that limits
the relative change of thermal energy of a particle within a single
timestep. Although the diffusion studied here is in principle very
similar to the conduction problem, an equivalent criterion
is not a good choice, simply because unlike for thermal conduction,
the relevant reservoir can often be empty. In fact, we typically start
simulations from initial conditions where all the cosmic ray particle
densities are identical to zero.

However, a closer look at the Green's function $G(\vec{x},t) = (4 \pi
\kappa t)^{-3/2} \exp [ -\vec{x}^2 /(4 \kappa t) ]$ of the diffusion
process shows that differences between two points with a distance of
$|\vec{x}|$ are diffused away with a characteristic timescale
$\vec{x}^2/\kappa$. Using the mean interparticle separation of SPH
particles for the distance, this suggests the definition of a
diffusion timescale in the form
\begin{equation}
  t_{\rm diff} = \frac{1}{\kappa} \left( \frac{m}{\rho} \right)^{2/3}.
\label{eqn:diffusive_timescale}
\end{equation}
We use this to limit the maximum timestep for particles to be
constrained by
\begin{equation}
  \Delta t < \varepsilon \ t_{\rm diff} = \varepsilon \frac{1}{\kappa}
  \left( \frac{m}{\rho} \right)^{2/3}
\end{equation}
where we used $\varepsilon = 0.1$ for the simulations presented in
this study. This has provided us with a
numerically stable cosmic ray diffusion while at the same time are
prevented from becoming impractically small.

As an additional refinement to the implementation of diffusion, we have
implemented the method proposed by \citet{Jubelgas04} to obtain a manifestly
conservative scheme for cosmic ray energy and particle number, even when
individual and adaptive timesteps are used. To this end, we rewrite equation
(\ref{eqn:energydiff}) as 
\begin{equation}
m_i\frac{\diff{\tilde{\varepsilon}_{i}}}{\diff{t}} = \sum_j \frac{{\rm
    d}E_{ij}}{{\rm d}t},
\end{equation}
where we have defined a pairwise exchange term of cosmic ray energy in the
form
\begin{equation}
\frac{{\rm
    d}E_{ij}}{{\rm d}t} = \frac{m_i m_j}{\rho_i \rho_j} \frac{2 \overline{\kappa}_{ij} ( \overline{D}_{\varepsilon,j} - D_{\varepsilon,i} )}{|\vec{x}_{ij}|^2} \vec{x}_{ij} \nablavec_i W_{ij}.
\end{equation}
In each system step, we now determine the change of the cosmic ray energy of
particle $i$
according to
\begin{equation}
m_i \Delta \tilde{\varepsilon}_i = \frac{1}{2} \sum_{jk} \Delta t_j (\delta_{ij} - \delta_{ik}) \frac{\diff{E_{jk}}}{\diff{t}}.
\end{equation}
The double sum on the right can be simply evaluated by the ordinary SPH sums
over the active particles, provided that for each neighbour $j$ found for a
particle $i$ one records a change of $\Delta t_iE_{ij}/2 $ for $i$, and a
change of $- \Delta t_iE_{ij}/2 $ for the particle $j$.  We then apply the
accumulated changes of the cosmic ray energy (or particle number) to {\em all}
particles at the end of the step, i.e.~not only to the ones that are active on
the current timestep. In this way, we arrive at a scheme that manifestly
conserves total cosmic ray energy and number density for each diffusive step.

Finally, we note that we have implemented a further refinement in order to
cope with technical problems associated with the situation in which isolated
CR-pressurized particles are embedded in a background of particles with zero
CR pressure. In the neighbourhood of such an isolated particle, the smoothed
cosmic ray energy field $\overline{D}_{\varepsilon,j}$ will be non-zero for
particles that have themselves no CR component (yet). This can then in turn
lead to exchange terms $E_{ij}$ between particles which both have zero cosmic
ray pressure, leading to unphysical negative values for the energy in one of
them. We found that this can be avoided if we limit the value of the
interpolant $\overline{D}_{\varepsilon,i}$ to be no more than a factor $\chi\simeq
2.0$ larger than the value ${D}_{\varepsilon,i}$ for the particle itself.
With this change, we recovered numerical stability of the diffusion in this
situation.

\section{Numerical details and tests}

\begin{figure}
\resizebox{8cm}{!}{\includegraphics{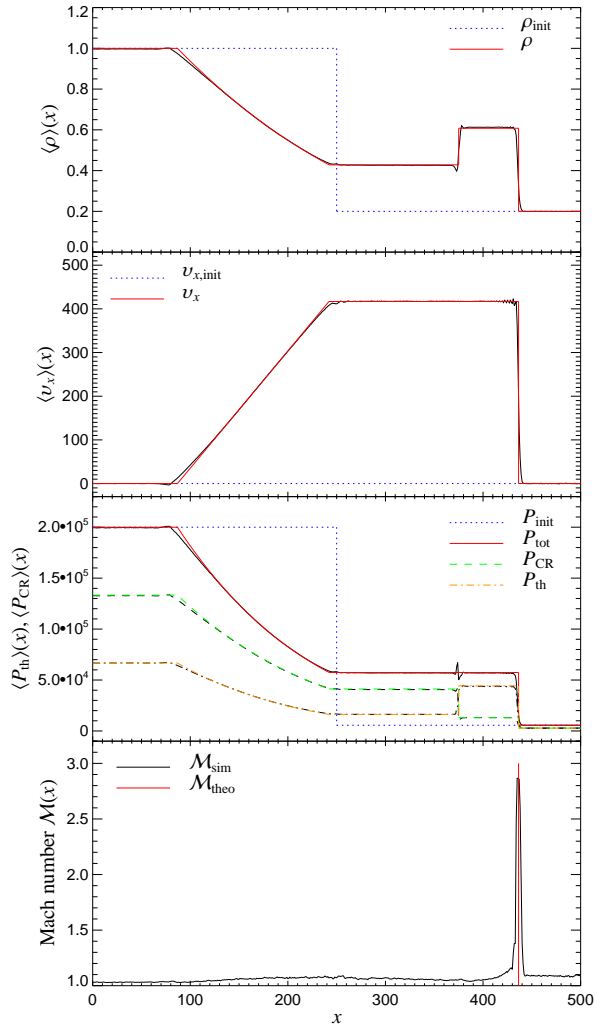}}
\caption{Shock-tube test for a gas with thermal and cosmic ray
  pressure components.  The simulation is carried out in a three-dimensional
  periodic box which is longer in the $x$-direction than in the other two
  dimensions.  Initially, the relative CR pressure is $X_{\rm CR} = P_{\rm CR}
  / P_{\rm th} = 2$ in the left half-space ($x<250$), while we assume pressure
  equilibrium between CRs and thermal gas for $x>250$. The evolution then
  produces a Mach number ${\mathcal M}=3$ shock wave.  The numerical result of
  the volume averaged hydrodynamical quantities $\bra \rho(x)\ket$, $\bra
  P(x)\ket$, $\bra v_x(x)\ket$, and $\bra {\mathcal M}(x)\ket$ within bins
  with a spacing equal to the interparticle separation of the denser medium is
  shown in black and compared with the analytic result shown in colour.  }
  \label{fig:shocktube}
\end{figure}

The numerical framework for cosmic ray physics presented above allows
for very complex dynamics that interacts in non-linear and rather
non-trivial manner with different aspects of ordinary hydrodynamics,
and in particular with the physics of radiative cooling and star
formation included in \gadget{} to describe galaxy formation. Together
with the nearly complete absence of analytic solutions, this makes a
direct validation of the numerical implementation of cosmic ray
physics in the simulation code particularly hard.

However, there are a few areas where the careful checks of
individual subroutines of the code that we carried out can be
augmented with tests of problems where analytical solutions are
known. One such area are hydrodynamical shock waves that involve a
cosmic ray pressure component.  This allows us to test one of the most
interesting dynamical aspects of cosmic ray physics which is the
introduction of a variable adiabatic index $\gamma$.  Note that the
pressure of a `hybrid gas' with a thermal and a cosmic ray energy
density can no longer be described with the simple parameterizations
used for a polytropic gas. In particular, both the relative
contribution of the cosmic ray pressure and the adiabatic index of the
CR pressure component itself will change during an adiabatic
compression. While more complex than for an ideal gas, the Riemann
problem for a shocks in such a composite gas can be solved
analytically \citep{Pfrommer2005}, and we will use this as a test for
our numerical treatment.

Another aspect which can be tested with simple toy set-ups is the
diffusion of cosmic rays. To this end we will consider a geometrically
simple initial cosmic ray distribution, together with the gas being
forced to be at rest.  This allows us to test the correct diffusion
speed, and the conservative properties of the diffusion process.

\subsection{Implementation issues} \label{SecNumDetails}

We note that the treatment of a new physical process in SPH often requires a
modification of the timestep criterion to ensure proper time integration of
the added physics. In the case of cosmic rays, it turns out that an additional
limit on the timestep is not really required, because the Courant criterion
for hydrodynamics is already general enough and automatically adjusts the
timestep if the cosmic ray dynamics requires it, as the latter then induces a
large sound speed due to extra cosmic ray pressure.  The non-adiabatic source
and sink terms on the other hand are essentially encapsulated in subresolution
models, and are therefore not particularly demanding with respect to the
integration timestep.

As can be seen in equations (\ref{eqenergy}) and (\ref{eqpressure}),
calculating physical properties like pressure and specific energy of the
cosmic ray component involves the evaluation of incomplete beta functions,
which can be rather expensive numerically.  Also note that in the individual
timestep scheme of \gadget{}, which is essential for simulations with a large
dynamic range, the pressure for all SPH particles needs to be set at every
system step, even if a particle is `passive' and does not receive a force
computation itself in the current step.  Clearly, a costly evaluation of
special functions for the pressure would therefore imply a significant burden
in terms of processor time.  We have therefore implemented a series of look-up
tables discretized in $\log q$ that allow us a fast evaluation of terms
involving costly incomplete beta functions. Interpolating from these tables
allows a rapid and accurate computation of all cosmic ray related quantities
without special function evaluations during the simulation.

\begin{figure*}
\begin{center}
  \includegraphics[width=16cm]{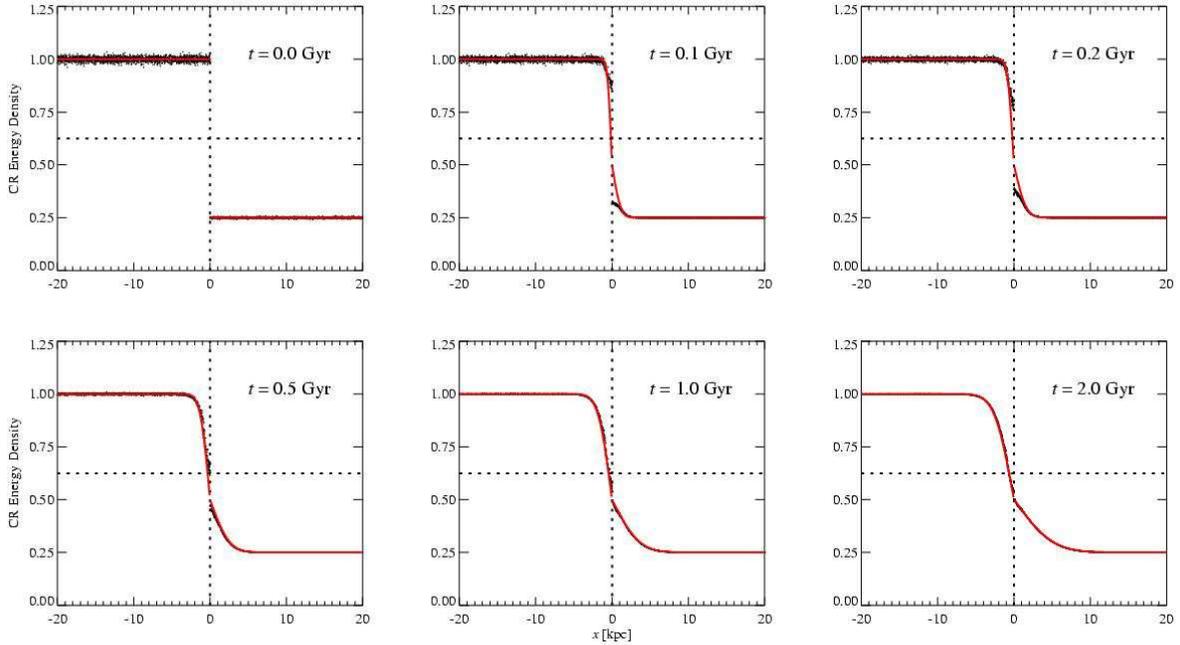}
\end{center}
  \caption{ Time evolution of a step function in cosmic ray energy
    density due to diffusion (the spectral cut-off and slope of the
    cosmic rays are constant throughout the volume). The population of
    SPH particles is kept at rest.  The times shown in the different
    panels are (from top left to bottom right): $t = 0$, $0.1$, $0.2$,
    $0.5$, $1.0$ and $2.0\Gyr$. Black dots give particle values at the
    corresponding time, while the red line shows the analytical
    solution. The diffusivity $\tilde \kappa$ is constant at $1 \,
    {\rm kpc^2 \, Gyr^{-1}}$ on the left hand side, and four times
    higher on the right hand side.}
  \label{fig:diffusion-test}
\end{figure*}

The same technique is also used to numerically invert the specific energy
$T_{\rm CR}$ of equation (\ref{eqenergy}) for $q$, a task that arises when the
mean CR energy is updated.  Here we again resort to a pre-computed look-up
table in which we locate the new mean CR energy, and then interpolate in the
table to determine the spectral cutoff $q$.  Once the spectral cutoff and
injected CR-to-baryon fraction is found, the new spectral normalization
$\tilde{C}$ can be computed from equation~(\ref{eqn:baryonfraction}). As a
final step, we can then update the effective hydrodynamic pressure due to
cosmic rays of the particle in question.
 
For treating Coulomb losses in an accurate and robust way, we use an implicit
scheme because of the very sensitive dependence of the cooling rate on the
spectral cut-off $q$. In fact, in order to ensure that this cooling process
leaves the normalization of the spectrum constant and only increases the
amplitude $q$, we solve the following implicit equation \be
\tilde{\varepsilon} (q', \tilde{ C}) = \tilde{\varepsilon} (q, \tilde{ C}) -
\frac{\tilde{\varepsilon} (q', \tilde{ C})} {\tau_{\rm C}(q')} \Delta t \ee
for a new spectral cut-off $q'$ when the cooling lasts for a time interval
$\Delta t$. This scheme is robust also in cases where $\Delta t$ exceeds
$\tau_{\rm C}(q)$.  Unlike the Coulomb losses, the timescale of hadronic
losses is comparatively long and varies little with the spectral cut-off, as
seen in Figure~\ref{FigCoolTimes}. It therefore is easier to integrate
accurately and does not require an implicit solver.

\subsection{Shocks in cosmic ray pressurized gas}

We performed a number of three-dimensional test simulations that
follow a shock wave in a rectangular slab of gas, which is extended
along one spatial dimension. Periodic boundary conditions in the
directions perpendicular to this axis are used to make sure that no
boundary-effects occur. This allows us to simulate a planar shock in
3D which can then be compared to a corresponding one-dimensional
analytic solution. The initial conditions of our shock-tube tests were
set-up with relaxed `glass' structures of particles initially at rest,
and by giving the two halves of the slab different temperatures and
cosmic ray pressures. The particle mass was constant and chosen such
that the mean particle spacing was $1$ length unit in the high-density
regime.  While a reduction of the particle mass in the low-density
region would have resulted in an increased spatial resolution there,
our constant particle mass set-up is more appropriate for the
conditions encountered in cosmological simulations.

In Figure~\ref{fig:shocktube}, we show the state of the system after a
time of $t=0.3$, for an initial density contrast of 5, a total
pressure ratio of 35.674, a homogeneous mixture of $1/3$ cosmic ray
pressure and $2/3$ thermal pressure contribution on the left-hand
side, and pressure equilibrium between CRs and thermal gas on the
right-hand side. A shock with Mach number $M=3$ travelling to the
right, a rarefaction wave to the left, and a contact discontinuity in
between develop for these initial conditions. The analytically
predicted shock front position and density distribution are matched
nicely by the simulation. Due to the smooth nature of SPH simulations,
the density jump at the shock at $x\simeq 430$ is not a sharp
discontinuity, but stretches over a small number of mean interparticle
spacings. The contact discontinuity at $x\simeq 375$ is reproduced
well, with only a small `blib' seen in both the density and the
pressure profile, which is characteristic for SPH in shock tube
tests. Note that cosmic ray pressure dominates over thermal pressure
on the left side of the contact discontinuity due to adiabatic
rarefaction of the initially CR-dominated state on the left-hand side,
while this is reversed on the other side because CRs are adiabatically
compressed at the shock while the thermal gas experiences entropy
injection. The rarefaction wave traveling to the left shows the
expected behaviour over most of its extent, only in the leftmost parts
at $x\simeq 100$ some small differences between the analytic and
numerical solution are seen. However, the overall agreement is very
reassuring, despite the fact that here a shock in a composite gas was
simulated.

This shows that the simulation code is able to correctly follow rapid
compressions and rarefactions in a gas with substantial cosmic ray
pressure support, including shocks that feed their dissipated energy
into the thermal component.  In cosmological simulations where
diffusive shock acceleration of cosmic rays is included, some of the
dissipated energy will instead be fed into the cosmic ray reservoir,
so that there the resulting shock behaviours can be yet more
complicated. We also note that our shock detection technique
\citep{Pfrommer2005} is able to correctly identify the shock location
on-the-fly during the simulation, and returns the right Mach number in
the peak of the shock profile, where most of the energy is
dissipated. We can use this to accurately describe the Mach-number
dependent shock-injection efficiency of cosmic rays in shocks.

\subsection{Cosmic ray diffusion}

In order to test the diffusion part of the code in a clean way, we use a
system of gas particles that are at rest, which avoids the complications that
would otherwise occur due to the motions of particles.  We achieve this by
setting all particle accelerations to zero, such that the gas in fact behaves
essentially like a solid body. As a side effect, the densities remain constant
over time, such that all variations in the distribution of cosmic rays are
entirely due to diffusive transport (we also switched of the Coloumb and
catastrophic losses for this test). For this idealized situation, analytic
solutions for the diffusion problem can be derived which can then be readily
compared with numerical results.

For definiteness, we set up a periodic slab of matter with density
$1\times 10^{10}\Msol\kpc^{-3}$, spanning a basic volume of $10 \times
10 \times 100\kpc$. The periodicity across the short dimensions
ensures the absence of boundary effects, such that we can compare the
numerical results to effectively one-dimensional analytic solutions.
The cosmic ray distribution was initialized such that the energy
density due to relativistic particles has a sharp step. The spectral
cutoff at both sides of the step was set equal to $q = 0.3$ in the
test discussed here, but we note that similar results are also
obtained for different choices. Again, we used an irregular glass-like
configuration as initial particle distribution in order to more
realistically model the noise properties in density fields encountered
in cosmological applications. Note that small-scale numerical noise
can be problematic for the treatment of diffusion
\citep[e.g.][]{Jubelgas04}, so this is an important aspect for testing
the robustness of the scheme.

In real physical applications, the diffusion implementation will have
to deal with a spatially varying diffusion coefficient. In particular,
there will be steep gradients in the diffusivity at phase transition
between the cold, dense gas and the hot, yet thin ambient
intergalactic and intra-cluster medium. It is therefore advisable to
verify that the implemented numerical scheme for the diffusion is well
behaved at sharp jumps of the diffusivity.  We incorporate this aspect
into our test scenario by setting up a fiducial temperature-dependent
cosmic ray diffusivity of
\begin{equation}
 \tilde \kappa = 1.0 \frac{\kpc^2}{\Gyr}
  \left(\frac{T}{1000\unit{K}}\right)
\end{equation}
and varying the gas temperature from $1000\unit{K}$ in the left half
of the matter slab to $4000\unit{K}$ in the right, causing an increase
of the diffusivity by a factor of 4 across the $x = 0$ plane.  Of
course, this particular choice of conductivity is arbitrary, but the
chosen values are not too dissimilar compared with what we will
encounter in cosmological simulations later on.

In Figure~\ref{fig:diffusion-test}, we present the time evolution for
diffusion of an initial step function.  The simulation was run over a
time span of $2\,\unit{Gyr}$, and for a number of times in between, we
compare the spatial distribution of the cosmic ray energy density
obtained numerically with the analytical solution~\citep{Pfrommer2005}
for the problem (shown in red).  The match of the numerical result and
the analytic solution is very good, especially at late times.  In
fact, after $t = 1\Gyr$, we no longer see any significant deviation
between the numerical solution and the analytical one. The code
reliably traces the flattening of the cosmic energy density jump over
time.  The largest differences occur in the very early phases of the
evolution, at around the initial discontinuity, as expected. Due to
the smoothing inherent in SPH and our diffusion formulation, sharp
gradients on very small-scales are washed out only with some delay,
but these errors tend to not propagate to larger scales, such that the
diffusion speed of large-scale gradients is approximately correct.

Note that small-scale noise present in the initial cosmic ray energy
distribution is damped out with different speeds in the left and right
parts of the slab. This is due to the different conductivities in the
low- and high-energy regimes, which give rise to characteristic
diffusion timescales of $t_{\rm diff} = 1\Gyr$ and $t_{\rm diff} =
0.25\Gyr$, respectively, for our mean interparticle separation of
$1\,\kpc$, consistent with Eqn.~(\ref{eqn:diffusive_timescale}).  We
note that we have verified the good accuracy of the diffusion results
for a wide range of matter densities and diffusivities, including also
cases with stronger spatial variations in diffusivity.  We are hence
confident that our numerical implementation should produce accurate
and robust results in full cosmological simulations, where the
diffusivity can show non-trivial spatial dependences.

\section{Simulations of isolated galaxies and halos}

We now turn to a discussion of the effects of our cosmic ray model on the
galaxy formation process.  Due to the complexity of the involved physics,
which couples radiative cooling, star formation, supernova feedback, cosmic
ray physics, self-gravity, and ordinary hydrodynamics, it is clear however
that our analysis cannot be fully exhaustive in this methodology paper.
Instead, our strategy is to provide a first exploration of the most important
effects using a set of simulations with idealized initial conditions, and a
restricted set of full cosmological simulations. This can then guide further
in-depth studies of the individual effects.

One of the possible effects of cosmic ray physics is that the injection of CRs
due to supernovae may alter the regulation of star formation by feedback,
which may directly translate into observable differences in forming galaxies.
Since CR-pressurized gas has a different equation of state than ordinary
thermal gas, it may rise buoyantly from star-forming regions, which could
perhaps help to produce outflows from galactic halos.  Also, because energy
stored in cosmic rays will be subject to different dissipative losses than
thermal gas, we expect that the radiative cooling of galaxies could be
altered.  Of special importance is also whether the strength of any of these
effects shows a dependence on halo mass, because a change of the efficiency of
galaxy formation as a function of halo mass is expected to modify the shape of
the resulting galaxy luminosity function.

\begin{figure*}
\begin{center}
\resizebox{7.5cm}{!}{\includegraphics{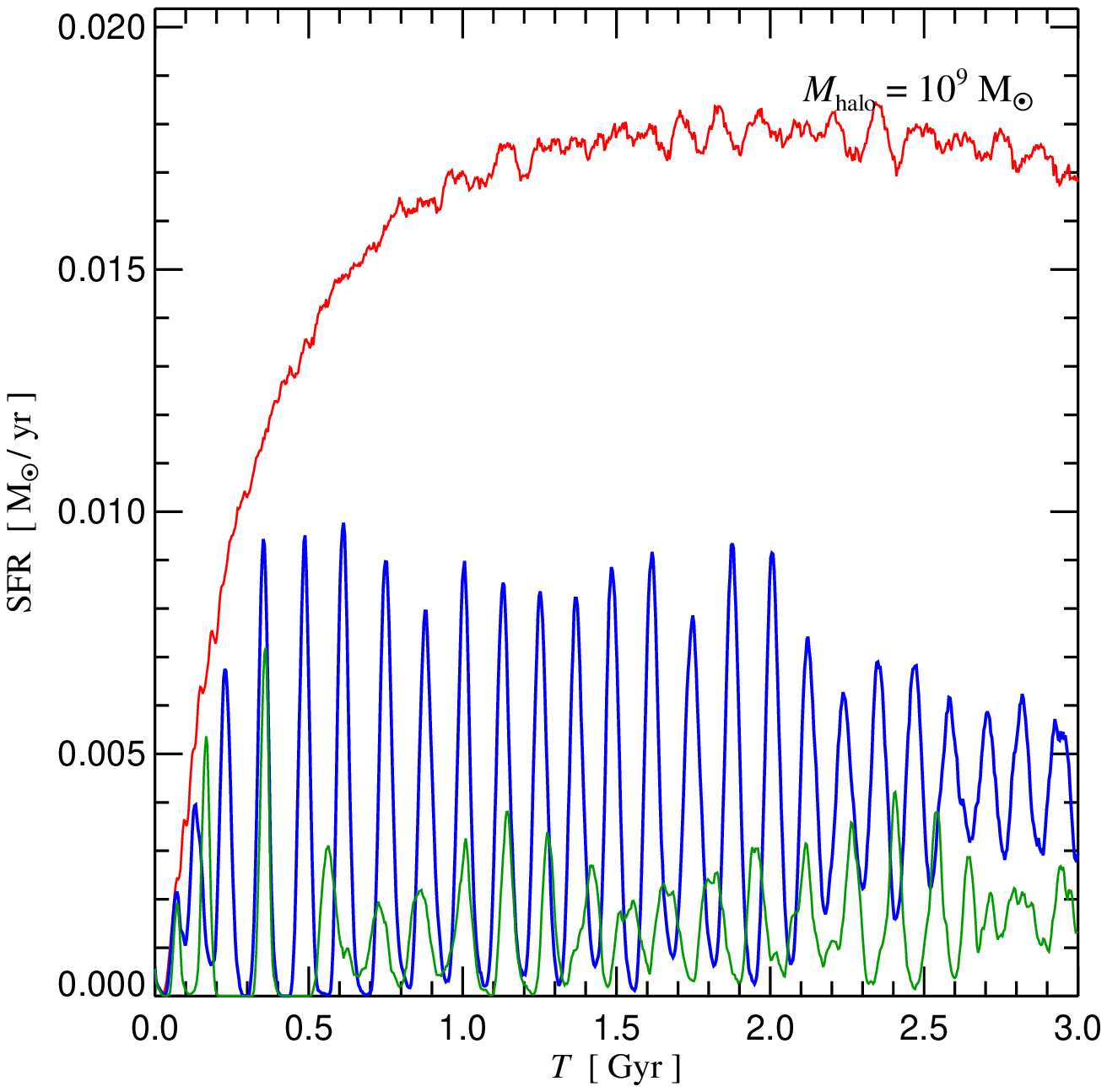}}%
\resizebox{7.5cm}{!}{\includegraphics{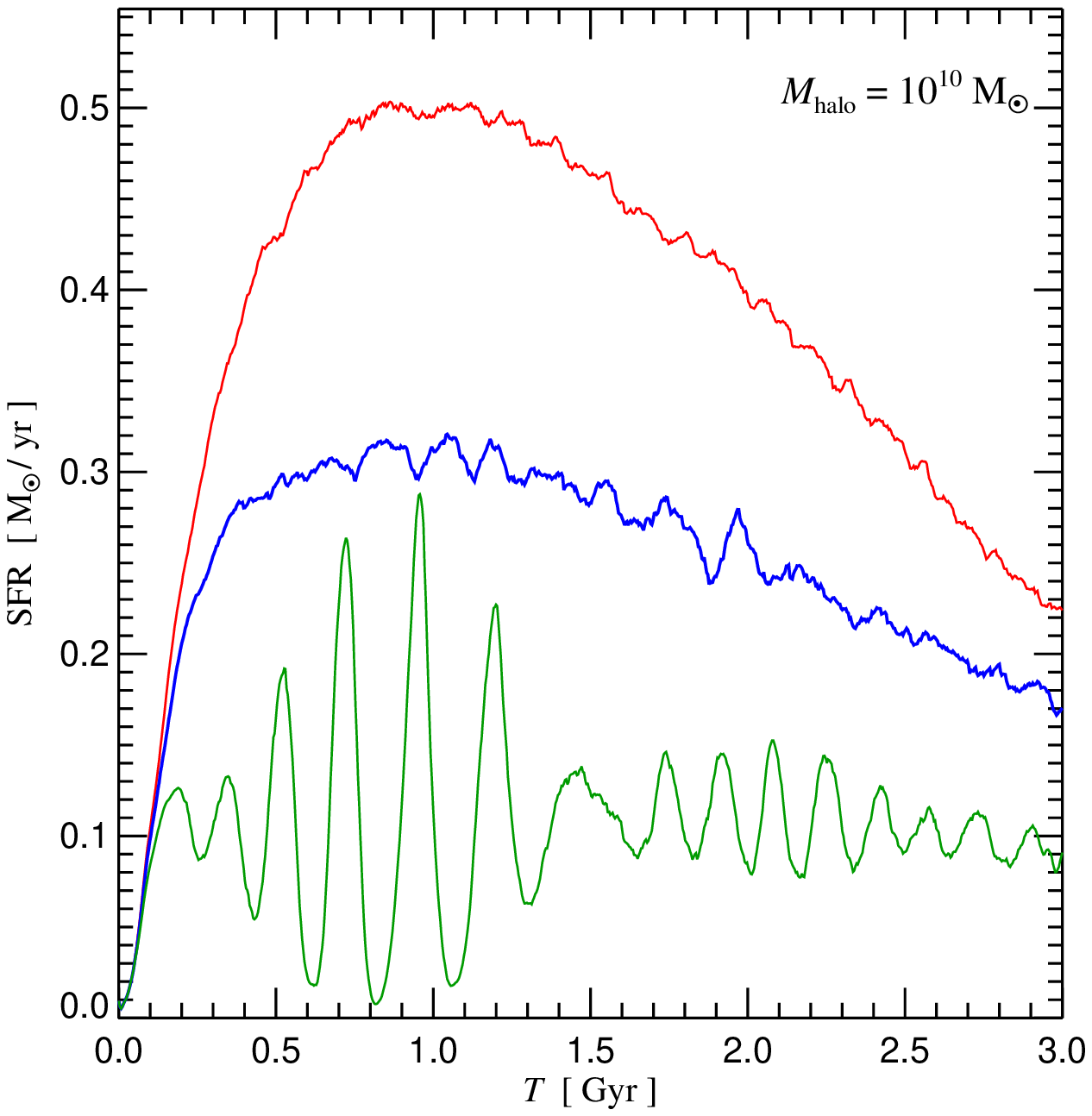}}\\
\resizebox{7.5cm}{!}{\includegraphics{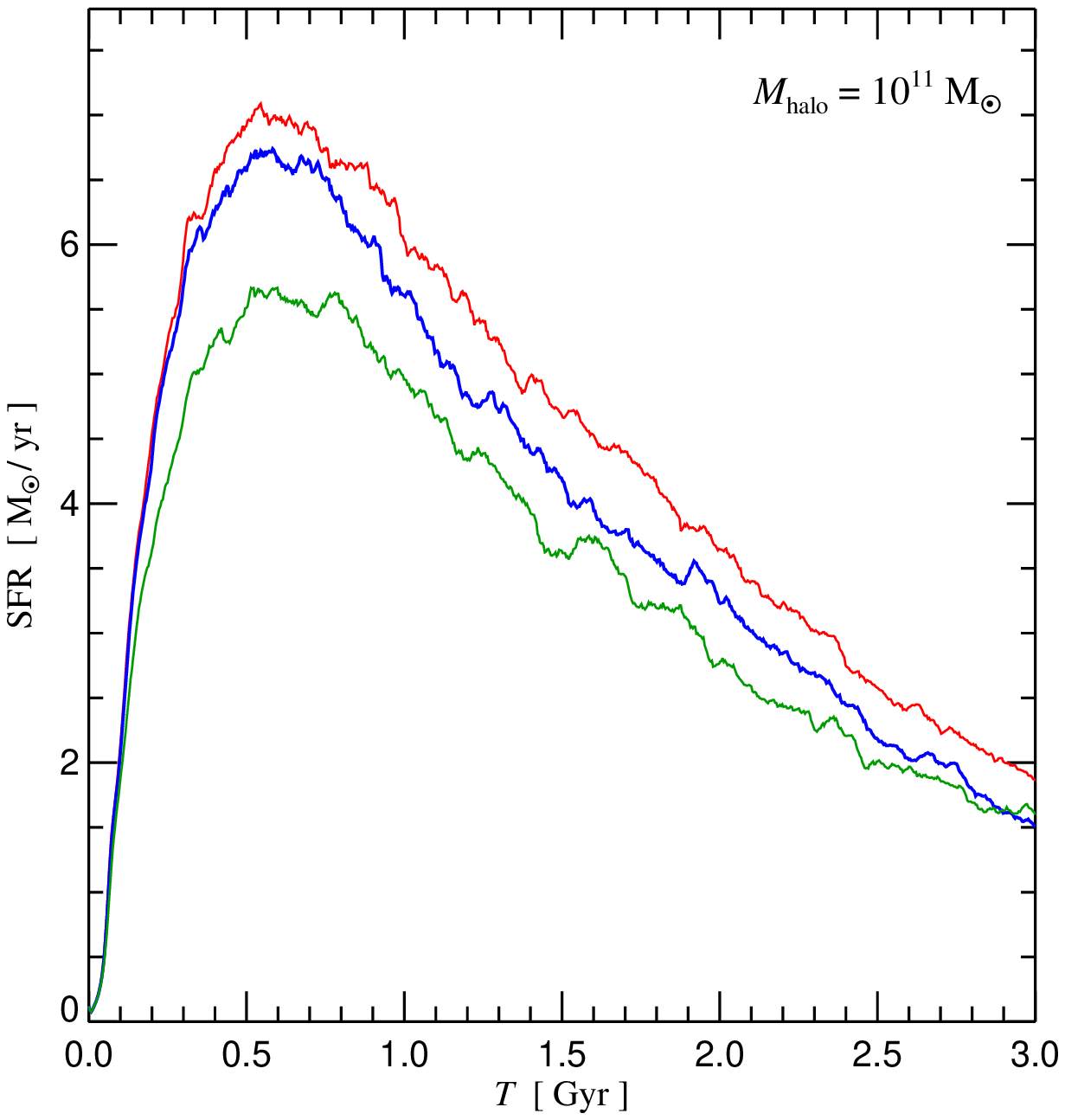}}%
\resizebox{7.5cm}{!}{\includegraphics{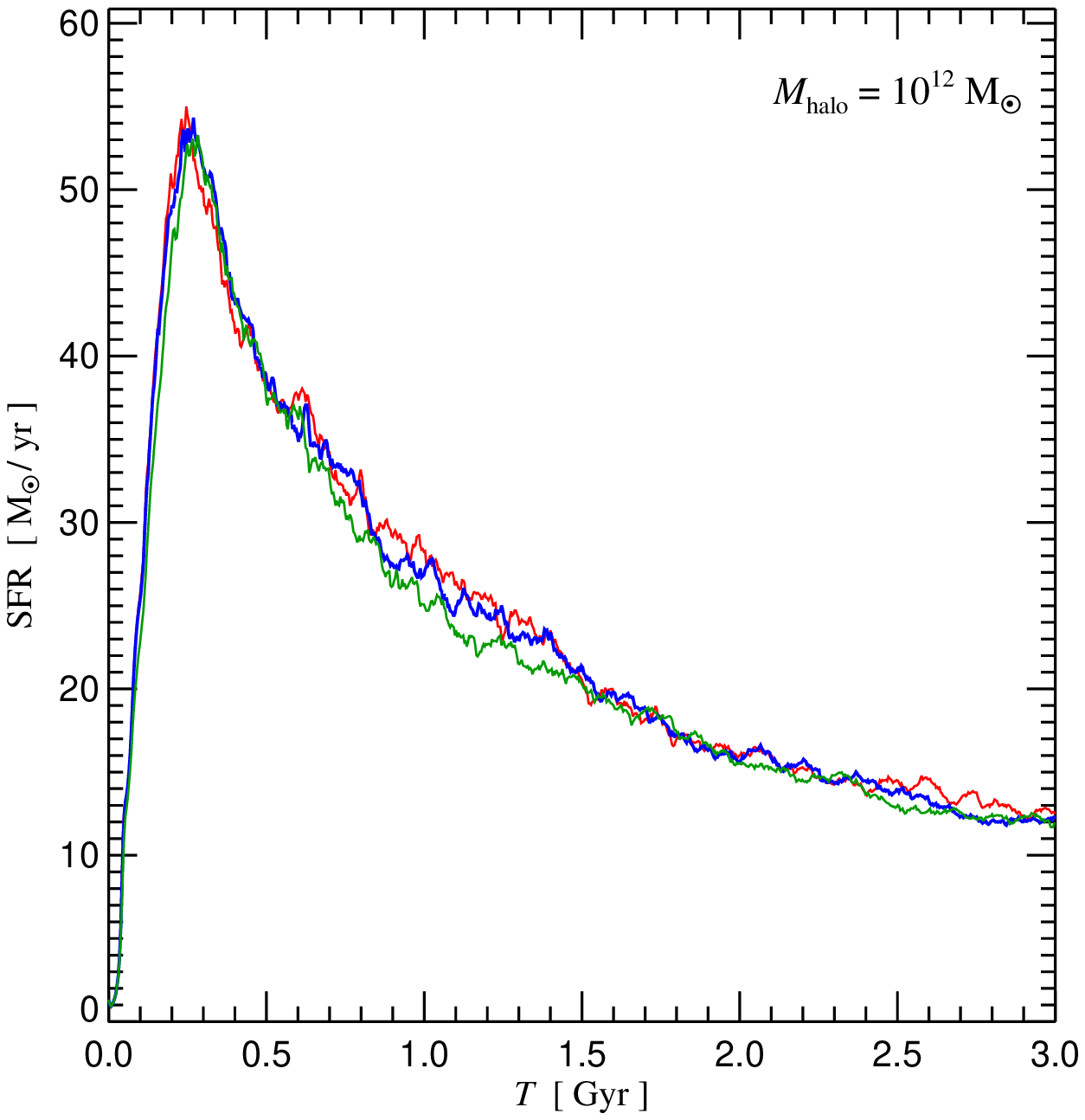}}\\
\end{center}
  \caption{Time evolution of the star formation rate in isolated halos
    of different mass which are initially in virial equilibrium. In each
    panel, we compare the star formation rate in simulations without cosmic
    ray physics (solid red line) to two runs with different injection
    efficiency of cosmic rays by supernovae, $\zeta_{\rm SN}=0.1$ (blue lines)
    and $\zeta_{\rm SN}=0.3$ (green lines), respectively. From top left to
    bottom right, results for halos of virial mass \mhalo{9} to \mhalo{12} are
    shown, as indicated in the panels. Efficient production of cosmic rays can
    significantly reduce the star formation rate in very small galaxies, but
    it has no effect in massive systems.\label{FigSFRs} }
\end{figure*}

\begin{figure*}
\begin{center}
\resizebox{8cm}{!}{\includegraphics{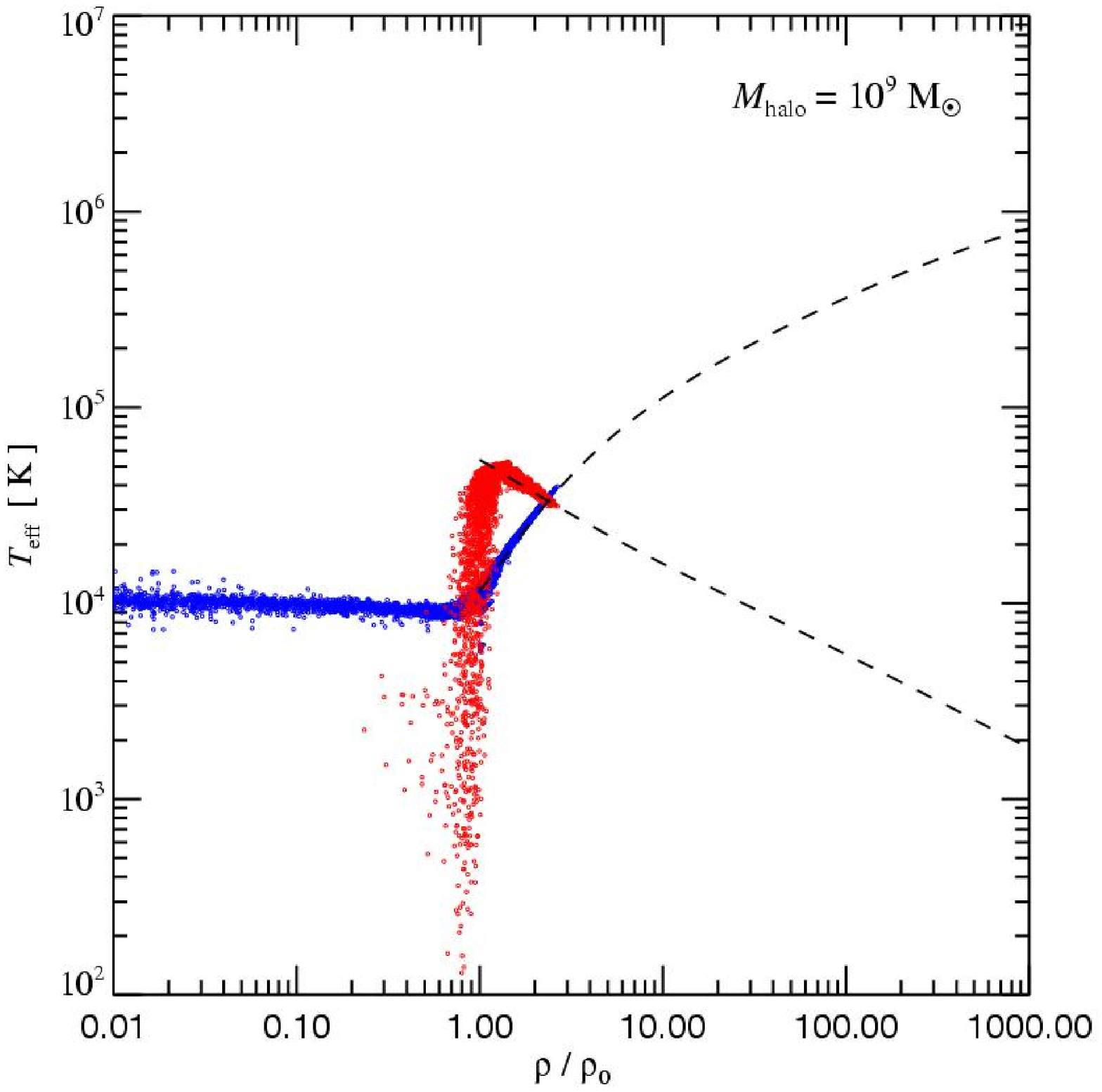}}%
\resizebox{8cm}{!}{\includegraphics{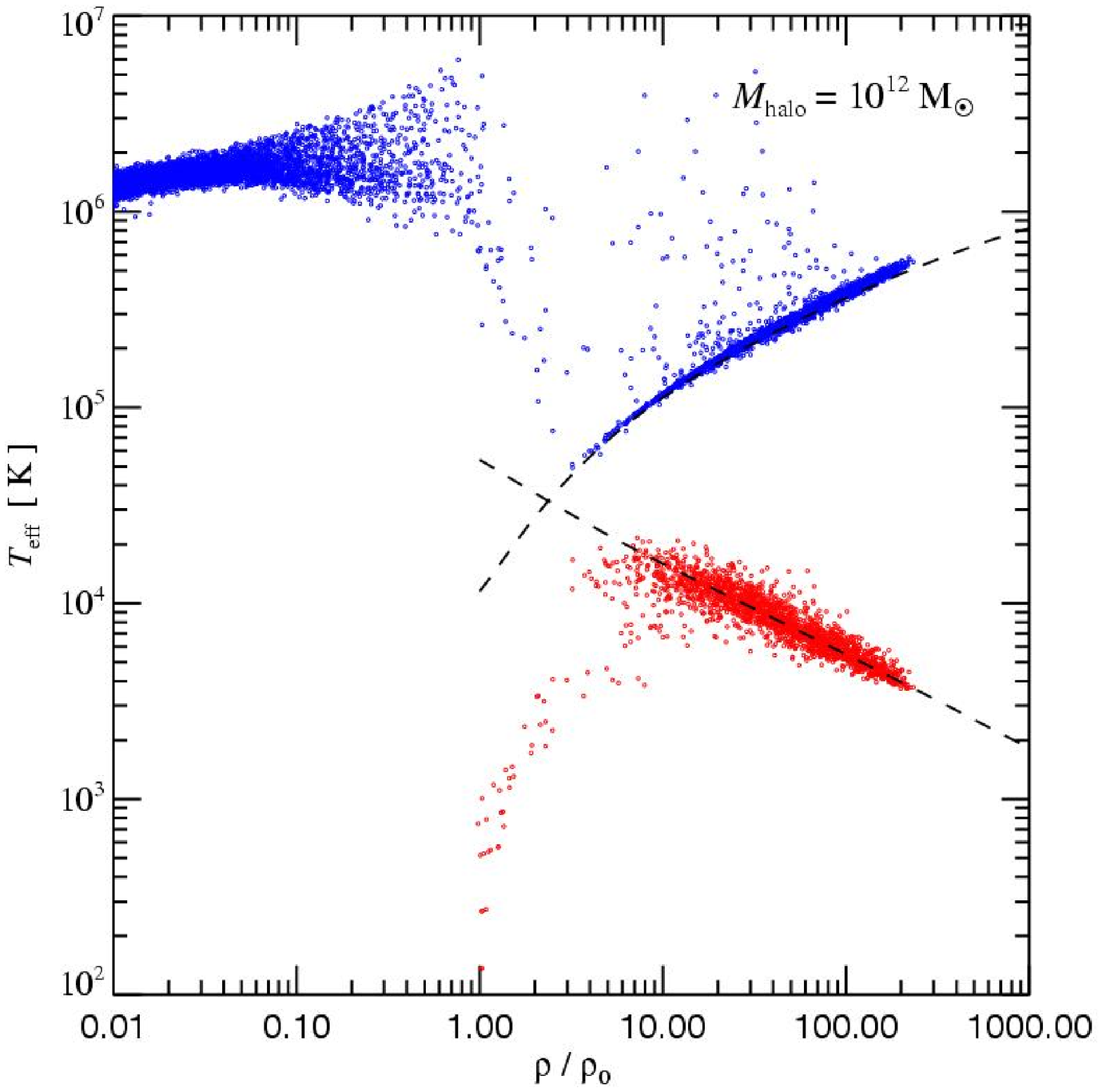}}\\
\end{center}
  \caption{Phase-space diagram of the star-forming phase in two
    simulations with halos of different mass. In these fiducial
    simulations, we included cosmic ray physics but ignored the cosmic
    ray pressure in the equations of motion, i.e.~there is no
    dynamical feedback by cosmic rays.  However, a comparison of the
    cosmic ray pressure and the thermal pressure allows us to clearly
    identify regions where the cosmic rays should have had an
    effect. For graphical clarity, we plot the pressures in terms of a
    corresponding effective temperature, $T_{\rm eff}= (\mu/k) P /
    \rho$.  Above the star formation threshold, the small galaxy of
    mass \mhalo{9} shown in the left panel has a lot of gas in the
    low-density arm of the effective equation of state, shown by the
    curved dashed line.  On the other hand, the massive \mhalo{12}
    galaxy shown on the right has characteristically higher densities
    in the ISM. As a result, the cosmic ray pressure is insufficient
    to affect this galaxy significantly. Note that the falling dashed
    line marks the expected location where cosmic ray loss processes
    balance the production of cosmic rays by supernovae.  We show the
    systems at time $t= 2.0\,{\rm Gyr}$ after the start of the
    evolution.
    \label{FigPhaseSpace}}
\end{figure*}

Another intriguing possibility is that the total baryonic fraction ending up
in galactic halos could be modified by the additional pressure component
provided by the relativistic particle population.  In particular, the softer
equation of state of the cosmic-ray gas component (its adiabatic index varies
in the range $4/3 < \gamma_{\rm CR} < 5/3$) could increase the concentration
of baryonic matter in dark matter potential wells, because the pressure
increases less strongly when the composite CR/thermal gas is compressed.  On
the other hand, a partial cosmic ray pressure support might reduce the overall
cooling efficiency of gas in halos, causing a reduction of the condensated
phase of cold gas in the centres.

To examine the non-linear interplay of all these effects, we will study them
in a number of different scenarios. We first use isolated galaxy models which
allow a precise control over the initial conditions and an easy analysis and
interpretation of the results.  Next, we use non-radiative cosmological
simulations to investigate the efficiency of CR production at structure
formation shock waves.  We then use high-resolution cosmological simulations
that include radiative cooling and star formation to study the formation of
dwarf galaxies, with the aim to see whether our identified mass trends are
also present in the full cosmological setting.  We also use these simulations
to investigate whether CRs influence the absorption properties of the
intergalactic medium at high redshift.  Finally, we use high-resolution
`zoomed' simulations of the formation of clusters of galaxies to study how CR
injection by accretion shocks and supernovae modifies the thermodynamic
properties of the gas within halos.

\subsection{Formation of disk galaxies in isolation \label{SecIsolatedDisks}}

As a simple model for the effects of cosmic ray feedback on disk galaxy
formation, we consider the time evolution of the gas atmospheres inside
isolated dark matter halos. The initial conditions consist of a dark matter
potential with a structure motivated from cosmological simulations, combined
with a hydrostatic gas distribution initially in equilibrium within the halo.
We then consider the evolution of this system under radiative cooling, star
formation and cosmic ray production by supernovae. We expect that the gas in
the centre looses its pressure support by cooling, and then collapses into a
rotationally supported disk that forms inside-out \citep{Fall1980}.

It is clear that this is a highly idealized model for disk galaxy formation,
which glosses over the fact that in a more realistic cosmological setting
galaxies originate in a hierarchical process from the gravitational
amplification of density fluctuation in the primordial mass distribution,
gradually growing by accretion and merging with other halos into larger
objects. However, the simplified approach we adopt here should still be able
to capture some of the basic processes affecting this hierarchy, and it does
so in a particular clean way that should allow us to identify trends due with
galaxy mass due to cosmic rays.

We model the dark matter and baryonic content of our isolated halos as NFW
density profiles \citep*{NFW}, which we slightly soften at the centre to
introduce a core into the gas density, with a maximum density value lying
below the threshold for star formation, allowing for a `quiet' start of the
simulations.  The velocity dispersion of the dark matter and the temperature
of the gas were chosen such that the halos are in equilibrium initially,
i.e.~when evolved without radiative cooling, the model halos are perfectly
stable for times of order the Hubble time. We also impart angular momentum
onto the halo with a distribution inside the halo consistent with results
obtained from full cosmological simulations \citep{Bull01}.

We simulated a series of host halos with masses varying systematically between
\mhalo{9} and \mhalo{12}. In all cases, we adopted a baryon fraction of
$\Omega_{\rm b}/\Omega_{\rm m} = 0.133$, and a matter density of $\Omega_{\rm
  m}=0.3$. We typically represented the gas with $10^5$ particles and the
dark matter with twice as many. In some of our simulations, we also replaced
the live dark halo with an equivalent static dark matter potential to speed up
the calculations. In this case, the contraction of the dark matter due to
baryonic infall is not accounted for, but this has a negligible influence on
our results.  We have kept the concentration of the NFW halos fixed at a value
of $c=12$ along the mass sequence, such that the initial conditions are scaled
versions of each other which would evolve in a self-similar way if only
gravity and ideal hydrodynamics were considered.  However, this
scale-invariance is broken by the physics of cooling, star formation and
cosmic rays.

When one of these halos is evolved forward in time, radiative cooling leads to
a pressure loss of the gas in the centre, which then collapses and settles
into a rotationally supported cold disk. In the disk, the gas is compressed by
self-gravity to such high densities that star formation ensues. Unfortunately,
the physics of star formation is not understood in detail yet, and we also
lack the huge dynamic range that would be necessary do directly follow the
formation and fragmentation of individual star-forming molecular clouds in
simulations of whole galaxies. In this study, we therefore invoke a
sub-resolution treatment for star formation, in the form described by
\citet{Springel2003}.  The model assumes that the dense interstellar medium
can be approximately described as a two-phase medium where cold clouds form by
thermal instability out of a diffuse gaseous phase. The clouds are the sites
of star formation, while the supernovae that accompany the star formation heat
the diffuse medium, and, in particular, evaporate some of the cold clouds. In
this way, a self-regulation cycle for star formation is established.

When our new cosmic ray model is included in our simulation code, a fraction
of the deposited supernova energy is invested into the acceleration of
relativistic protons, and hence is lost to the ordinary feedback cycle. While
this energy no longer directly influences the star formation rate, it has an
indirect effect on the star-forming gas by providing a pressure component that
is not subject to the usual radiative cooling.  If this pressure component
prevails sufficiently long, it can cause the gas to expand and to lower its
density, thereby reducing the rate of star formation.

In Figure~\ref{FigSFRs}, we show the time evolution of the star formation rate
for four different halos masses, ranging from $10^9\,h^{-1}{\rm M}_\odot$ to
$10^{12}\,h^{-1}{\rm M}_\odot$. For each halo mass, we compare three different
cases, a reference simulation where the ordinary model of \citet{Springel2003}
without cosmic rays was used, and two simulations where cosmic ray production
by supernovae was included (without allowing for diffusion), differing only in
the assumed efficiency of $\zeta_{\rm SN}=0.1$ and $\zeta_{\rm SN}=0.3$ for
this process, respectively.  Interestingly, the simulations with cosmic rays
show a substantial reduction of the star formation rate in the two small mass
systems, but already for the $10^{11}\,h^{-1}{\rm M}_\odot$ halo the effects
becomes comparatively small, while for the massive halo of mass
$10^{12}\,h^{-1}{\rm M}_\odot$, no significant differences can be detected.
Clearly, the ability of cosmic ray feedback to counteract star formation shows
a rather strong mass dependence, with small systems being affected most.  For
higher efficiencies $\zeta_{\rm SN}$ of CR-production by supernovae, the
reduction of the star formation rate becomes larger, as expected.

Figure~\ref{FigPhaseSpace} provides an explanation for this result, and also
elucidates the origin of the oscillatory behaviour of the SFR in the
CR-suppressed cases. In the figure, we show phase-space diagrams of the gas
particles of the $10^{9}\,h^{-1}{\rm M}_\odot$ and $10^{12}\,h^{-1}{\rm
  M}_\odot$ halos, respectively, in a plane of effective temperature versus
density. We plot the thermal pressure and the cosmic ray pressure separately.
In order to cleanly show whether a dynamical effect of cosmic rays can be
expected, we here use a fiducial simulation where the cosmic ray pressure is
ignored in the equations of motion, but is otherwise computed with the full
dynamical model.  As Figure~\ref{FigPhaseSpace} demonstrates, the bulk of the
star-forming gas in the massive halo lies at much higher density and higher
effective pressure than in the low mass halo.  Because the cosmic ray pressure
exceeds the effective thermal pressure of the multi-phase ISM only for
moderate overdensities relative to the star formation threshold, most of the
gas in the $10^{12}\,h^{-1}{\rm M}_\odot$ halo is simply too dense to be
affected by the cosmic ray pressure. We note that the relative sizes of the
two pressure components are consistent with the analytic expectations shown in
Figure~\ref{FigPessure}. In fact, these expectations are replicated as dashed
lines in Figure~\ref{FigPhaseSpace} and are traced well by the bulk of the
particles.  Because the shallower potential wells in low-mass halos cannot
compress the gas against the effective pressure of the ISM to comparably high
overdensities as in high-mass halos, it is therefore not surprising that the
cosmic ray pressure becomes dynamically important only in small systems.

Figure~\ref{FigPhaseSpace} also makes it clear that in the regime where cosmic
ray pressure may dominate we cannot expect a dynamically stable
quasi-equilibrium with a quiescent evolution of the star formation rate.  This
is simply due to the decline of the effective cosmic ray pressure with increasing
density of the ISM, a situation which cannot result in a stable equilibrium
configuration where self-gravity is balanced by the cosmic ray pressure.
Instead, the system should be intrinsically instable in this regime. When some
gas becomes dense enough to start star formation, it will first have no cosmic
ray pressure support but it will be stabilized against collapse by the thermal
pressure of the ISM that is quickly established by supernova feedback.  After
some time, the ongoing star formation builds up a cosmic ray pressure
component, which eventually starts to dominate, at which point the gas is
driven to lower density.  As a result, the star formation rate declines
strongly.  After some time, the CR pressure is dissipated such that the gas
can collapse again.  Star formation will then start again and the `cycle' can
repeat. This scenario schematically describes the origin of the oscillations
in the star formation rate seen in the results for the
$10^{9}\,h^{-1}{\rm M}_\odot$ and $10^{10}\,h^{-1}{\rm M}_\odot$ halos when
cosmic rays are included.

\begin{figure}
\begin{center}
\resizebox{8.5cm}{!}{\includegraphics{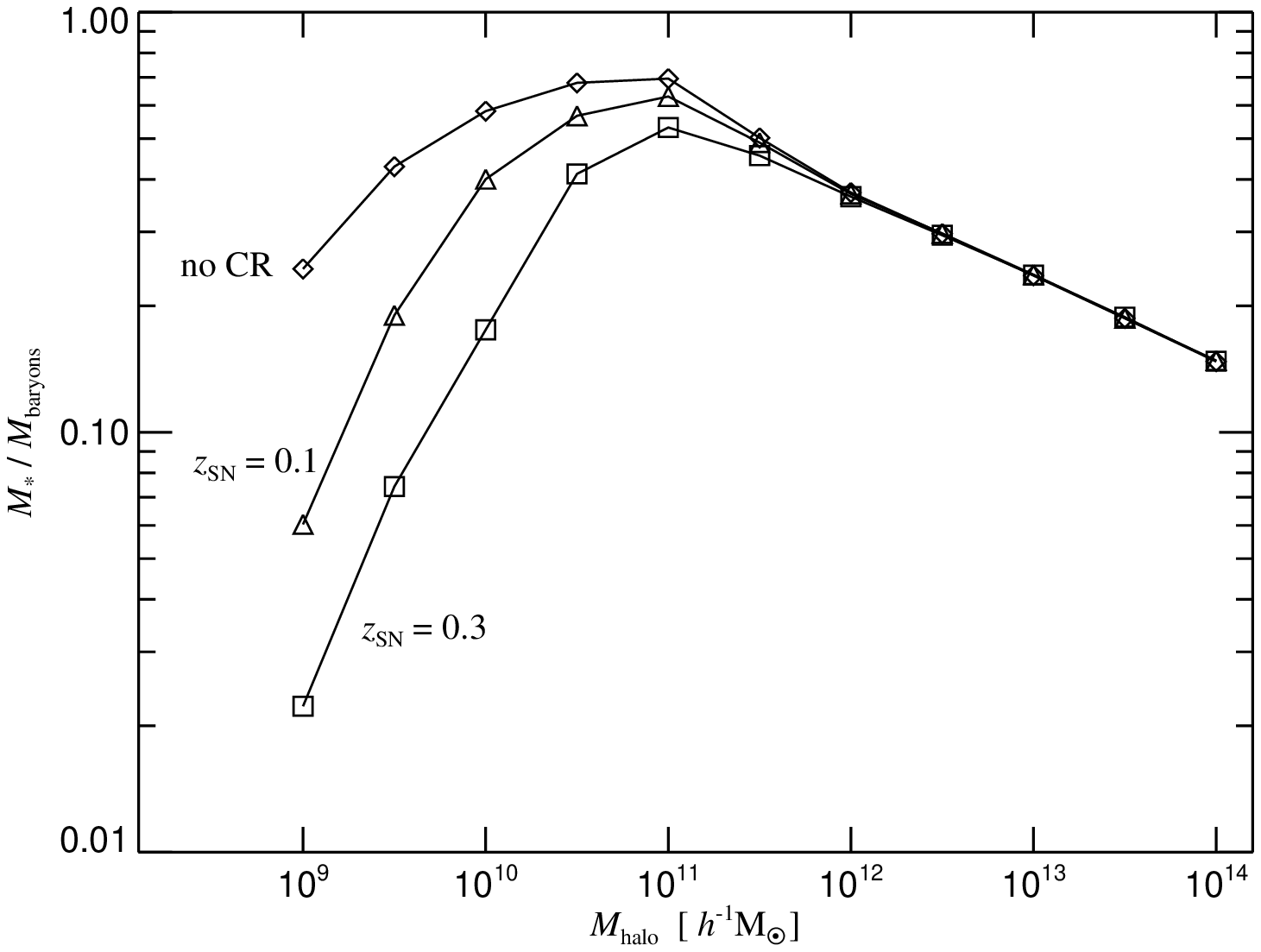}}
\end{center}
  \caption{Efficiency of star formation as a function of halo mass in
    our isolated disk formation simulations. We show the ratio of the stellar
    mass formed to the total baryonic mass in each halo, at time $t=3.0\,{\rm
      Gyr}$ after the start of the simulations, and for two different
    efficiencies of cosmic ray production by supernovae. Comparison with the
    case without cosmic ray physics shows that star formation is strongly
    suppressed in small halos, by up to a factor $\sim 10-20$, but large
    systems are essentially unaffected. \label{FigEffSFRHaloMass}}
\end{figure}

Another view of the halo mass dependence of the effects of cosmic ray feedback
on star formation is given by Figure~\ref{FigEffSFRHaloMass}. Here we show the
integrated stellar mass formed up to time $t=3\,{\rm Gyr}$, normalized by the
total baryonic mass.  Again, we compare two different injection efficiencies
($\zeta_{\rm SN}=0.1$ and $\zeta_{\rm SN}=0.3$) with a reference case where no
cosmic ray physics is included.  In general, star formation is found to be
most efficient at intermediate mass scales of $\sim 10^{11} {\rm M}_\odot$ in
these simulations. However, the simulations with cosmic ray production show a
clear reduction of their integrated star formation rate for halos with mass
below a few times $10^{11}h^{-1} {\rm M}_\odot$, an effect that becomes {\em
  progressively stronger} towards lower mass scales. For the $
10^{9}\,h^{-1} {\rm M}_\odot$ halo, the suppression reaches more than an order
of magnitude for $\zeta_{\rm SN}=0.3$.

\begin{figure*}
\begin{center}
        \includegraphics[width=5cm]{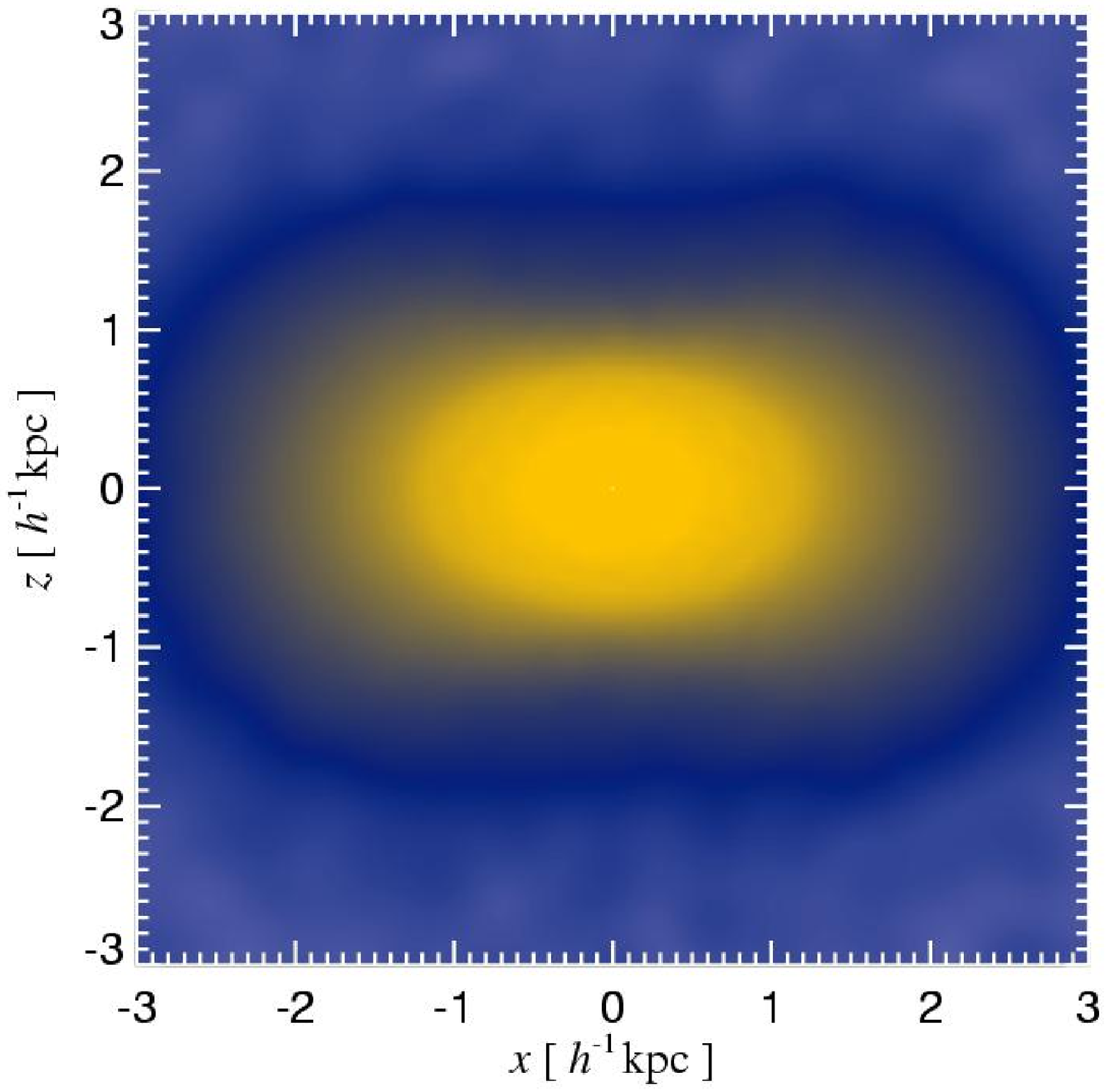}%
        \includegraphics[width=5cm]{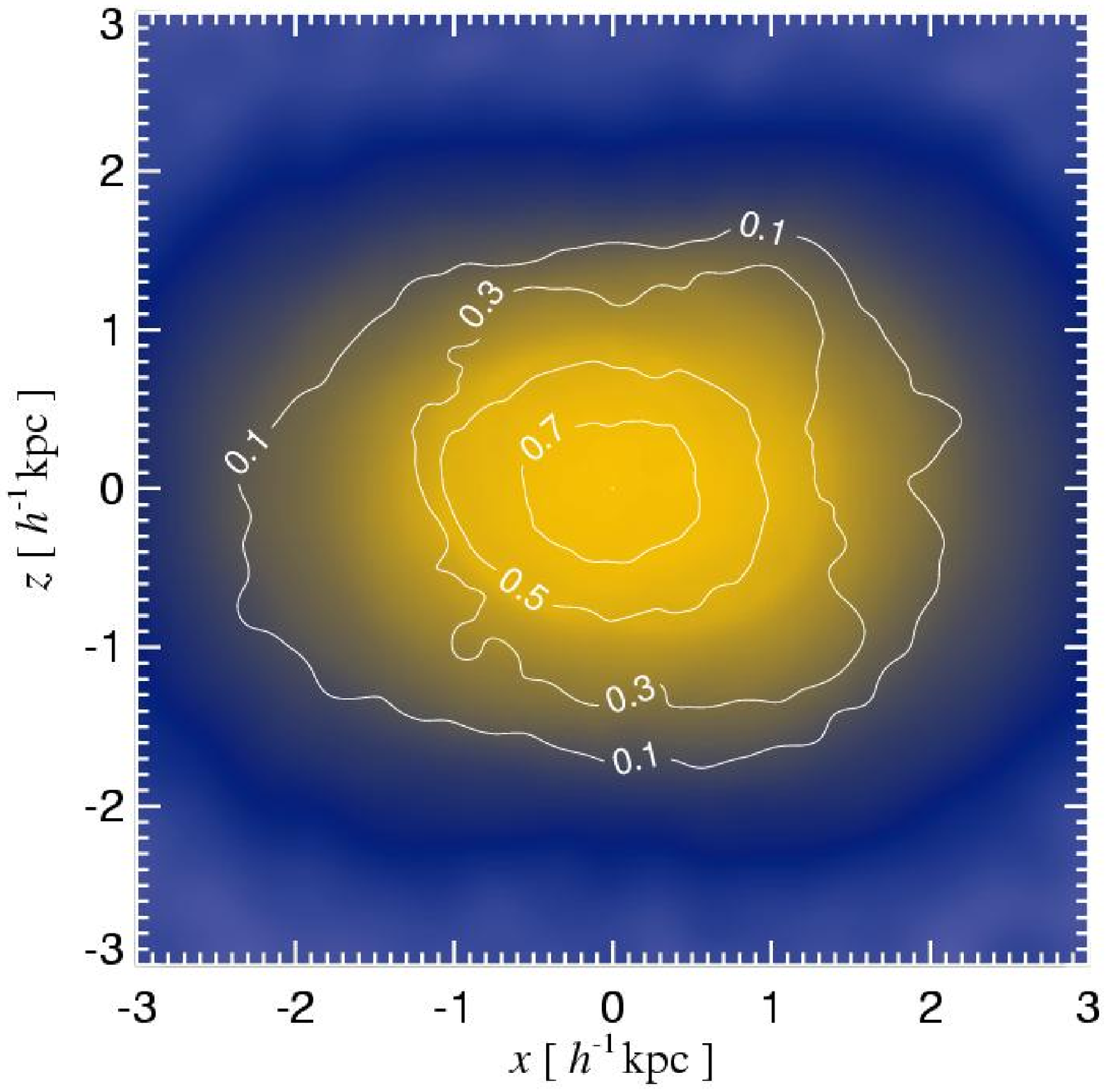}%
        \includegraphics[width=5cm]{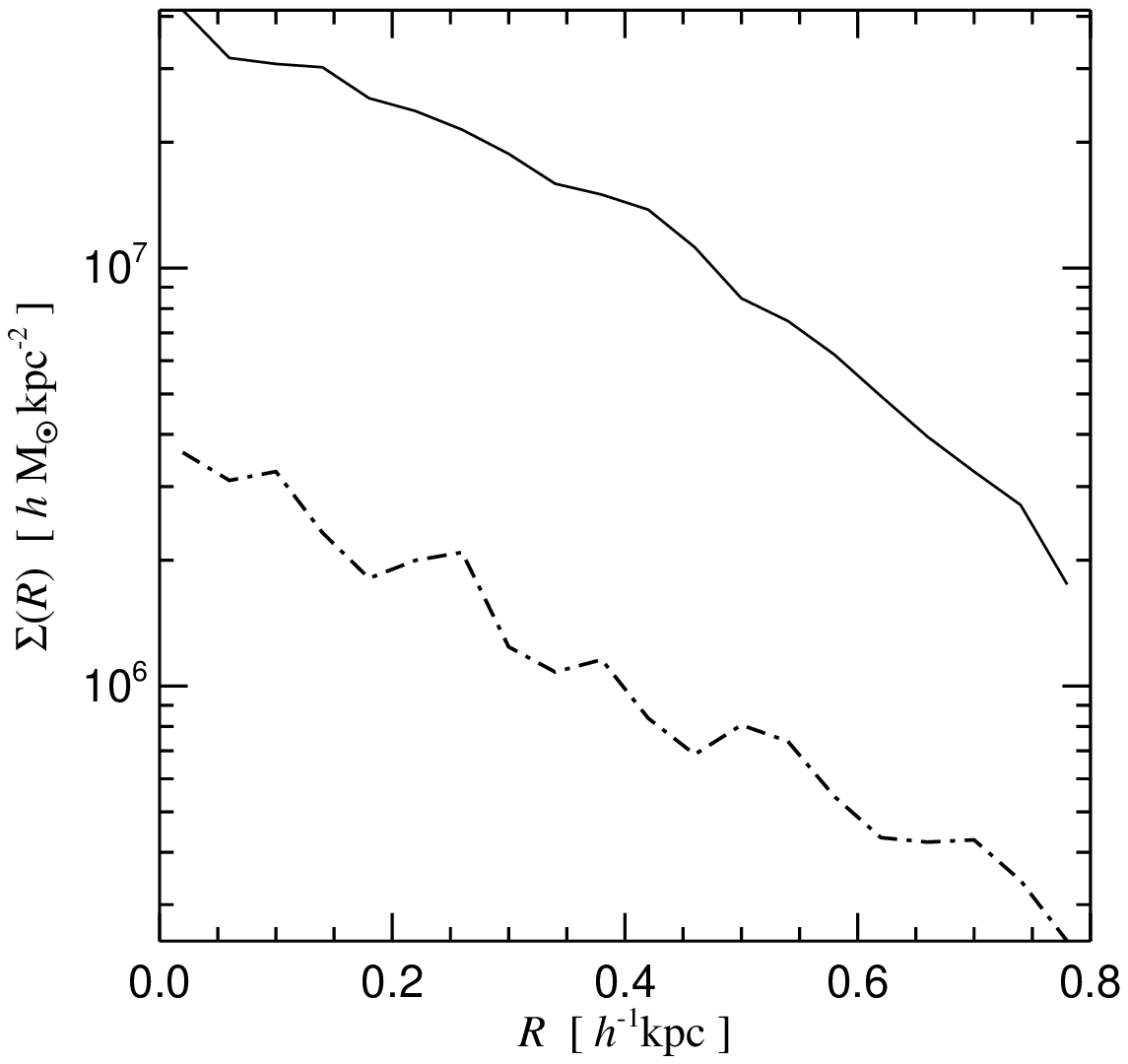}\\
        \includegraphics[width=5cm]{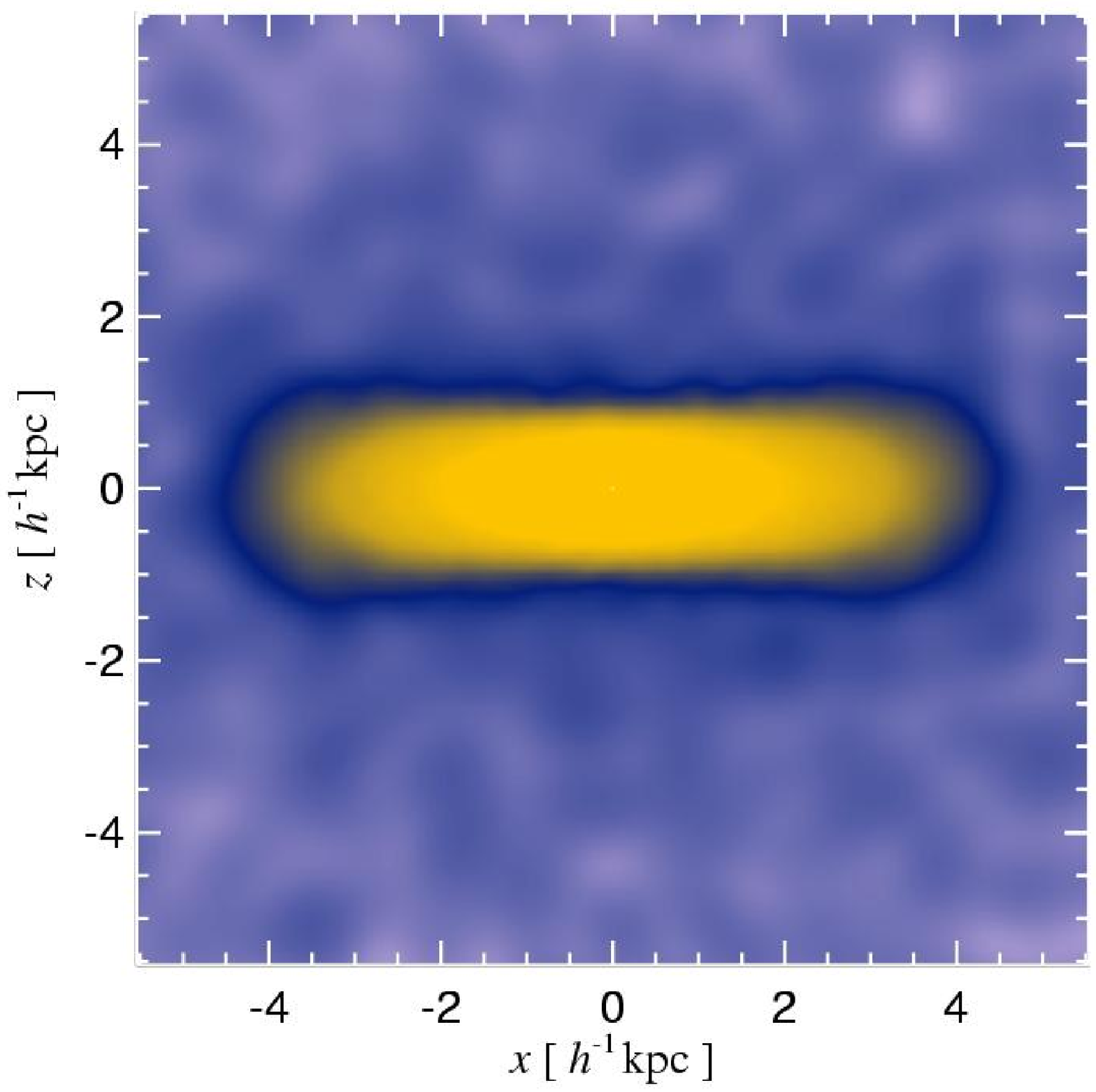}%
        \includegraphics[width=5cm]{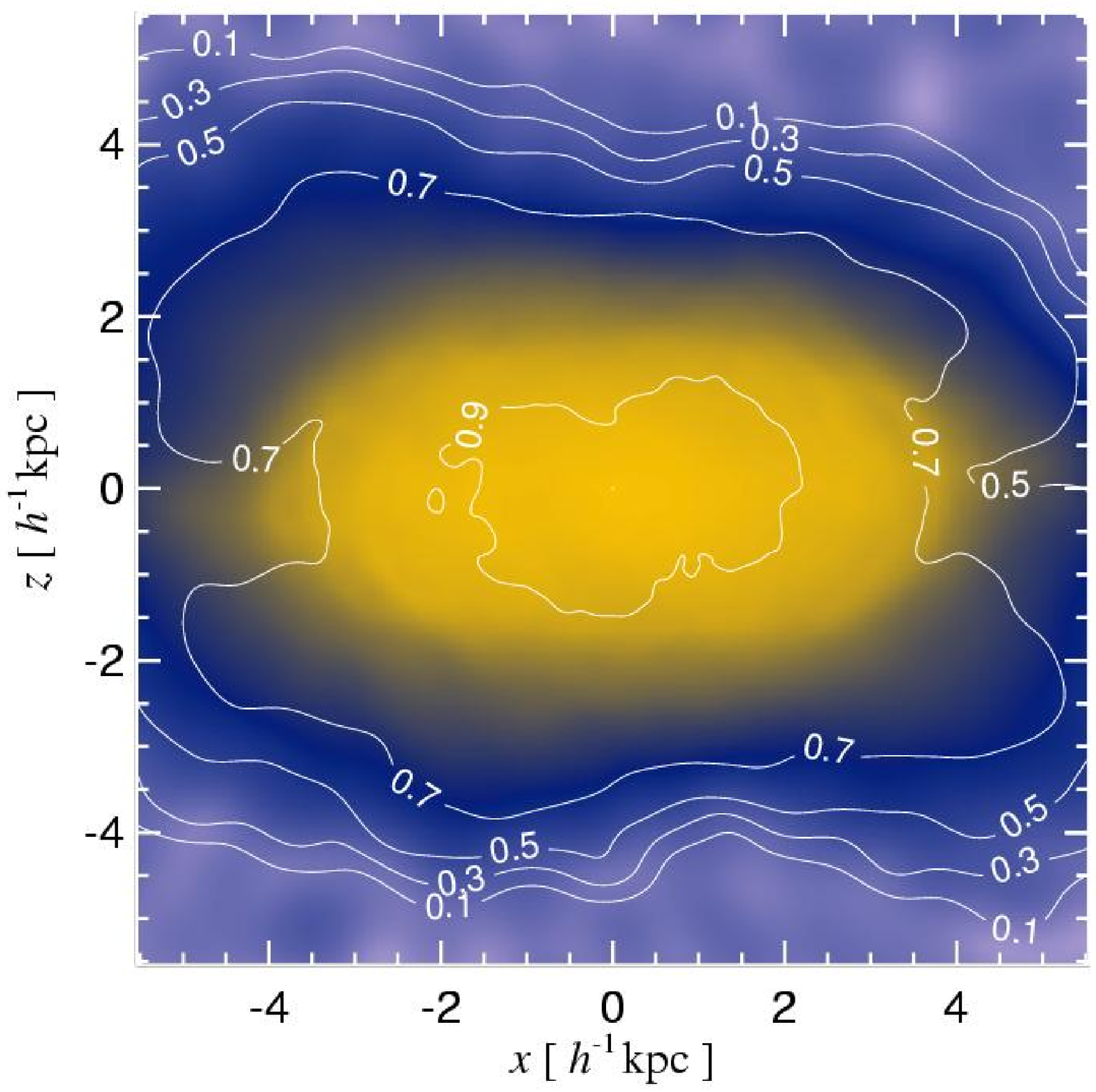}%
        \includegraphics[width=5cm]{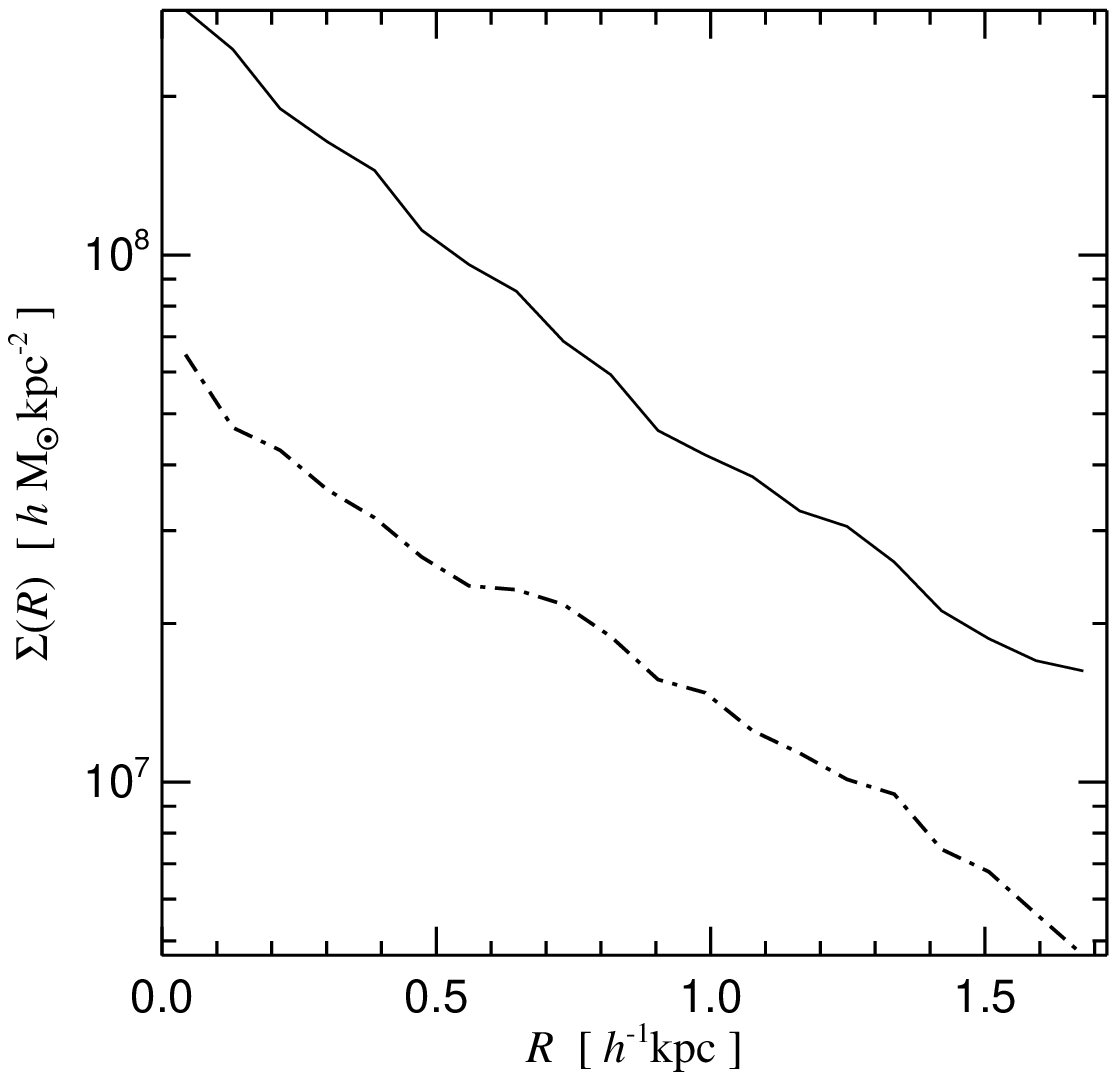}\\
        \includegraphics[width=5cm]{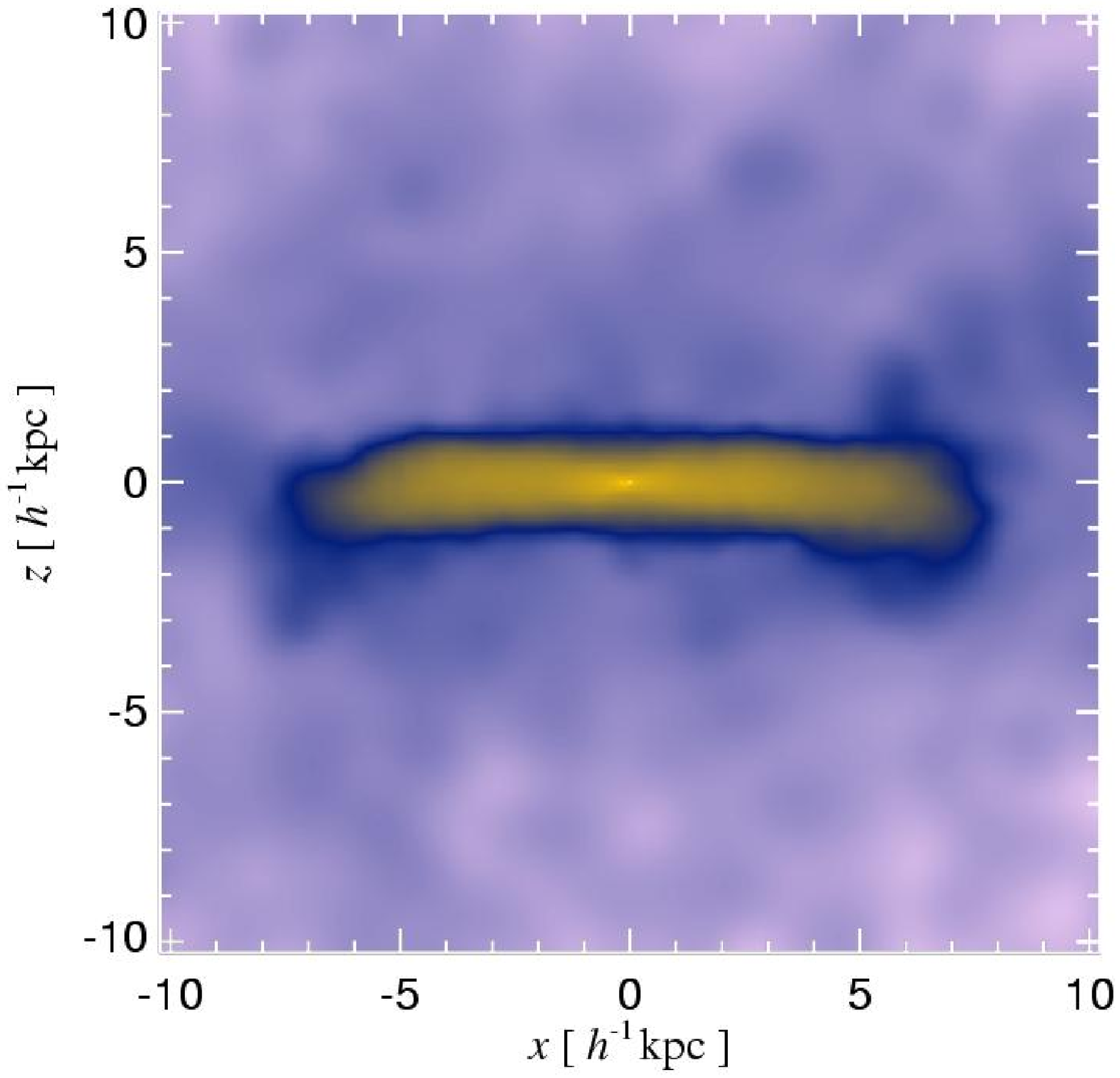}%
        \includegraphics[width=5cm]{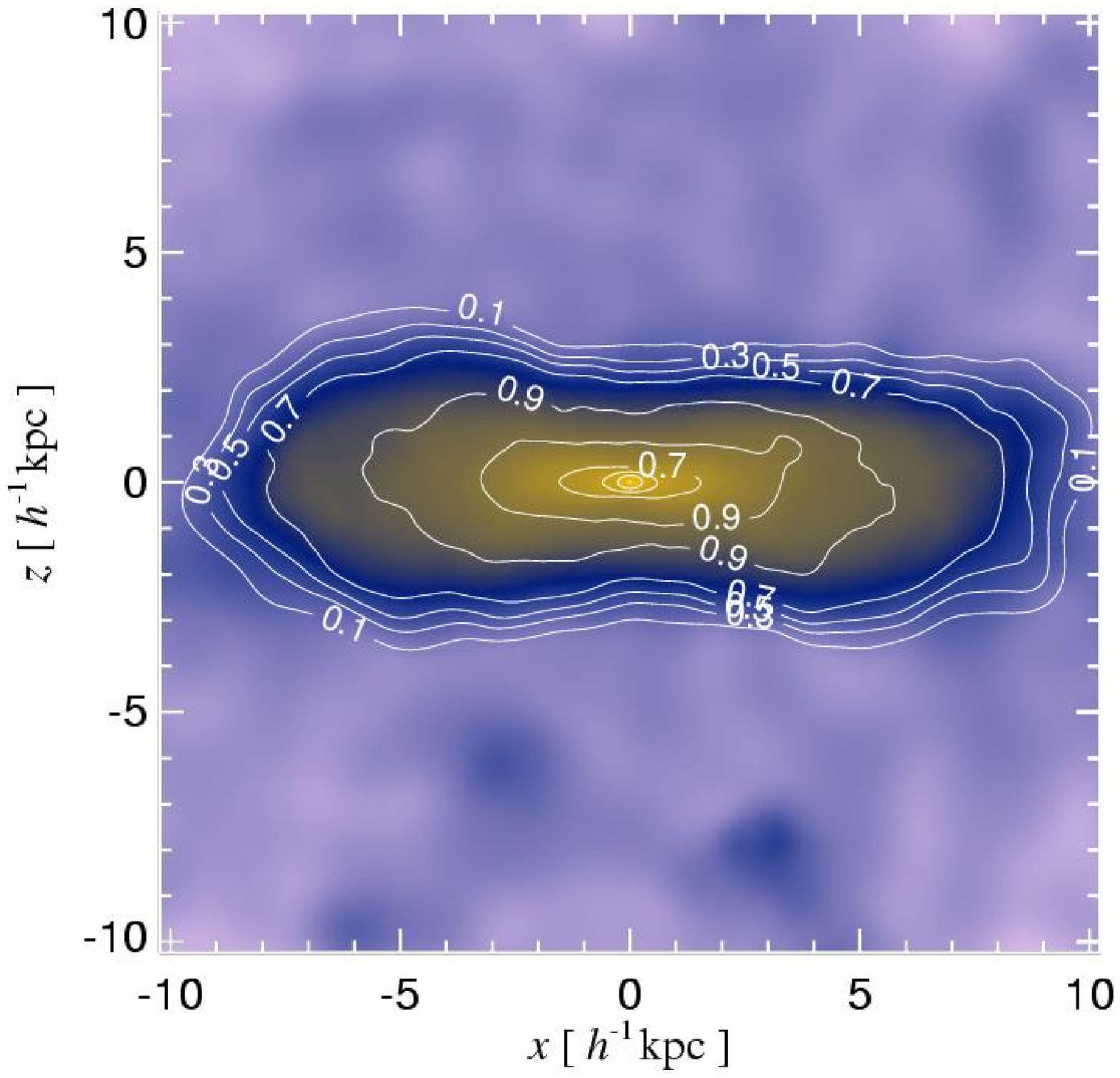}%
        \includegraphics[width=5cm]{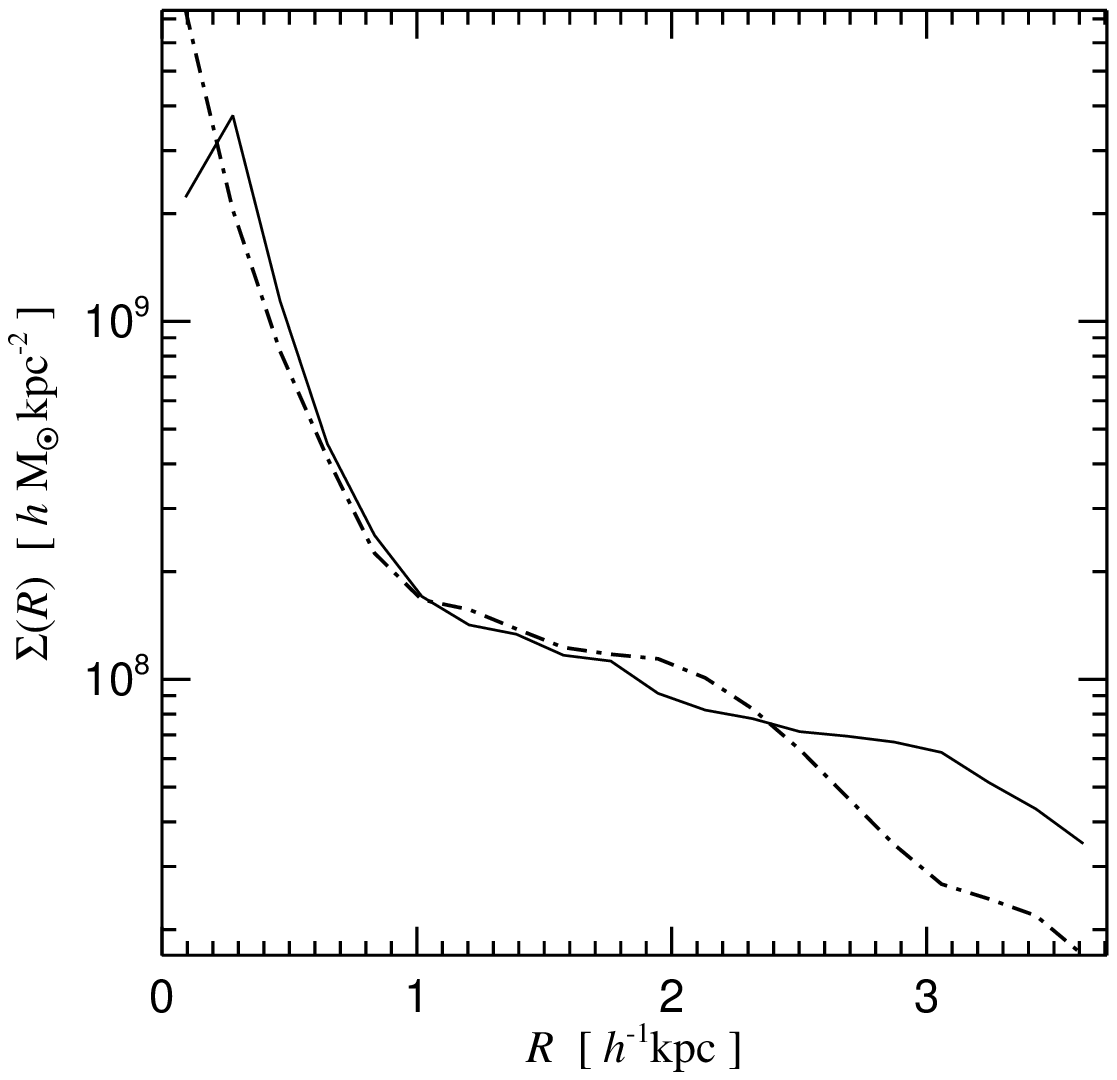}\\
        \includegraphics[width=5cm]{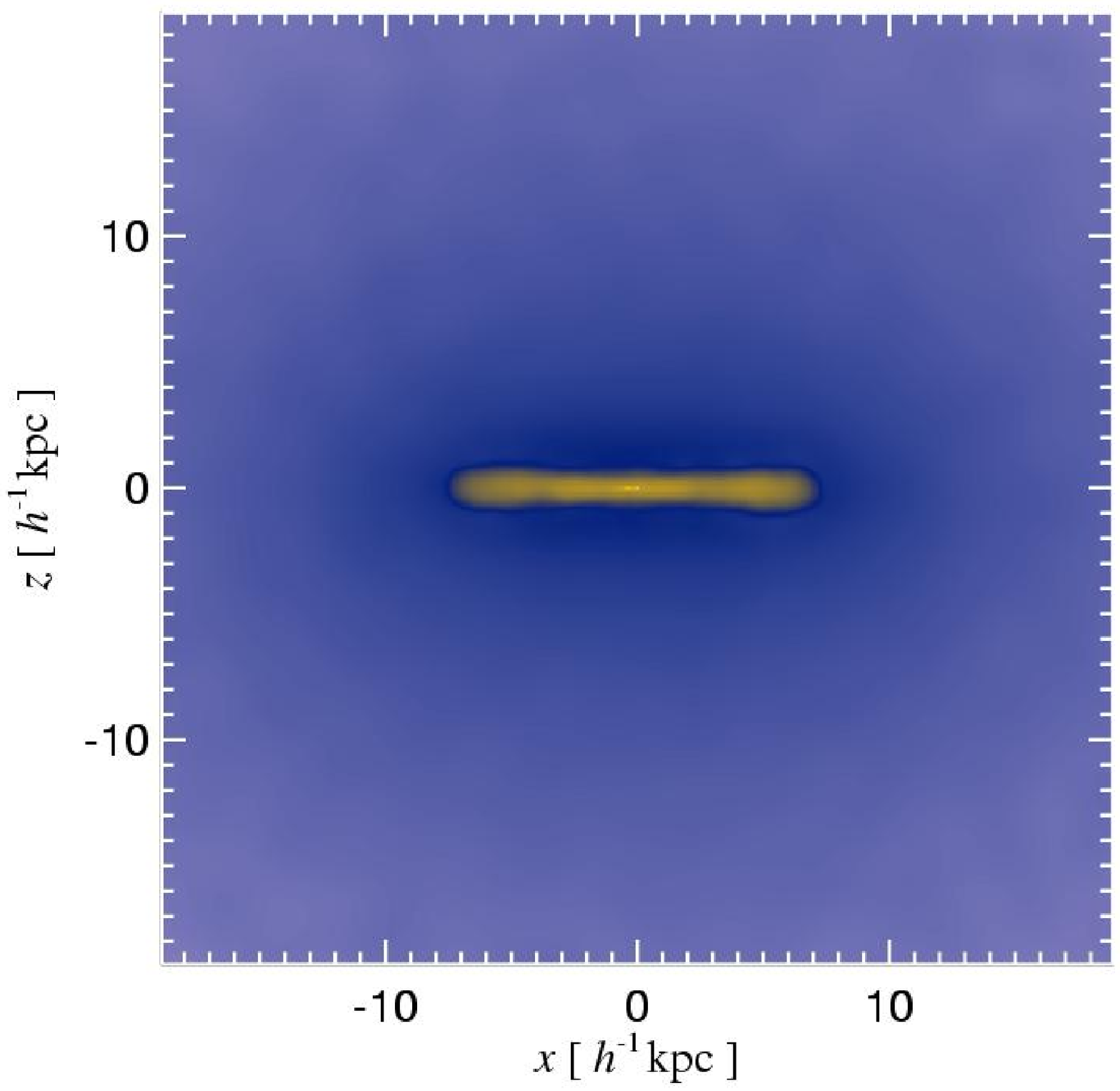}%
        \includegraphics[width=5cm]{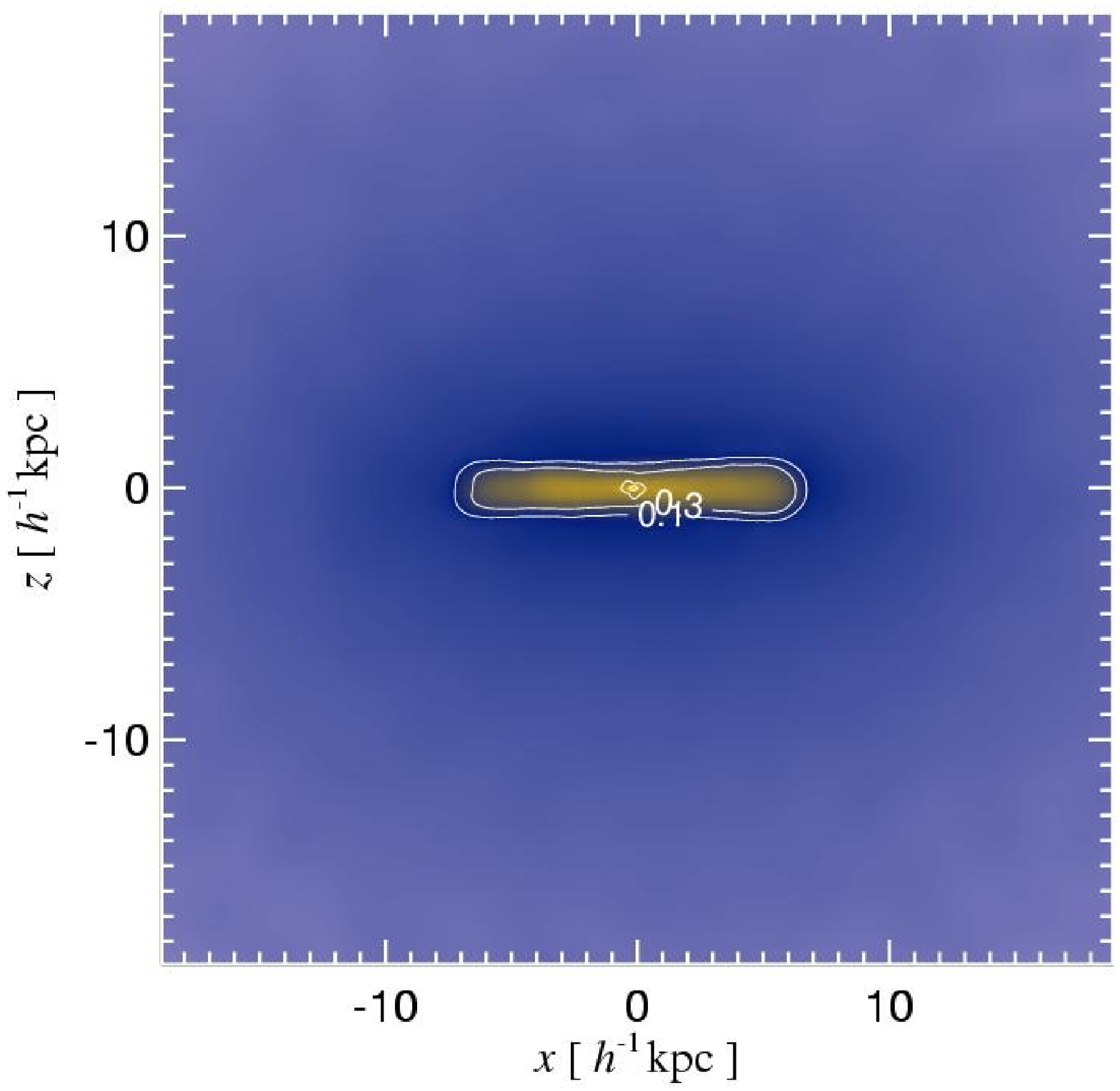}%
        \includegraphics[width=5cm]{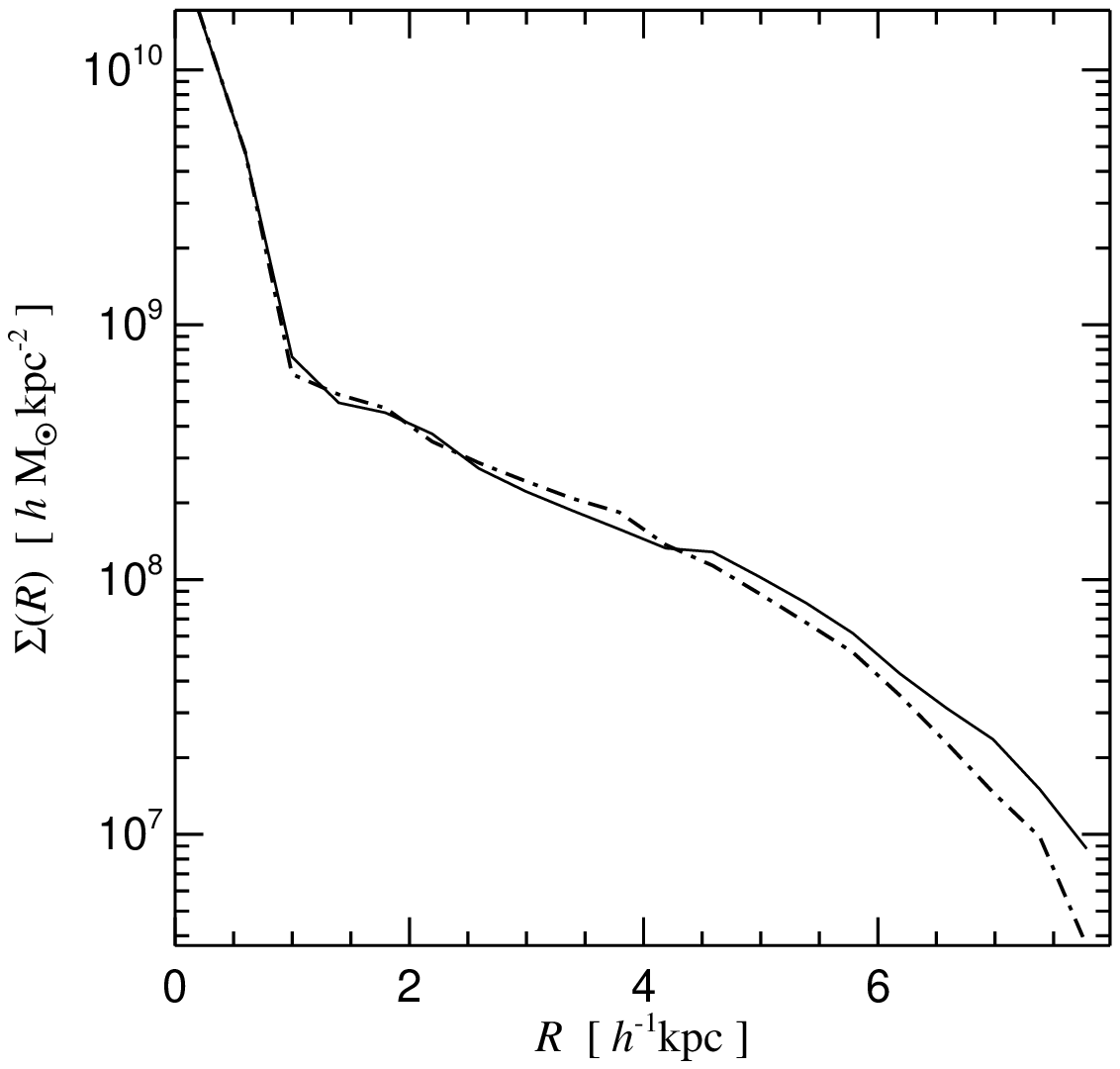}\\
\end{center}
        \caption{Effect of cosmic ray feedback on 
          star formation in simulations of isolated disk galaxy formation.
          Each row shows results for a different halo mass, for $M_{\rm halo}
          = 10^9$, $10^{10}$, $10^{11}$ and \mhalo{12} from top to bottom.  We
          compare the projected gas density fields at time $t=2.0\Gyr$ of runs
          without cosmic ray feedback (left column) to that of runs with
          cosmic ray production by supernovae (middle column). The gas density
          field is colour-coded on a logarithmic scale. For the simulation
          with cosmic rays, we overplot contours that show the contribution of
          the projected cosmic ray energy density to the total projected
          energy density (i.e. thermal plus cosmic rays), with contour levels
          as indicated in the panels. Finally, the right column compares the
          azimuthally averaged stellar surface density profiles at time
          $t=2.0\Gyr$ for these runs. Results for simulations without cosmic
          ray physics are shown with a solid line, those for simulations with
          CR feedback with a dot-dashed line.}
        \label{fig:Galaxy:Slices}
\end{figure*}

In Figure~\ref{fig:Galaxy:Slices}, we take a closer look at the spatial
distribution of the cosmic ray pressure in the different cases, and the
profiles of the stellar disks that form.  To this end, we show the projected
gas density distribution in an edge-on projection at time $t=2.0\Gyr$,
comparing the case without cosmic rays (left column) to the case with cosmic
rays (middle column), for a range of halo masses from \mhalo{9} to \mhalo{12}.
For the simulation with cosmic rays, we overlay contours for the relative
contribution of the projected cosmic ray energy to the total projected energy
density. This illustrates, in particular, the spatial extent the cosmic ray
pressure reaches relative to the star-forming region.  Finally, the panels on
the right compare surface density profiles of the stellar mass that has formed
up to this time.

Consistent with our earlier results, the stellar density profiles of the low
mass halos show a significant suppression when cosmic rays are included, while
they are essentially unaffected in the high mass case. Interestingly, we also
see that the gaseous disks in the low mass halos appear to be ``puffed up'' by
the additional pressure of the cosmic rays. It is remarkable that in the two
low-mass systems there is substantial CR pressure found significantly above
the star-forming regions, at densities much below the star formation
threshold.  This is despite the fact the acceleration of relativistic
particles only occurs in star-forming regions of high density within the
galactic disk in these simulations.  Presumably, some of the CR-pressurized
gas buoyantly rises from the star-forming disk into the halo, a process that
is suppressed by the stronger gravitational field in the high mass systems.

\begin{figure}
\resizebox{8.5cm}{!}{\includegraphics{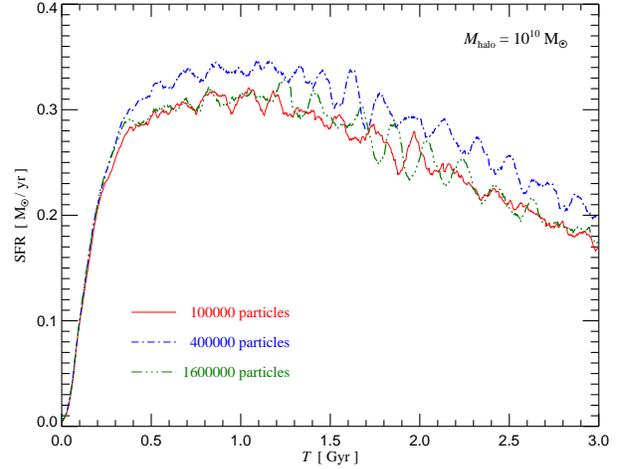}}%
        \caption{Resolution study of the star formation rate during the
          formation of a galactic disk in a halo of mass \mhalo{10}, including
          production of CRs with an efficiency of $\zeta_{\rm SN}=0.1$. We
          compare results computed with $10^5$, $4\times 10^5$, and $1.6\times
          10^6$ gas particles, respectively.
        \label{fig:SFR:Resolution}}
\end{figure}

As a final analysis of our isolated disk simulations, we examine how well our
simulation methodology for cosmic ray feedback converges when the numerical
resolution is varied.  To this end we repeat one of the simulations with
cosmic ray feedback ($\zeta_{\rm SN}=0.1$) of the \mhalo{11} halo using a
higher number of gas particles, namely $4\times 10^5$ and $1.6 \times 10^6$,
respectively. In Figure~\ref{fig:SFR:Resolution}, we compare the resulting
star formation rates. While there are some small fluctuations when the
resolution is varied, there is clearly no systematic trend with resolution,
and the results appear to be quite robust. In particular, the star formation
rates for the simulations with $10^{5}$ and $1.6 \times 10^6$ particles are in
very good agreement with each other despite a variation of the mass resolution
by a factor of 16. Note also that the oscillations are reproduced by all three
resolutions, but they are not exactly in phase.  Overall, this resolution test
is very promising and suggests that the numerical model is well posed and can
be applied to cosmological simulations where the first generation of galaxies
is typically only poorly resolved. We can still expect meaningful results under
these conditions.

\subsection{Cooling in isolated halos}

The comparison of the maximum cosmic ray cooling timescale with the
thermal cooling time in the bottom panel of Figure~\ref{FigCoolTimes}
has shown that for a relatively wide temperature range, the lifetime
of CRs is larger than the thermal cooling time. In a composite gas
with a substantial cosmic ray pressure component, this could
potentially stabilize the gas temporarily and reduce the rate at which
gas cools and accumulates at the bottom of the potential well of a
halo. Also, models have been conjectured where the temperature
structure of the intracluster medium, with its characteristic observed
minimum of one-third of the ambient cluster temperature, could be
explained by a strong CR component in the intracluster medium
\cite{Cen2005}.

\begin{figure}
\begin{center}
\resizebox{8.3cm}{!}{\includegraphics{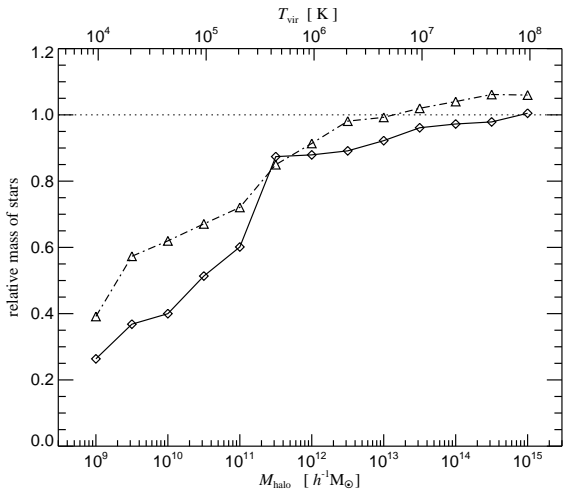}}%
\end{center}
  \caption{Relative suppression of
    star formation in simulations of isolated halos when a fraction of $0.3$
    of the initial thermal pressure is replaced by a CR component of equal
    pressure. We show results as a function of halo virial mass for two
    different times after the simulations were started, for $t=1.0\, {\rm
      Gyr}$ (solid line), and for $t=3.0\,{\rm Gyr}$ (dot-dashed). For halos
    of low mass, the cosmic ray pressure contribution can delay the cooling in
    the halos.
 \label{FigCoolingFlowEff} }
\end{figure}

We here want to get an idea about the potential strength of this
effect, and examine to this end a small set of toy simulations. To
this end we consider a series of self-similar dark matter halos with a
gas component that is initially in hydrostatic equilibrium, just as
before in Section~\ref{SecIsolatedDisks}.  In fact, we use the same
initial conditions as before, except that we replace a fraction of the
initial thermal pressure with cosmic ray pressure, keeping the total
pressure constant. For definiteness, we choose a spectral cut-off of
$q=1.685$ and a spectral index $\alpha=2.5$ for the initial CR
population. We then evolve the halos forward in time, including
radiative cooling processes for the thermal gas as well as cosmic ray
loss processes, but we disregard any sources of new cosmic ray
populations.  We are interested in the question whether the cooling
flows that develop in these halos are modified by the presence of the
cosmic rays. Note that some studies have predicted cosmic ray pressure
contributions of up to $\sim 50$ per cent in clusters of galaxies
\citep{Miniati2001,Ryu2003}. These fiducial test simulations can give
a first indication of the size of the change of the cooling rates if
these claims are indeed realistic.

In Figure~\ref{FigCoolingFlowEff}, we show the results of these simulations as
a function of halo mass, in the form of the integrated star formation rate
relative to an equivalent simulation without initial CR population.  The
cumulative star formation activity can here be taken as a proxy for the
integrated strength of the cooling flow in the halo.  We see that the total
star formation for cluster halos of mass \mhalo{15} is essentially unchanged
in the first $1-2\,{\rm Gyr}$ of evolution, while at late times, it is even
slightly increased.  For systems of significantly lower mass, the star
formation rates are however reduced in the CR case, by up to $\sim 40$ per
cent. This can be qualitatively understood based on a comparison of the
thermal radiative cooling time with the CR dissipative cooling time. As the
lower panel of Fig.~\ref{FigCoolTimes} has shown, the timescales are
comparable at the virial temperature corresponding to the \mhalo{15} halo, but
are quite different for lower temperatures, where the radiative cooling is
significantly faster. In fact, a naive comparison of these timescales would
perhaps suggest an even stronger suppression of the cooling efficiency in
systems of intermediate mass. In reality, the effect turns out to be much more
moderate.  This can be understood based on the softer equation of state of the
CR component, combined with the fact that its cooling timescale typically
declines faster than that of thermal gas when a composite gas is compressed.
As a result, the ability of a CR pressure component to delay thermal collapse
for a long time is quite limited, unless perhaps active sources for new
populations of CR particles are present.

\section{Cosmological Simulations}

Previous simulation work on the effects of cosmic rays on structure formation
has not accounted for the dynamical effects due to cosmic ray pressure,
i.e.~the effectiveness of cosmic ray production has only been estimated
passively. Interestingly though, these works suggested that the cosmic ray
production may be quite efficient, with up to $\sim 50\%$ of the pressure
being due to CRs \citep{Miniati2001,Ryu2003,RyuKang2003,RyuKang2004}.  Here we present
the first self-consistent cosmological simulations of CR production that also
account for the dynamical effects of cosmic rays. We first study the global
efficiency of cosmic ray production at structure formation shocks. We then
study the influence of cosmic ray feedback on star formation in cosmological
simulations, and on possible modifications of the Lyman-$\alpha$ forest.
Finally, we study the modification of the thermodynamic properties of the
intracluster medium in high-resolution simulations of the formation of a
cluster of galaxies.

\subsection{Cosmic ray production in structure formation shocks} \label{secshocks}

In this subsection, we examine the efficiency of cosmic ray production at
structure formation shock waves. To this end, we use simulations that include
cosmic ray production injection at shocks and the cosmic ray loss processes
(i.e.~Coulomb cooling and hadronic losses), but we disregard radiative cooling
and star formation. The cosmological model we have simulated is a concordance
$\Lambda$CDM model with parameters $\Lambda$CDM model, with parameters
$\Omega_0=0.3$, $\sigma_8=0.84$, baryon density $\Omega_b=0.04$. We have
picked a comoving box of side-length $L = 100\,h^{-1}{\rm Mpc}$, and simulated
each of our models at two resolutions, with $2\times 128^3$ and $2\times
256^3$ particles, respectively. The results of the two resolutions are in good
agreement with each other, so we restrict ourselves to reporting the results
of the higher resolution runs with $2\times 256^3$ particles in the following.

We compare three simulations that differ in the treatment of the cosmic ray
physics.  In our `full model', we account for shock injection
self-consistently, i.e.~we use the Mach number estimator of our companion
paper \citep{Pfrommer2005} to determine the energy content and the slope of
the proton populations accelerated at each shock front. This simulation hence
represents our best estimate for the overall efficiency of the CR production
process due to structure formation shocks. We contrast this simulation with a
model where the CR injection has been artificially maximized by adopting a
fixed efficiency $\zeta_{\rm inj}=0.5$ and a fixed injection slope
$\alpha=2.5$ for {\em all shocks}. Note that this high efficiency is normally
only reached as limiting case for high Mach number shocks, so that this model
also allows us to assess the importance of the dependence of the shock
injection efficiency on Mach number.  Finally, we compare these two models
with a reference simulation where no cosmic ray physics was included. This
reference simulation is hence a standard non-radiative calculation where only
shock-heating is included and the gas behaves adiabatically otherwise.

\begin{figure}
\begin{center}
\resizebox{8.4cm}{!}{\includegraphics{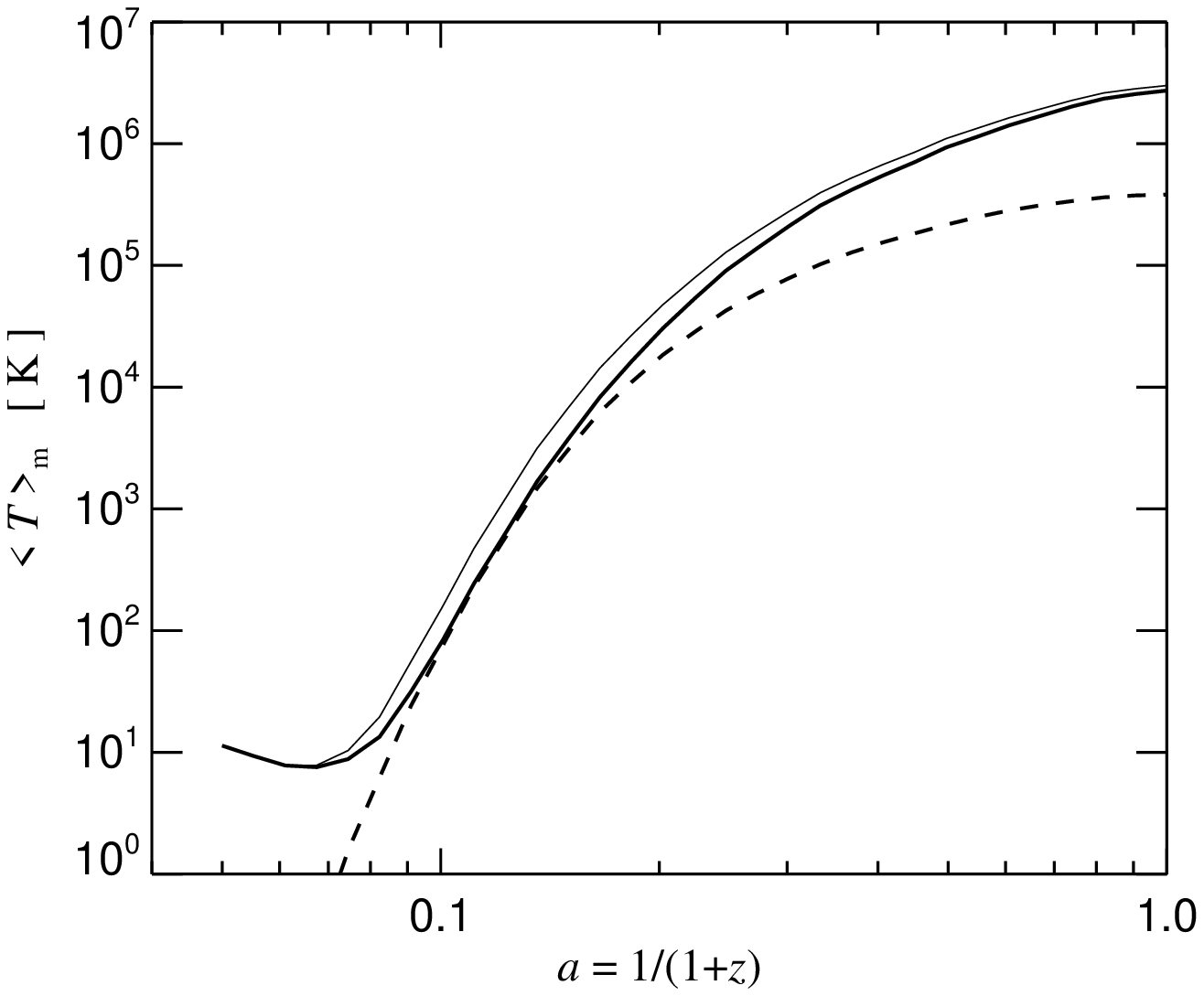}}\\%
\resizebox{8.4cm}{!}{\includegraphics{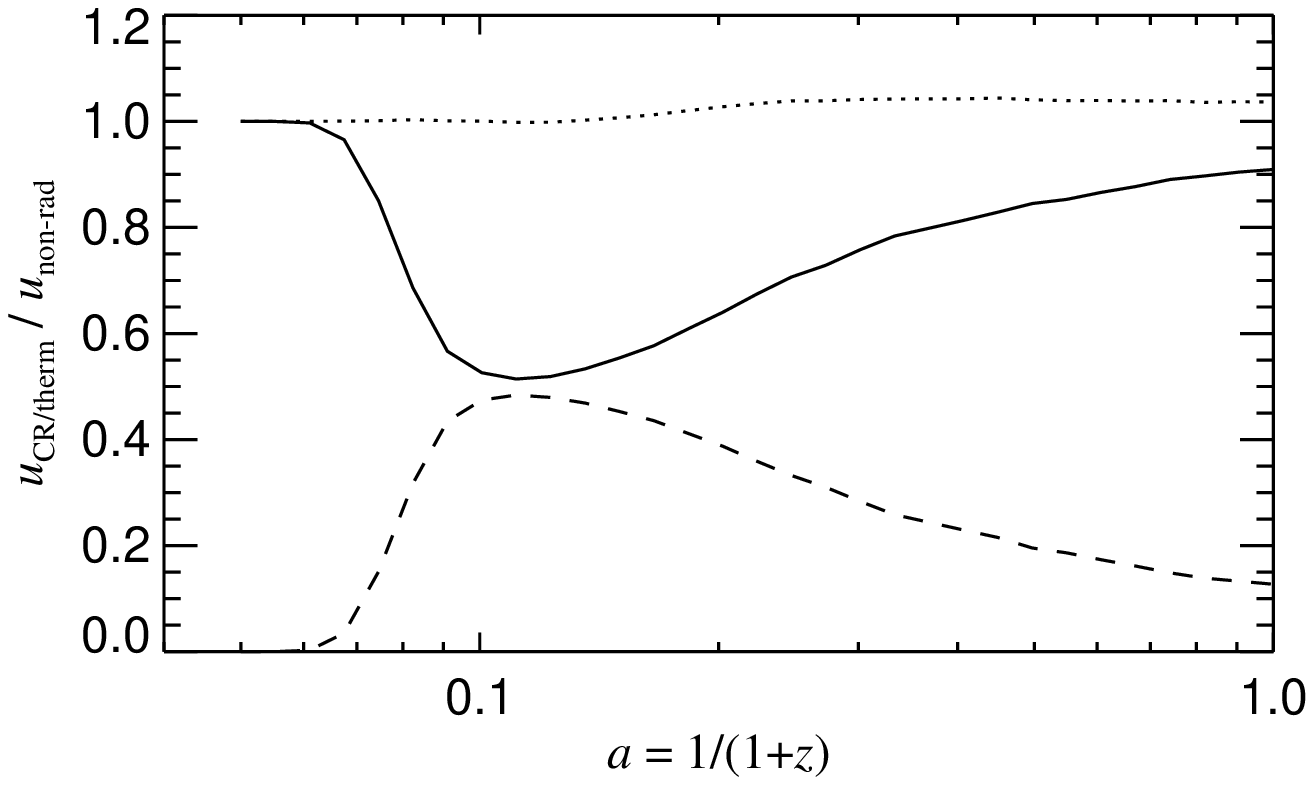}}%
\end{center}
  \caption{Time evolution of the mean thermal energy and the cosmic
    ray energy content of the gas in non-radiative cosmological simulations.
    In the top panel, the solid thick line shows the mass-weighted temperature
    for a simulation where the efficiency of cosmic ray production at
    structure formation shocks is treated self-consistently based on our
    on-the-fly Mach number estimator.  The dashed line is the corresponding
    mean cosmic ray energy, which we here converted to a fiducial mean
    temperature by applying the same factor that converts thermal energy per
    unit mass to temperature. For comparison, the thin solid line shows the
    evolution of the mean mass-weighted temperature in an ordinary
    non-radiative simulation without cosmic ray physics. In the bottom panel,
    we show the ratio of the mean thermal energy in the cosmic ray case
    relative to the energy in the simulation without cosmic rays (solid line),
    while the dashed line is the corresponding ratio for the cosmic ray
    component. Finally, the dotted line gives the total energy in the cosmic
    ray simulation relative to the ordinary simulation without cosmic rays.
 \label{FigTempVsTime}}
\end{figure}

In Figure~\ref{FigTempVsTime}, we compare the time evolution of the mean
mass-weighted temperature of the full cosmic ray model to that in the ordinary
non-radiative simulation. We also include a measurement of the mean energy in
cosmic rays, converted to a fiducial temperature using the same factor that
converts thermal energy per unit mass to temperature. Interestingly, at high
redshift the cosmic ray energy content evolves nearly in parallel to the
thermal energy, and both are roughly half what is obtained in the simulation
without cosmic rays. Apparently, the thermalization of gas is dominated by
strong shocks which reach the asymptotic injection efficiency of 50 percent.
At late times, however, the CR energy does not grow as quickly as the thermal
energy content any more, and the thermal energy in the CR simulation becomes
closer to the thermal energy in the ordinary simulation.

These trends become more explicit when the energy content in CRs and in the
thermal reservoir of the full CR simulation is divided by the thermal energy
content of the reference simulation, as shown in the bottom panel of
Figure~\ref{FigTempVsTime}.  Around redshifts $z\sim 6-10$, the CR energy
content nearly reaches the same value as the thermal energy in the full
CR-model, and their sum is essentially equal to the thermal energy in the
simulation without cosmic rays. Over time, the relative importance of the
cosmic rays declines, however, and the thermal energy in the full CR model
slowly climbs back to the value obtained in the non-radiative reference
simulation. At the same time, the sum of cosmic ray and thermal energy
obtained in the full model becomes a few percent higher at the end than that
in the simulation without cosmic rays, despite the fact that the CR simulation
loses some energy to radiation via the hadronic decay channels.  Apparently,
the simulation with cosmic rays extracts slightly more energy out of the
gravitational potential wells of the dark matter. An explanation for this
behaviour could derive from the fact that more energy needs to be invested
into CRs to reach the same pressure compared with ordinary thermal gas. This
should allow the gas in CR simulations to fall deeper into gravitational
potential wells before it is stopped by shocks and pressure forces, such that
more gravitational energy is liberated overall.

\begin{figure*}
\begin{center}
\resizebox{8.3cm}{!}{\includegraphics{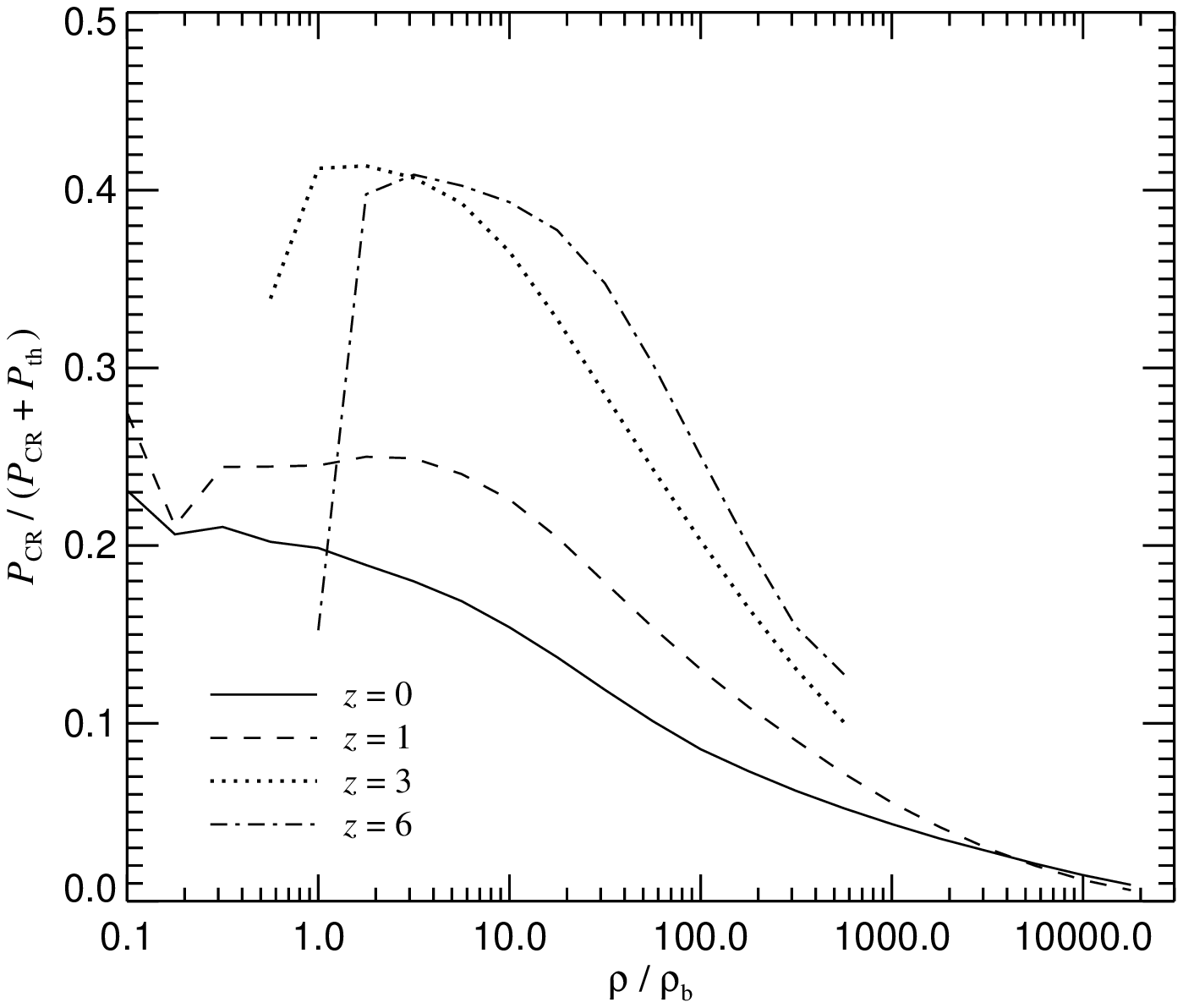}}%
\resizebox{8.3cm}{!}{\includegraphics{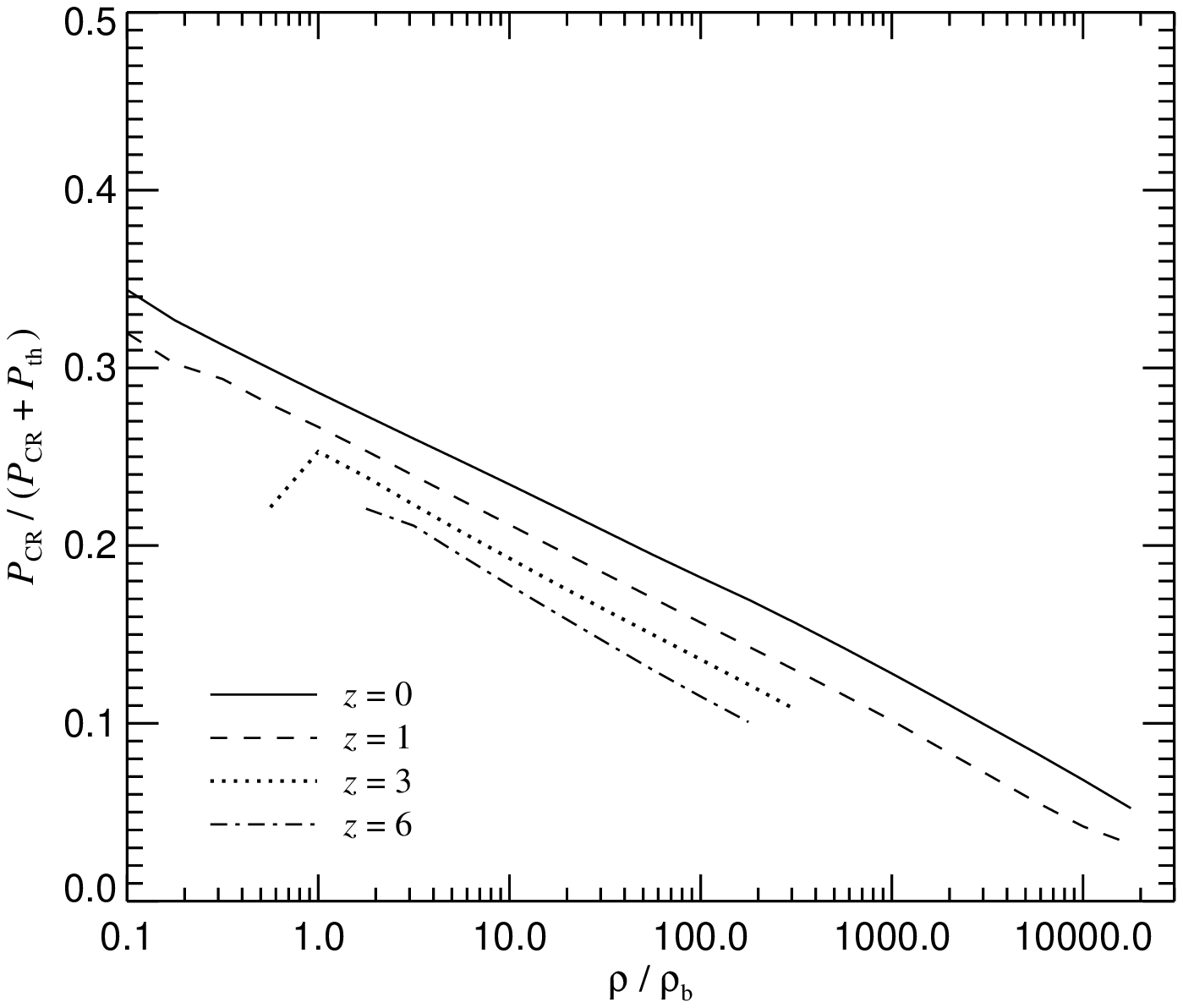}}%
\end{center}
  \caption{Mean relative contribution of the cosmic ray pressure to the
    total pressure, as a function of gas overdensity in non-radiative
    cosmological simulations. We show measurements at epochs $z=0$, 1, 3, and
    6. The panel on the left gives our result for a simulation where the
    injection efficiency and slope of the injected cosmic ray spectrum are
    determined based on our on-the-fly Mach number estimation scheme. For
    comparison, the panel on the right shows the result for a simulation with
    a fixed injection efficiency $\zeta_{\rm lin}=0.5$ and a soft spectral
    injection index of $\alpha_{\rm inj}=2.9$. Clearly, the relative
    contribution of cosmic ray pressure becomes progressively smaller towards
    high densities. It is interesting that the trends with redshift
    are reversed in the two cases.
    \label{FigCRPressVsDensity} }
\end{figure*}

It is also interesting to ask how the relative importance of cosmic
rays depends on gas density. This is addressed by
Figure~\ref{FigCRPressVsDensity}, where we show the relative
contribution of cosmic rays to the total gas pressure, as a function
of baryonic overdensity, separately for different redshifts.  We give
results both for the simulation with self-consistent shock injection
(left panel), and for the one where we imposed a constant injection
efficiency (right panel). In general, the importance of cosmic rays is
largest for densities around the mean cosmic density, and declines
towards higher densities. This is consistent with the expectation that
the strongest shocks occur at low to moderate overdensities in the
accretion regions around halos and filaments
\citep{Kang1996,Quilis1998,Miniati2002,Ryu2003,Pfrommer2005}, and also
with the growing importance of cosmic ray loss processes at high
densities.  Another interesting trend found in our self-consistent
simulation (the `full model) is that cosmic rays are particularly
important at high redshift, with a gradual decline towards lower
redshift, suggesting that the mean Mach number of shocks becomes lower
at later times, as is indeed confirmed by studies of the cosmic Mach
number distribution \citep{Ryu2003,Pfrommer2005}.  Note that this
trend is reversed in our fiducial simulation with a fixed shock
injection efficiency, where at all overdensities the relative
importance of CRs grows with cosmic time.  This emphasizes that a
correct accounting of the shock strengths is highly important even at
a qualitative level to correctly model the evolution of the cosmic ray
pressure distribution.  We note that an implicit assumption we made in
the above analysis is that weak magnetic fields are ubiquitous in the
universe, even at low density. Whether they really exist and where
they ultimately come from is an open question however.

\begin{figure*}
\begin{center}
\resizebox{8.3cm}{!}{\includegraphics{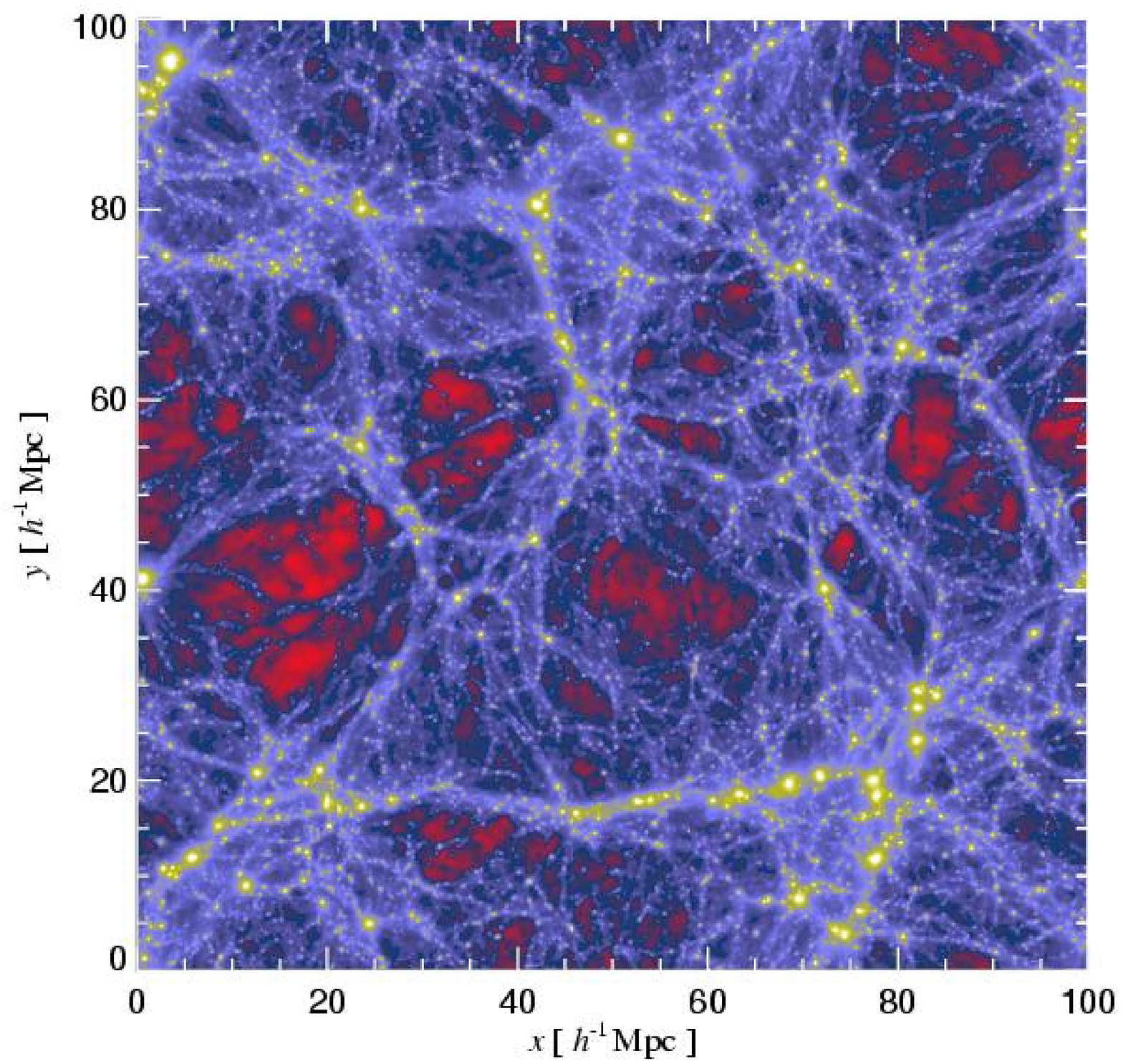}}%
\resizebox{8.3cm}{!}{\includegraphics{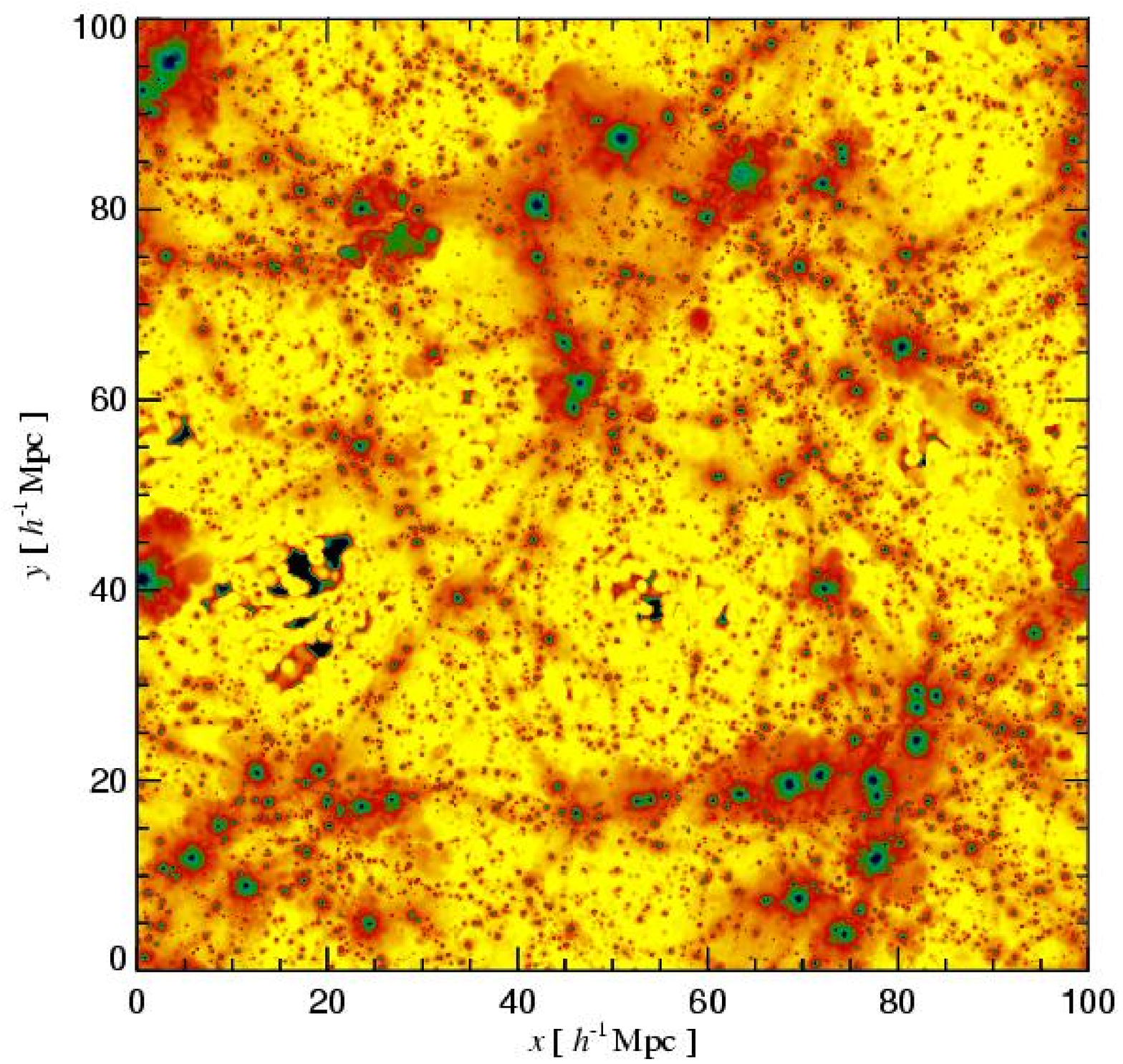}}%
\resizebox{!}{8.3cm}{\includegraphics{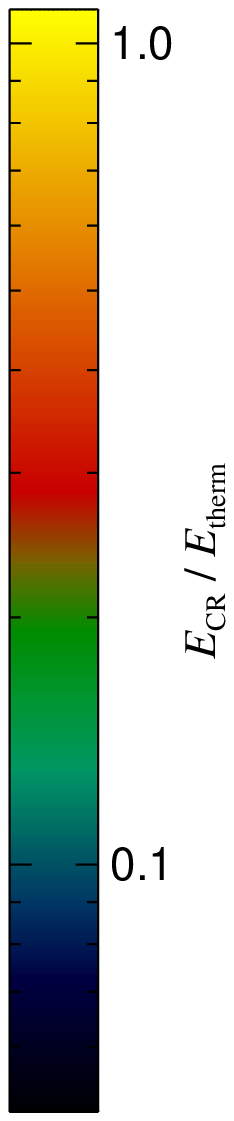}}
\end{center}
  \caption{Projected gas density field (left panel) in a slice of thickness 
    $20\,h^{-1}{\rm Mpc}$ through a non-radiative cosmological simulation at
    $z=0$. The simulation includes cosmic ray production at structure
    formation shocks using self-consistent efficiencies based on an on-the-fly
    Mach number estimation scheme. The panel on the right shows the ratio of
    the projected cosmic ray energy density to the projected thermal energy
    density. We clearly see that the local contribution of cosmic rays is
    largest in voids. It is also still large in the accretion regions around
    halos and filaments, but is lower deep inside virialized objects.
\label{FigCRTempMap}}
\end{figure*}

In Figure~\ref{FigCRTempMap}, we show the projected gas density field in a
slice through the simulation box at $z=0$.  To highlight the relative
importance of cosmic rays, the panel on the right shows the ratio of the
projected cosmic ray energy density to the projected thermal energy density.
The relative importance of CRs is clearly highest in the volume-filling gas at
low density. In the accretion regions around halos and filaments, the CR
contribution is still comparatively large, but the high-density regions inside
massive halos are avoided, in agreement with the results of
Figure~\ref{FigCRPressVsDensity}.  This raises the interesting question
whether cosmic rays may perhaps modify the state of the intergalactic medium
to the extent that the properties of Lyman-$\alpha$ absorption systems are
modified. The latter arise primarily in gas that is largely unshocked, so that
the effects might be weak even though the CR pressure contributions are
predicted to be on average large exactly at overdensities of a few. We will
come back to this question in Section~\ref{SecForest}.

\begin{figure}
\begin{center}
\resizebox{8.3cm}{!}{\includegraphics{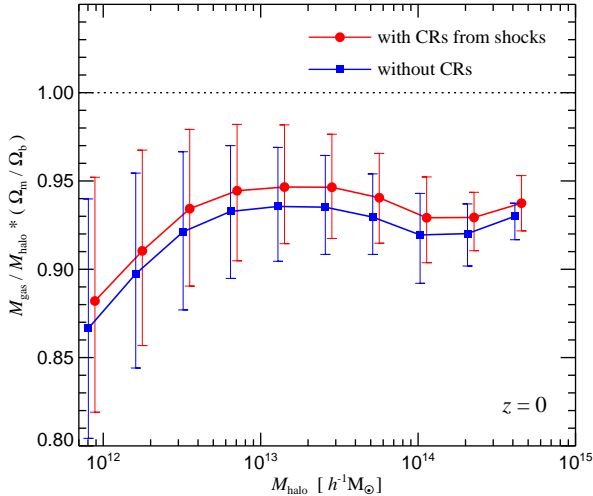}}%
\end{center}
  \caption{Mean baryon fraction within the virial radius as a function of halo
    mass, normalized by the universal baryon fraction. We compare results for
    two non-radiative simulations, one with cosmic ray production by shocks,
    the other without cosmic ray physics. The bars show the $1\sigma$
    scatter among the halos in each bin. When cosmic rays are included, the
    compressibility of the gas in halos becomes larger, leading to a slight
    increase of the enclosed baryon fraction.\label{FigBaryonFrac} }
\end{figure}

Another interesting question is whether the bulk properties of halos
are modified by the CR production at large-scale structure shocks. We
are for example interested in the question whether the concentration
of gas in halos is changed, which could manifest itself in a
modification of the mean gas mass inside dark matter halos.  To
examine this question we determine halo catalogues for our simulations
by means of the FOF algorithm with a standard linking-length of 0.2,
and measure the virial radii and masses by means of the spherical
overdensity algorithm.  In Figure~\ref{FigBaryonFrac}, we show the
mean baryonic mass fraction in halos as a function of halo mass for
the simulation with self-consistent CR injection and for the run
without cosmic ray physics. In both cases, the baryonic fraction
within the virial radius lies slightly below the universal baryon
fraction, reaching $\sim 0.91-0.94$ of it, and for poorly resolved
halos it drops a bit further.  Such a depression of the universal
baryon fraction is generally found in non-radiative SPH simulations
\citep[e.g][]{Frenk1999}.  However, a comparison of the two
simulations shows that the halos in the run with CRs have
systematically increased their baryonic fraction, albeit by only about
1-2 per cent of the universal baryon fraction. This is consistent with
expectations based on the higher compressibility of a composite gas
with thermal and CR components.

\begin{figure}
\begin{center}
\resizebox{8.3cm}{!}{\includegraphics{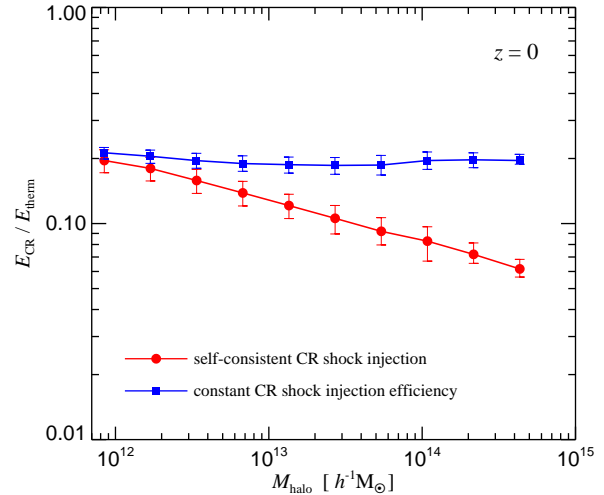}}%
\end{center}
  \caption{Ratio of energy in cosmic rays to thermal energy within the
    virialized region of halos, shown as a function of halo mass.  We compare
    results for two different non-radiative simulations, one treating the
    production of cosmic ray at shocks fronts using a self-consistent Mach
    number estimator, the other invoking a constant injection efficiency.  The
    bars give the $1\sigma$ scatter among the halos in each bin.
    Interestingly, the self-consistent injection scheme predicts a lower CR
    energy content in more massive systems. In contrast, a constant shock
    injection efficiency produced no significant trend of the CR energy
    content with halo mass.
 \label{FigCREnergyFrac}
}
\end{figure}

Using the group catalogues, we can also measure the mean cosmic ray energy
content inside the virial radius of halos. In Figure~\ref{FigCREnergyFrac}, we
show the ratio of cosmic ray to thermal energy as a function of halo mass. For
the simulation with a fiducial shock injection efficiency of $\zeta_{\rm
  inj}=0.5$ at {\em all shocks}, the ratio we obtain is $\sim 0.2$,
independent of halo mass.  Loss processes in the CR population and the
shallower adiabatic index of the CR component make this value much smaller
than expected in this case for the post-shock region of a single shock where
one would expect $\sim 0.5$.  In the simulation with self-consistent injection
of CRs, we find an interesting mass dependence where the ratio of
CR-to-thermal energy drops from about 0.2 for \mhalo{12} halos to $\sim 0.05$
for \mhalo{15} halos.  Apparently, for building up the thermal energy of
clusters of galaxies, weaker shocks and adiabatic compression are
comparatively more important than for galaxy-sized halos.  We note that the
value of $\sim 5-10\%$ we predict here for the CR energy content due to
structure formation shocks in clusters of galaxies is quite a bit lower than
previous estimates \citet{Miniati2001}. However, it is in good agreement with
CR constraints from gamma ray and radio observations
\citep{Pfrommer2003,Pfrommer2004}.

\subsection{Dwarf galaxy formation}   \label{secigm}

We now turn to studying the effects of cosmic ray feedback on galaxy formation
in cosmological simulations. We have already found that small galaxies should
be affected most. We hence expect small dwarf galaxies to be most susceptible
to sizable effects of CR feedback from star formation. To reach a reasonably
good mass resolution, we simulate periodic boxes of side-length
$10\,h^{-1}{\rm Mpc}$, using $2\times 256^3$ particles. This gives a mass
resolution of $6.62\times 10^{5}\,h^{-1}{\rm M}_\odot$ and $4.30\times
10^{6}\,h^{-1}{\rm M}_\odot$ in the gas and dark matter, respectively.  We
limit ourselves to evolving the simulations to a redshift of $z=3$, because at
lower redshift the fundamental mode of the small simulation volume would start
to evolve non-linearly, at which point the simulation as a whole could not be
taken as representative for the universe any more. We are hence restricted to
studying the high-redshift regime, but we expect that our results are
indicative for the trends that would be seen in the dwarf galaxy population
at lower redshifts as well, provided sufficiently well resolved simulations are
available.

We have simulated the same initial conditions three times, varying the cosmic
ray physics included. The first simulation is a reference run where we only
included radiative cooling and star formation but no cosmic ray physics. The
second simulation is a model where we also considered cosmic ray production by
the supernovae associated with star formation, using an efficiency of
$\zeta_{\rm SN}=0.35$.  Finally, our third simulation is a model where we in
addition included cosmic ray production by structure formation shocks, using
the self-consistent efficiencies derived from our on-the-fly Mach number
estimation scheme. The latter simulation hence represents our best estimate
for the total effect of cosmic rays on dwarf galaxy formation.

\begin{figure}
\begin{center}
\resizebox{8.3cm}{!}{\includegraphics{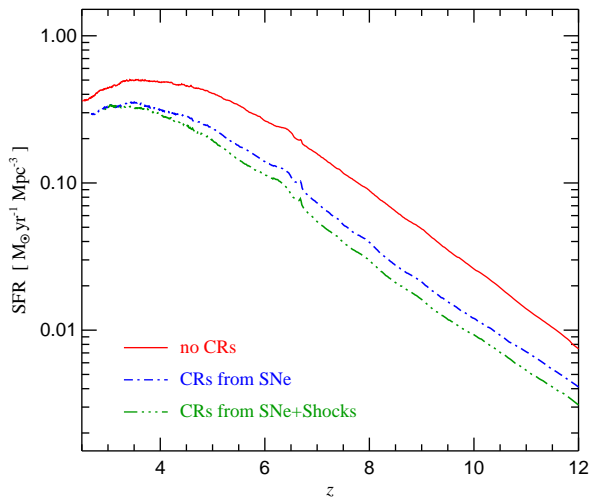}}%
\end{center}
  \caption{Evolution of the cosmic star formation rate density in
    simulations of galaxy formation at high redshift.
We compare results for three simulations that include different physics, a
reference simulation without cosmic ray physics, a simulation with CR
production by supernovae, and a third simulation which in addition accounts
for CR acceleration at structure formation shocks with an efficiency that
depends on the local Mach number.
\label{FigSFRCosmoEvolv}}
\end{figure}

In Figure~\ref{FigSFRCosmoEvolv}, we compare the cosmic star formation
histories of the three simulations up to a redshift of $z \sim 2.9$. The
incorporation of cosmic ray production by supernovae leads to a significant
reduction of the high redshift star formation activity, but the shape of the
star formation history, in particular its exponential rise, is not changed
significantly.  At these high redshifts, the star formation rate is dominated
by small halos which are affected strongly by CR feedback, so this result is
not unexpected given our previous findings. If CR production by structure
formation shocks is included as well, the star formation is reduced further,
although by only a small factor. This indicates that the cosmic ray pressure
component injected into forming halos indeed tends to slightly slow down the
cooling rates, consistent with the results we found for isolated halos.
Towards redshift $z\sim 3$, the differences in the star formation rates become
noticeably smaller however, suggesting that the low redshift star formation
histories will differ at most by a small amount.  Since the bulk of the star
formation shifts to ever larger mass scales at low redshift
\citep{SpringelSFRHistory2003}, this can be easily understood in terms of the
smaller influence of CR feedback on large halos.

\begin{figure}
\begin{center}
\resizebox{8.3cm}{!}{\includegraphics{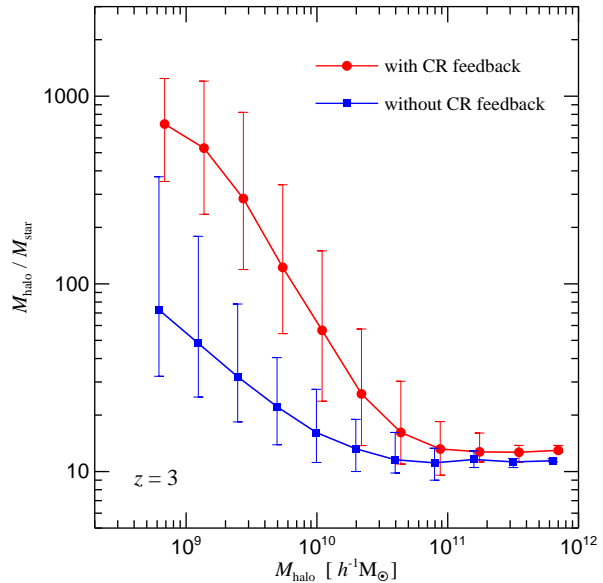}}\\%
\end{center}
\caption{Comparison of the averaged total mass-to-light-ratio 
  within the virial radius of halos formed in two high-resolution cosmological
  simulations up to $z=3$.  Both simulations follow radiative cooling and star
  formation, but one also includes CR-feedback in the form of cosmic
  production by supernovae, with an efficiency of $\zeta_{\rm SN}=0.35$ and an
  injection slope of $\alpha_{\rm SN}=2.4$. The  bars indicate the
  scatter among halos in the logarithmic mass bins (68\% of the objects lie
  within the range marked by the bars). Clearly, for halo masses below
  \mhalo{11}, CR feedback progressively reduces the overall star formation
  efficiency in the halos.
  \label{fig:cosmo:masslight}}
\end{figure}

In order to make the effects of CR feedback on small halos more explicit, we
determine halo catalogues in the simulations using a group finder.  We are
especially interested in the question how the efficiency of star formation is
changed by the inclusion of cosmic rays as a function of halo mass.
In~Figure~\ref{fig:cosmo:masslight}, we show the total-to-stellar mass ratios
of these groups as a function of halo mass, both for the simulation with CR
production by supernovae, and for the simulation without cosmic ray feedback.
The simulation where also CR production by shocks is included is quite similar
on this plot to the simulation that only accounts for CR from supernovae, and
is therefore not shown. The symbols show the mean total-to-stellar mass ratio
in small logarithmic mass bins, while the  bars indicate the scatter by
marking the central 68\% percentile of the distribution of individual halos.
The results show clearly that the star formation is significantly reduced by
CRs for low-mass halos, by factors of up to $\approx 10$ for host halo masses
of $\sim 10^{10}\Msol h^{-1}$ and below. On the other hand, the amount of
stars produced in massive halos is hardly changed. It is particularly
interesting that the effect of CRs manifests itself in a gradual rise of the
total-to-stellar mass ratio towards lower masses. This can be interpreted as a
prediction for a steeply rising `mass-to-light' ratio towards small halo
masses, which is exactly what appears to be needed to explain the observed
luminosity function of galaxies in the $\Lambda$CDM concordance model. The
problem is here that the halo mass function increases steeply towards low mass
scales. If the mass-to-light ratio is approximately constant for low masses,
this leads to a steeply rising faint end of the galaxy luminosity function, in
conflict with observations. However, a steeply rising mean mass-to-light ratio
towards low mass halos could resolve this problem and provide a suitable
`translation' between the halo mass function and the galaxy luminosity
function.

We note that conditional luminosity function analysis of the 2 Degree Field
Galaxy Redshift Survey (2dFGRS) has shown \citep{vdB03ML} that there appears
to be a minimum in the observed mass-to-light ratio of galaxies around a halo
mass of $\approx 3\times$\mhalo{11}. This feature is reproduced surprisingly
well in our simulations, although even with CR feedback included, the rise of
the stellar mass to light ratio towards low masses appears to be not as sharp
as required based on their analysis.

\begin{figure}
\begin{center}
\resizebox{8.3cm}{!}{\includegraphics{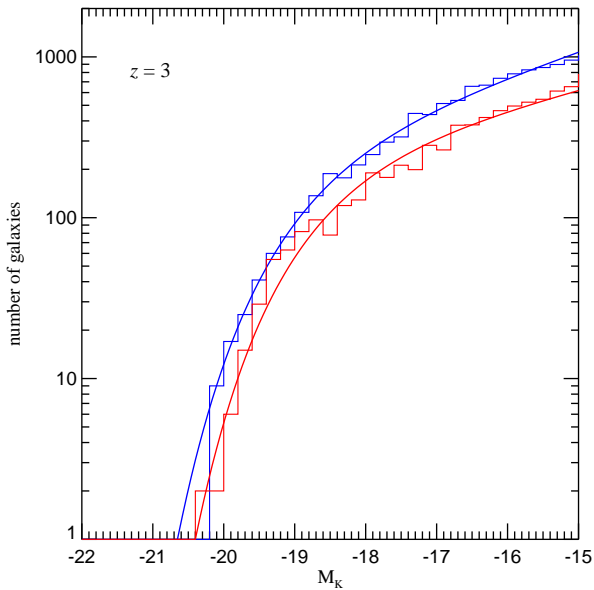}}\\%
\end{center}
  \caption{The K-band galaxy luminosity functions at $z=3$ in  two 
    high-resolution cosmological simulations. One of the simulations follows
    ordinary radiative cooling and star formation only (blue), the other additionally
    includes cosmic ray production by supernovae (red). The latter reduces the
    faint-end slope of the Schechter function fit (solid lines) to the data
    measured from the simulations (histograms). It is reduced from $-1.15$ to
    $-1.10$ in this case.
           \label{FigLumFunc}}
\end{figure}

However, one needs to caution that the results of
Fig.~\ref{fig:cosmo:masslight} cannot be naively translated into changes of
the faint-end slope of the luminosity function, as seen when we directly
compare the K-band luminosity functions at $z=3$. To determine those, we
identify individual groups of stars as galaxies using a modification of the
{\small SUBFIND} algorithm \citep{Springel2001} for detecting bound
substructures in halos. For each of the galaxies, we estimate magnitudes in
standard observational band based on \citet{Bruzual2003} population synthesis
models. In Figure~\ref{FigLumFunc}, we compare the resulting restframe K-band
luminosity functions at $z=3$ for the simulations with CR feedback by
supernovae and the simulation without any cosmic ray physics. We see that both
luminosity functions are well fit by Schechter functions, with faint-end
slopes of $\alpha=1.15$ and $\alpha=1.10$, respectively, for the cases without
and with CR feedback. We hence find that CRs only mildly reduce the faint-end
slope despite their differential reduction of the star formation efficiency
towards low mass scales. The result needs to be taken with a grain of salt
though, as the faint-end slope could still be influenced by resolution effects
in these simulations. A final assessment of the importance of CR feedback in
shaping the faint-end of the galaxy luminosity function needs therefore await
future simulations with substantially increased resolution.

\subsection{Cosmic ray effects on the intergalactic medium\label{SecForest}}

As the Mach number distribution is dominated by strong shocks at high
redshift, we expect that cosmic ray production is particularly efficient at
early epochs and at the comparatively low densities where the strongest shocks
occur, provided sufficient magnetization of the IGM existed to allow CR
acceleration to operate. Also, the thermalization time scales of cosmic rays
are quite long at low densities.  Figure~\ref{FigCRPressVsDensity} has shown
that the mean energy content of cosmic rays can reach a sizable fraction of
the thermal energy content at around redshift $z\sim 3$, suggesting a
potentially important influence on the intergalactic medium at this epoch.
Note however that in computing the results of Figure~\ref{FigCRPressVsDensity}
we had neglected cosmic reionization, which will boost the thermal energy
relative to the cosmic ray content. Also, large parts of the IGM at $z=3$,
particularly those responsible for the absorption seen in the Lyman-$\alpha$
forest, consist largely of unshocked material.  Whether the Lyman-$\alpha$
forest might show any trace of the influence of cosmic rays is therefore an
interesting and open question.

\begin{figure}
\begin{center}
\resizebox{8.3cm}{!}{\includegraphics{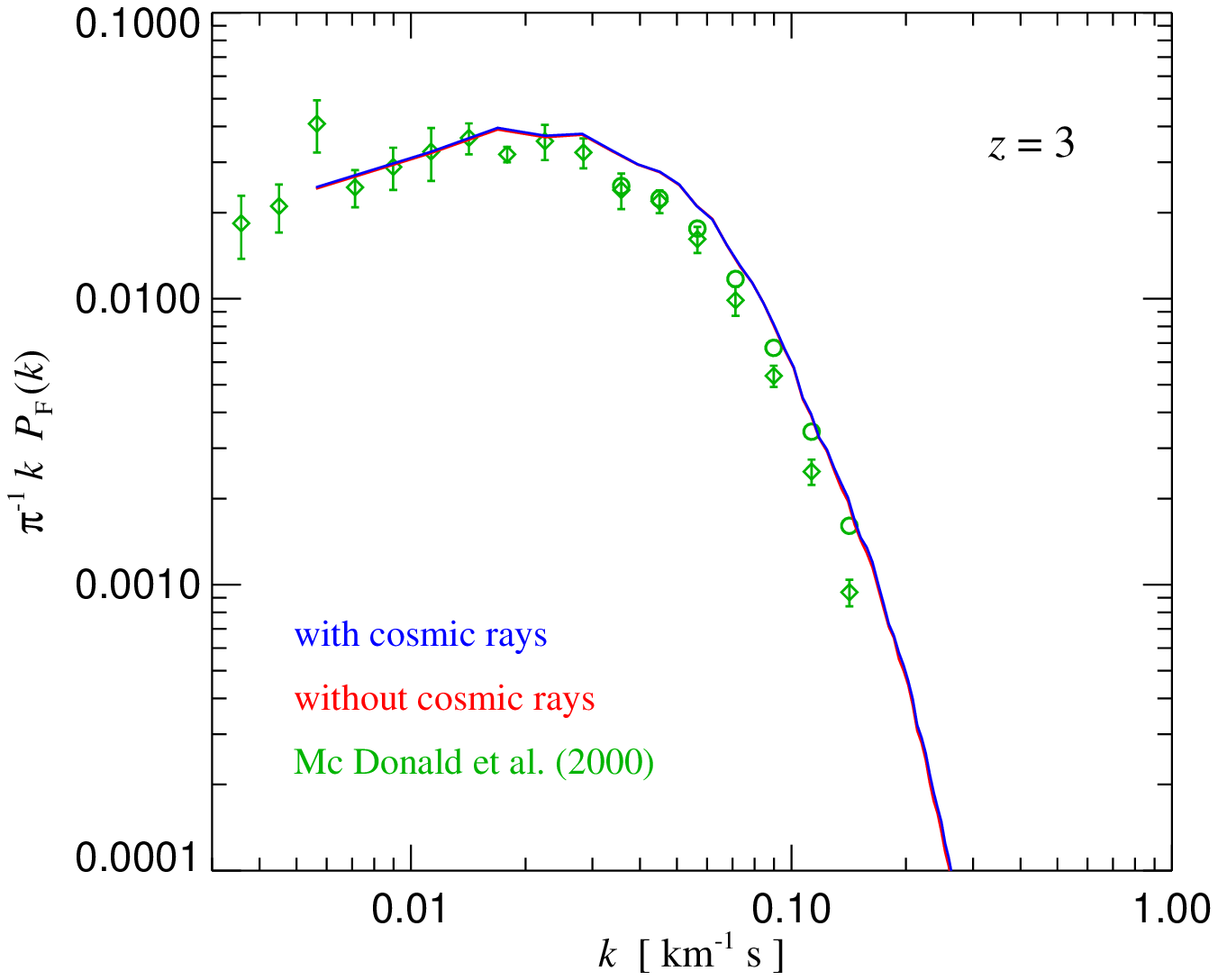}}\\%
\resizebox{8.3cm}{!}{\includegraphics{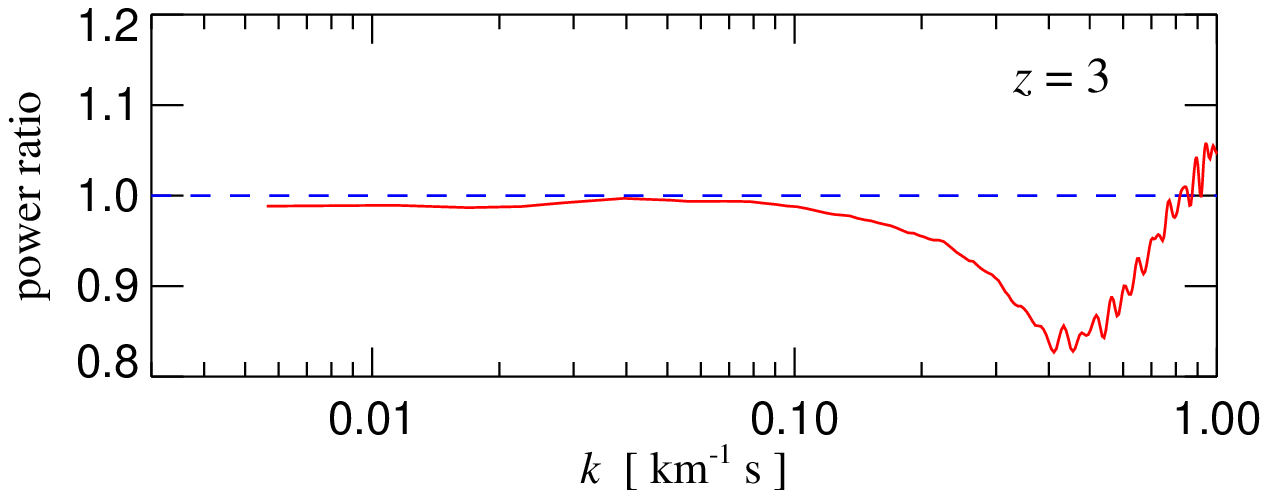}}%
\end{center}
  \caption{Ly-$\alpha$ flux power spectrum (top) at $z=3$ in simulations with and
    without cosmic ray production in structure formation shocks. The results
    lie essentially on top of each other, and only by plotting their ratio
    (bottom panel), it is revealed that there are small differences. In the
    simulation with cosmic rays, the power is suppressed by up to $\sim 15\%$
    on scales $0.1\,{\rm km^{-1}\, s} < k < 0.7\,{\rm km^{-1}\,s}$, while
    there is an excess on still smaller scales. However, on large scales $k<
    0.1\,{\rm km^{-1}\,s}$, which are the most relevant for determinations of
    the matter power spectrum from the Ly-$\alpha$ forest, the power spectrum
    is not changed by including CR physics. For comparison, we have also
    included observational data from \citet{McDonald2000} in the top panel
    (the open symbols are corrected by removing metal lines). A slightly
    warmer IGM in the simulations could account for the steeper thermal
    cut-off observed in the data.
           \label{FigLyAlph}}
\end{figure}

To investigate this question further, we have computed Ly-$\alpha$ absorption
spectra for the cosmological simulations with $10\, h^{-1}{\rm Mpc}$ boxes
analysed in the previous section. The two simulations we have picked both
include radiative cooling, star formation, and heating by a spatially uniform
UV background bases on a slightly modified \citet{Haardt1996} model, with
reionization at redshift $z=6$. While one of the simulations did not account
for any cosmic ray physics, the other included cosmic ray production by
large-scale structure shocks and supernovae, as well as dissipative loss
processes in the CR population.

For both simulations, we computed Lyman-$\alpha$ absorption spectra for 2048
lines of sights, along random directions parallel to the principal axes of the
simulation boxes. By slightly adjusting the UV intensity, we have renormalized
the spectra to the same mean transmission of $\left<\tau\right>=0.68$. A
direct comparison of the spectra along the same lines-of-sight through the two
simulations shows essentially perfect agreement, with very small residuals.
This already indicates that any systematic difference between the simulations
must be quite subtle, if present. To objectively quantify this, we have
computed the average 1-d flux power spectra for the two cases and compare them
in Figure~\ref{FigLyAlph}. The top panel compares the two flux spectra
directly with each other, and to observational data of \citet{McDonald2000}.
The results for the two simulations lie essentially on top of each other in
this representation. The agreement with observational data is good, apart from
a small excess of power on small scales, which can however be understood as a
consequence of the too cool temperature of the IGM in our simulations compared
with observations.

\begin{figure}
\begin{center}
\resizebox{8.3cm}{!}{\includegraphics{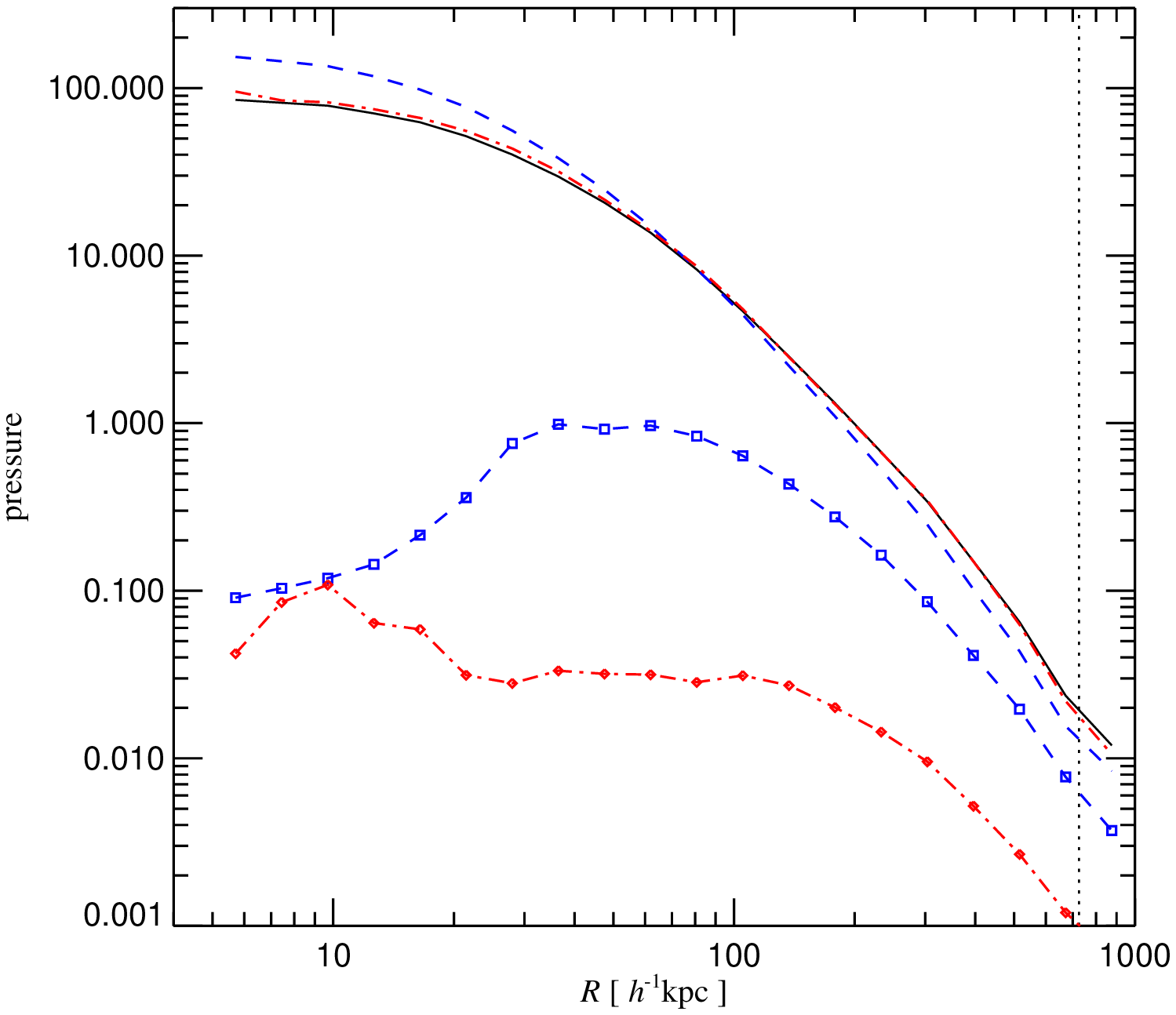}}\vspace*{-0.3cm}\\%
\resizebox{8.3cm}{!}{\includegraphics{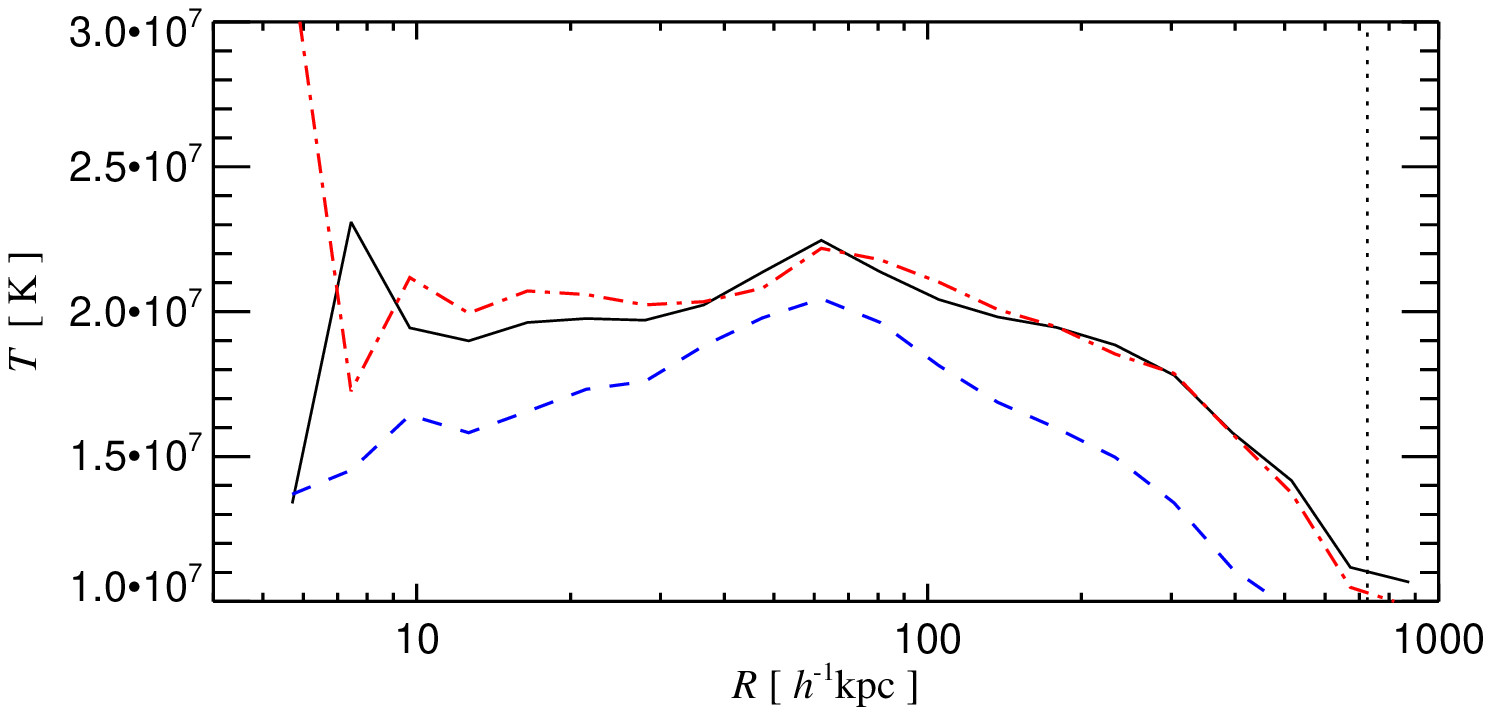}}\vspace*{-0.3cm}\\%
\resizebox{8.3cm}{!}{\includegraphics{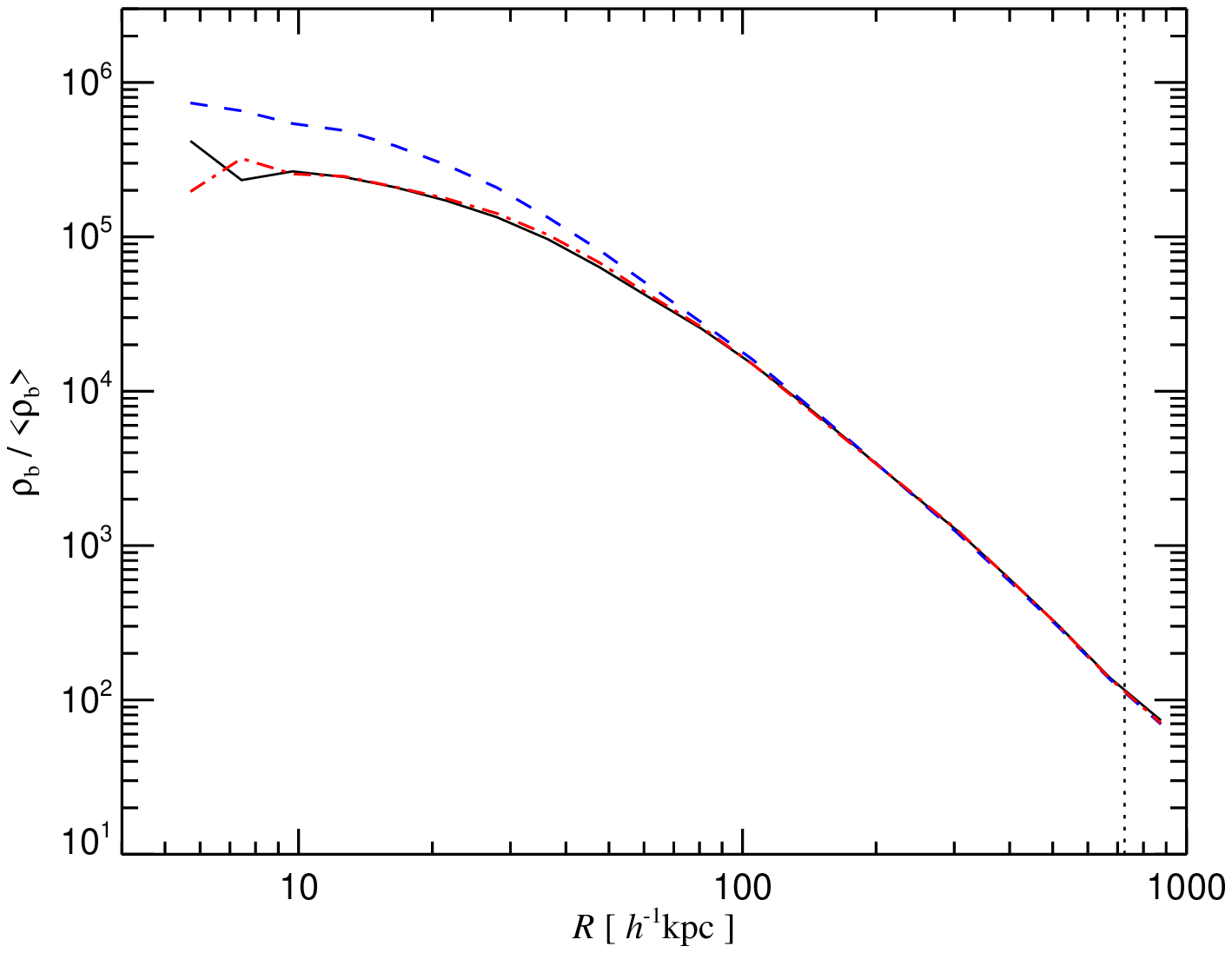}}\vspace*{-0.5cm}\\%
\end{center}
  \caption{Spherically averaged radial profiles 
    of thermodynamic gas properties in three re-simulations of the same
    cluster of galaxies. All three simulations were not following radiative
    cooling and star formation, and the reference simulation shown with a
    solid line does not include any CR physics. However, the simulation shown
    with a dashed line accounted for CR production at structure formation
    shocks with a fixed efficiency ($\zeta_{\rm inj}=0.5$, $\alpha_{\rm
      inj}=2.9$) while for the simulation shown with dot-dashed lines, the
    shock injection efficiency was determined self-consistently based on our
    Mach number estimation scheme. The panel on top compares the thermal
    pressure in the three simulations. For the two simulations with cosmic
    rays, we additionally plot the CR-pressure, marked with symbols. The panel
    in the middle compares the gas temperature, while the panel on the bottom
    shows the radial run of the gas density. The vertical dotted line marks
    the virial radius of the cluster.
           \label{FigClusterNonRad}}
\end{figure}

More interesting is perhaps an examination of the ratio of the flux power
spectrum with cosmic rays to that without cosmic rays, as shown in the bottom
panel of Figure~\ref{FigLyAlph}.  While for large-scale modes with
$k<0.1\,{\rm km^{-1}s}$, no noticeable differences are seen, there is a
5-15\% reduction of power in the wave-length range $0.1\,{\rm km^{-1}s} < k
< 0.7\, {\rm km^{-1}s}$, and at still smaller scales, the difference
changes sign and turns into a growing excess of power in the CR simulation.
These effects of CRs on the Ly-$\alpha$ therefore lie in a regime that is
normally not used to constrain the matter power spectrum with Lyman-$\alpha$
forest data, at least in conservative treatments that focus on $k < 0.03\,
{\rm km^{-1}s}$ \citep{Viel2004}.  In general we hence find that the
effects on the Lyman-$\alpha$ forest are very small and subtle; the forest
survives CR injection by large-scale structure shocks essentially unaltered,
even though they contribute a sizable fraction to the mean energy content of
the gas due to shock dissipation at densities at and around the mean density
of the universe. Note that our simulations did not allow for a possible
diffusion of CRs, but it seems unlikely that including this effect could
change this conclusion.

\subsection{Formation of clusters of galaxies}

In this section, we study in more detail the influence of cosmic rays on
individual halos formed in cosmological simulations. We focus on
high-resolution `zoom' simulations of the formation of a massive cluster of
galaxies. Such `zoom' simulations are resimulations of an object identified in
a cosmological structure formation simulation with large box-size. Once the
object of interest has been selected, its particles' are traced back through time
to their origin in the unperturbed initial conditions. The Lagrangian region of the
cluster thus identified is then populated with many more particles of lower
mass, thereby increasing the local resolution, while in regions further away,
the resolution is progressively degraded by using ever more massive particles.
In this way, the computational effort can be concentrated in the object of
interest, while at the same time its cosmological environment is still
accounted for accurately during its formation.

We have computed 6 resimulations of the same cluster of galaxies, using
different models for the physics of radiative cooling, star formation, and
cosmic rays. The cluster has been selected from a set of zoomed cosmological
initial conditions originally constructed by \cite{Dolag04} and has a virial
mass of $\approx$\mhalo{14} at redshift $z=0$.  The gas particle mass is
$1.6\times$\mhalo{8} in the high resolution region, implying that the cluster
is resolved with roughly 300000 gas and 300000 dark matter particles within
the virial radius. The gravitational softening length for the simulations was
kept fixed in comoving units at redshifts $z\ge 5$, and then held constant in
physical units at a value of $5\, h^{-1}{\rm kpc}$ at lower redshifts.

Our 6 simulations fall into two groups of 3 each. In the first group,
we have not included radiative cooling processes and star
formation. Here we use a non-radiative (`adiabatic') simulation as a
reference run, and compare it with two simulations that include cosmic
ray production at large-scale structure formation shocks, one of them
using the self-consistent Mach-number dependent injection efficiency,
and the other a fixed efficiency of $\zeta_{\rm inj}=0.5$ with
$\alpha_{\rm inj}=2.5$ for all shocks.  This set hence parallels the
types of simulations analyzed in section~\ref{secshocks}.  In the
second set of 3 simulations, we include radiative cooling and star
formation. Again, we consider one reference simulation without any
cosmic ray physics, and compare it with two simulations where cosmic
rays are included. These two consist of one run where cosmic rays are
only injected by supernovae associated with star formation (using
$\zeta_{\rm SN}=0.35$, and $\alpha_{\rm SN}=2.4$), while the other
also accounts for cosmic rays produced at shock waves. The second set
of simulations hence corresponds to the types of simulations analyzed
in section~\ref{secigm}. In all simulations with cosmic rays, we have
assumed and index $\alpha=2.5$ and included Coulomb cooling and
hadronic losses for the CR populations.

In Figure~\ref{FigClusterNonRad}, we compare spherically averaged radial
profiles of pressure, temperature, and gas density for the three non-radiative
simulations. For the pressure, we show the thermal as well as the cosmic ray
pressure for the two runs with cosmic ray physics. Interestingly, the cosmic
ray pressure component is substantially below the thermal one, even in the
fiducial case of a constant shock injection efficiency of $\zeta_{\rm inj}=0.5$
for all shocks.  However, in this case the thermal pressure is substantially
elevated compared to the run without cosmic rays. This goes along with an
increase of the gas density in the inner parts, and a reduction of the
thermal temperature throughout the cluster volume. This is the expected
behaviour based on the higher compressibility of the gas in this
case.

\begin{figure}
\begin{center}
\resizebox{8.3cm}{!}{\includegraphics{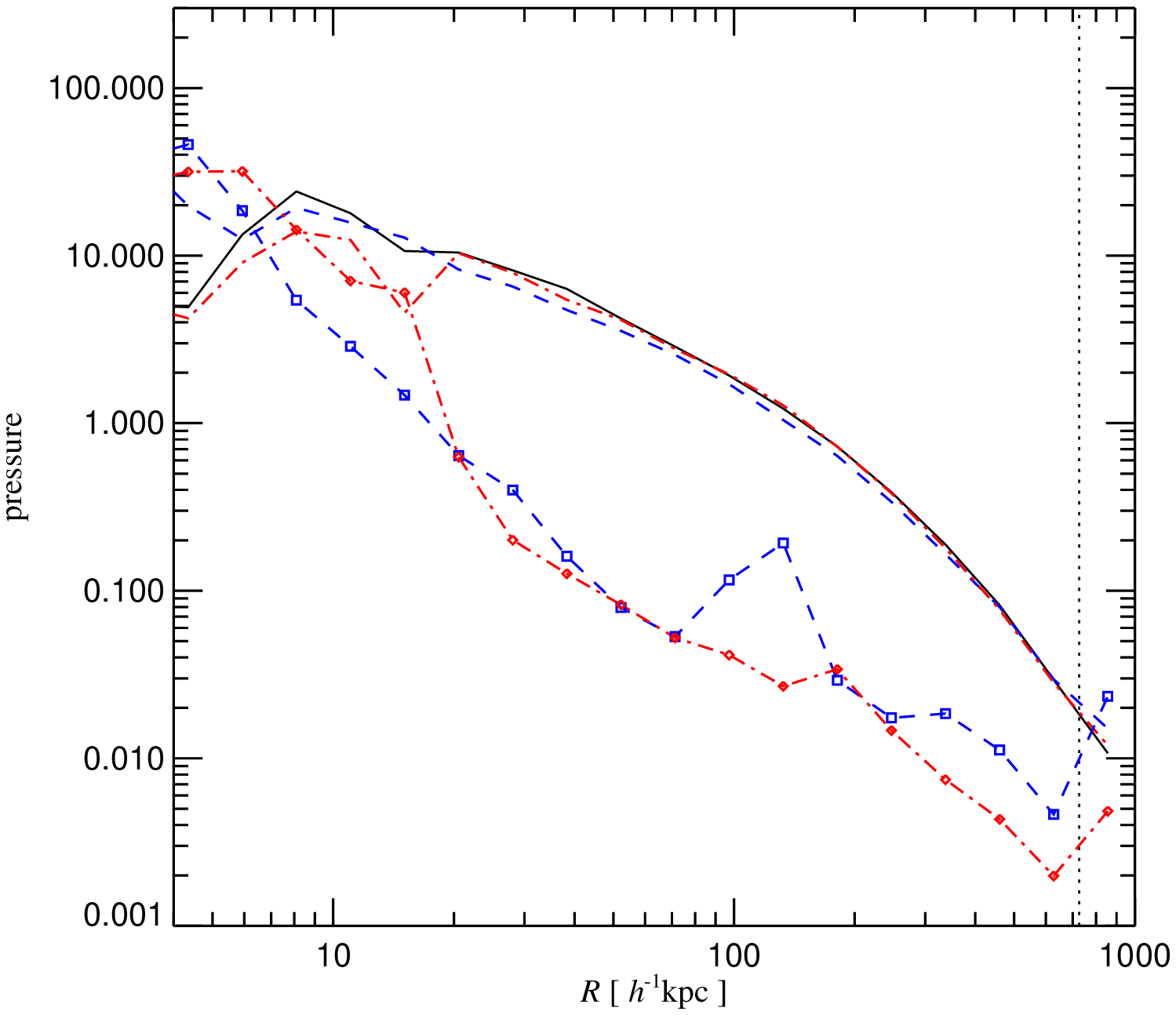}}\vspace*{-0.3cm}\\%
\resizebox{8.3cm}{!}{\includegraphics{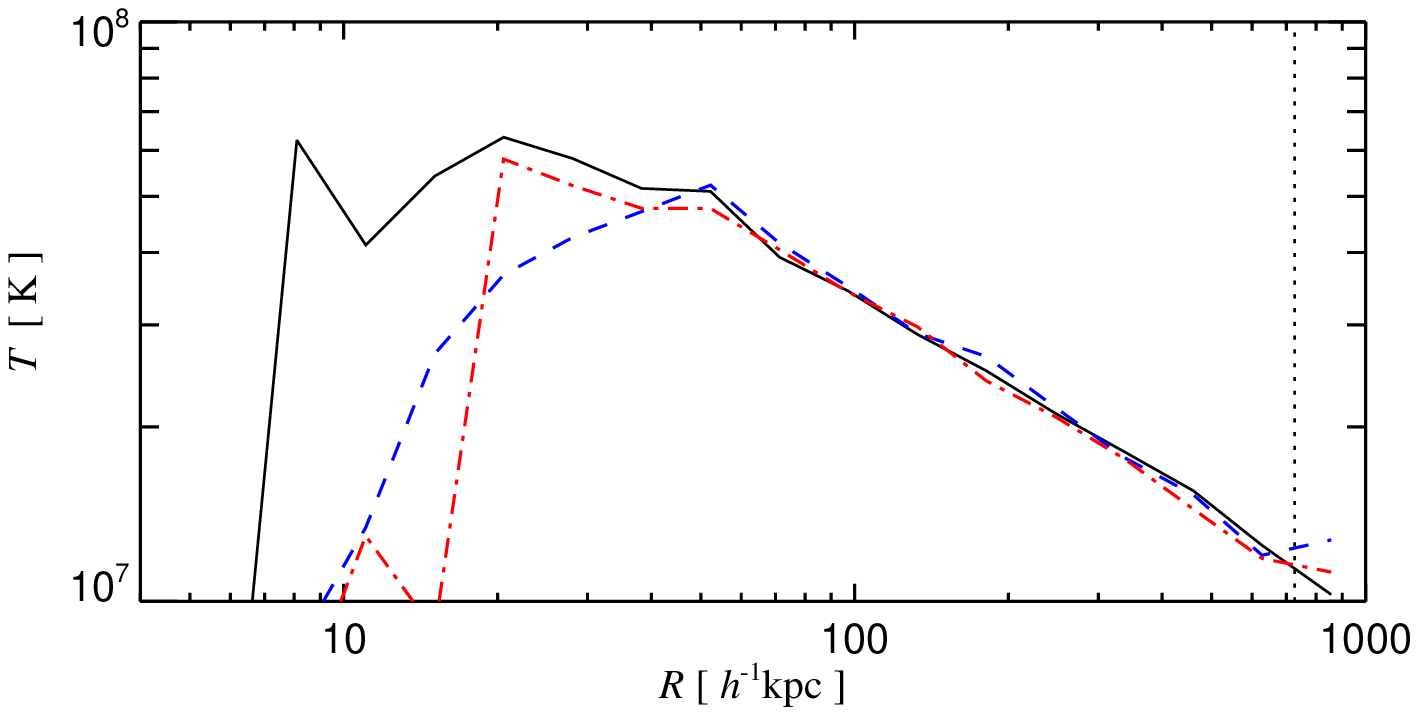}}\vspace*{-0.3cm}\\%
\resizebox{8.3cm}{!}{\includegraphics{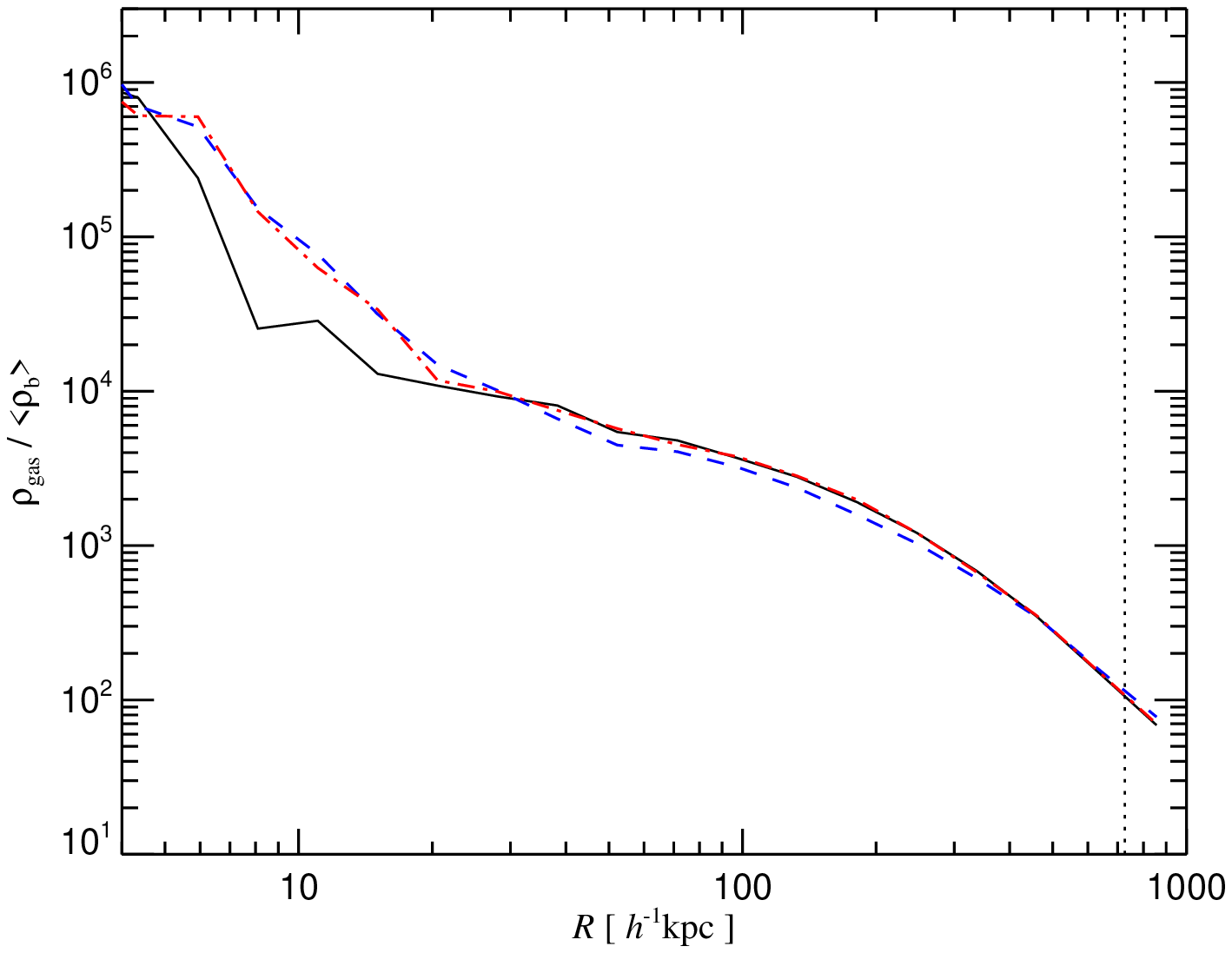}}\vspace*{-0.5cm}\\%
\end{center}
  \caption{Spherically averaged radial profiles of thermodynamic gas properties in
    three re-simulations of the same cluster of galaxies. All three
    simulations included radiative cooling of the gas, star formation and
    supernova feedback. The solid lines show the results of a reference
    simulation which did not include any cosmic ray physics. The other two
    simulations included CR production by supernovae, and the one shown with
    dot-dashed lines in addition accounted for CR injection at structure
    formation shocks, using self-consistent efficiencies based on our Mach
    number estimation scheme.  The panel on top compares the thermal pressure
    in the three simulations. For the two simulations with cosmic rays, we
    additionally plot the CR-pressure marked with symbols. The panel in the
    middle compares the gas temperature for the three cases, and the panel on
    the bottom shows the radial run of the gas density. 
           \label{FigClusterRad}}
\end{figure}

\begin{figure}
\begin{center}
\resizebox{8.3cm}{!}{\includegraphics{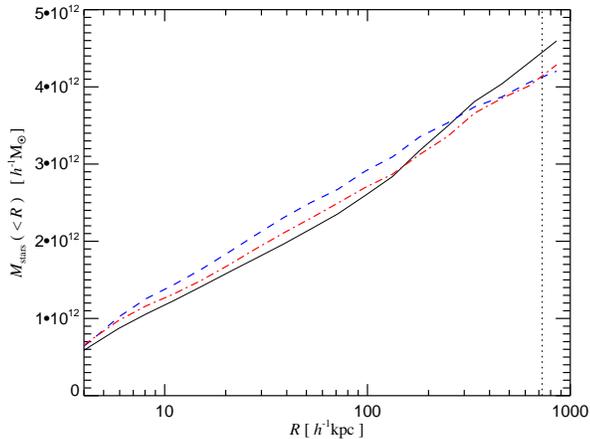}}\\%
\end{center}
  \caption{Cumulative radial stellar mass profile 
    in three re-simulations of the same cluster of galaxies. The simulations
    are the same ones also shown in Figure~\ref{FigClusterRad}. The solid line
    gives the result for a reference simulation without CR physics, the dashed
    line includes CR production by supernovae, and the dot-dashed line
    additionally accounts for CR injection at structure formation shocks. The
    vertical dotted line marks the virial radius of the cluster.
           \label{FigClusterStars}}
\end{figure}

However, the cosmic ray pressure in the simulation with a self-consistent
injection efficiency is substantially smaller, and even at the virial radius
is at most $\sim 10$ percent of the thermal pressure, while for much of the
inner parts, $r\le 100\,h^{-1}{\rm kpc}$, the cosmic ray pressure contribution
drops to the percent level and below. Obviously, cosmic rays are not produced
efficiently enough at shocks to fill much of the ICM with a dynamically
significant cosmic ray pressure component, consistent with the results of
Figure~\ref{FigCREnergyFrac}. Consequently we find that in this case the
profiles of gas density, temperature and thermal pressure are very similar to
the corresponding results for the simulation without cosmic ray physics.

\begin{figure*}
\begin{center}
\resizebox{8.0cm}{!}{\includegraphics{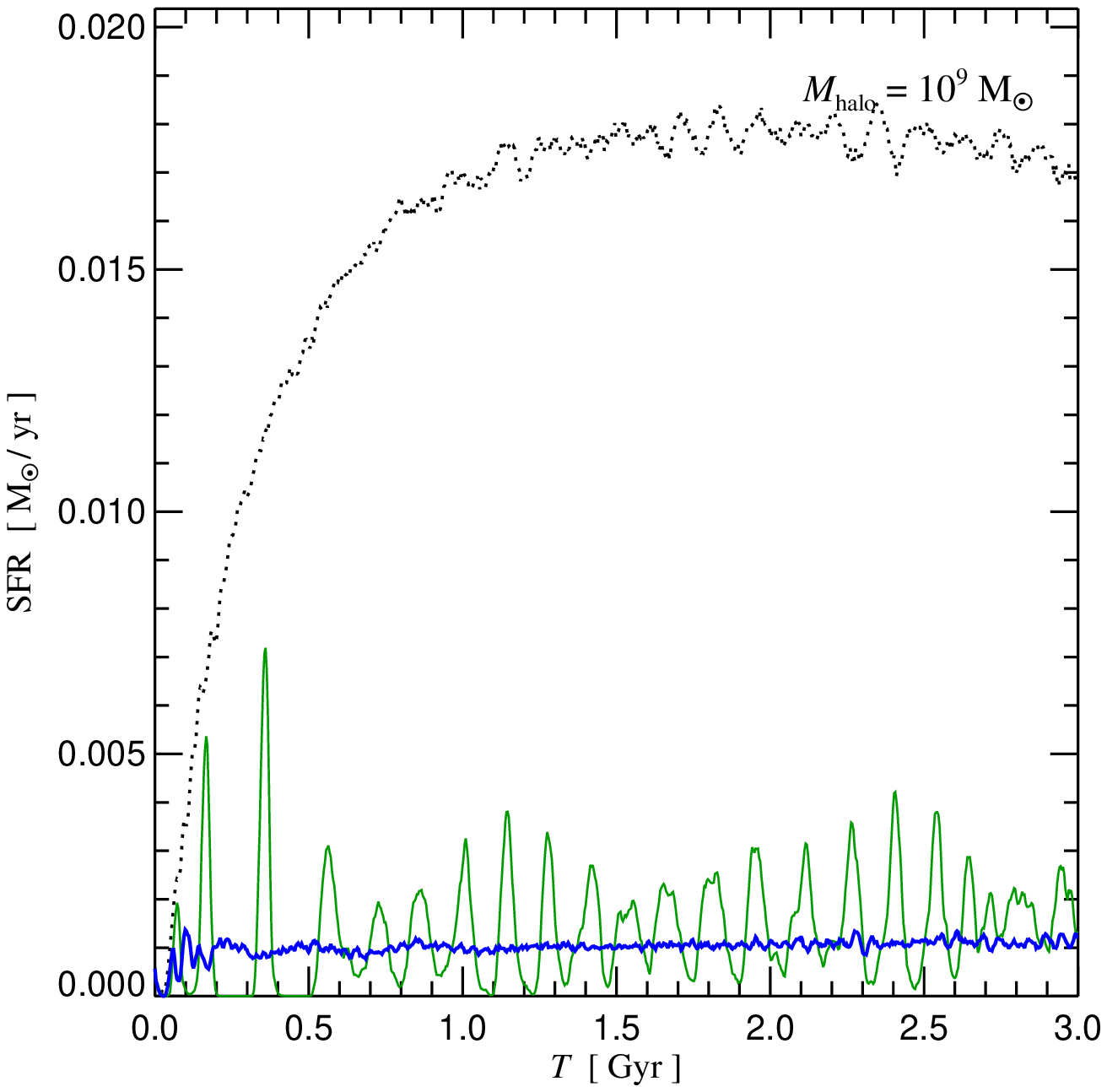}}%
\resizebox{8.0cm}{!}{\includegraphics{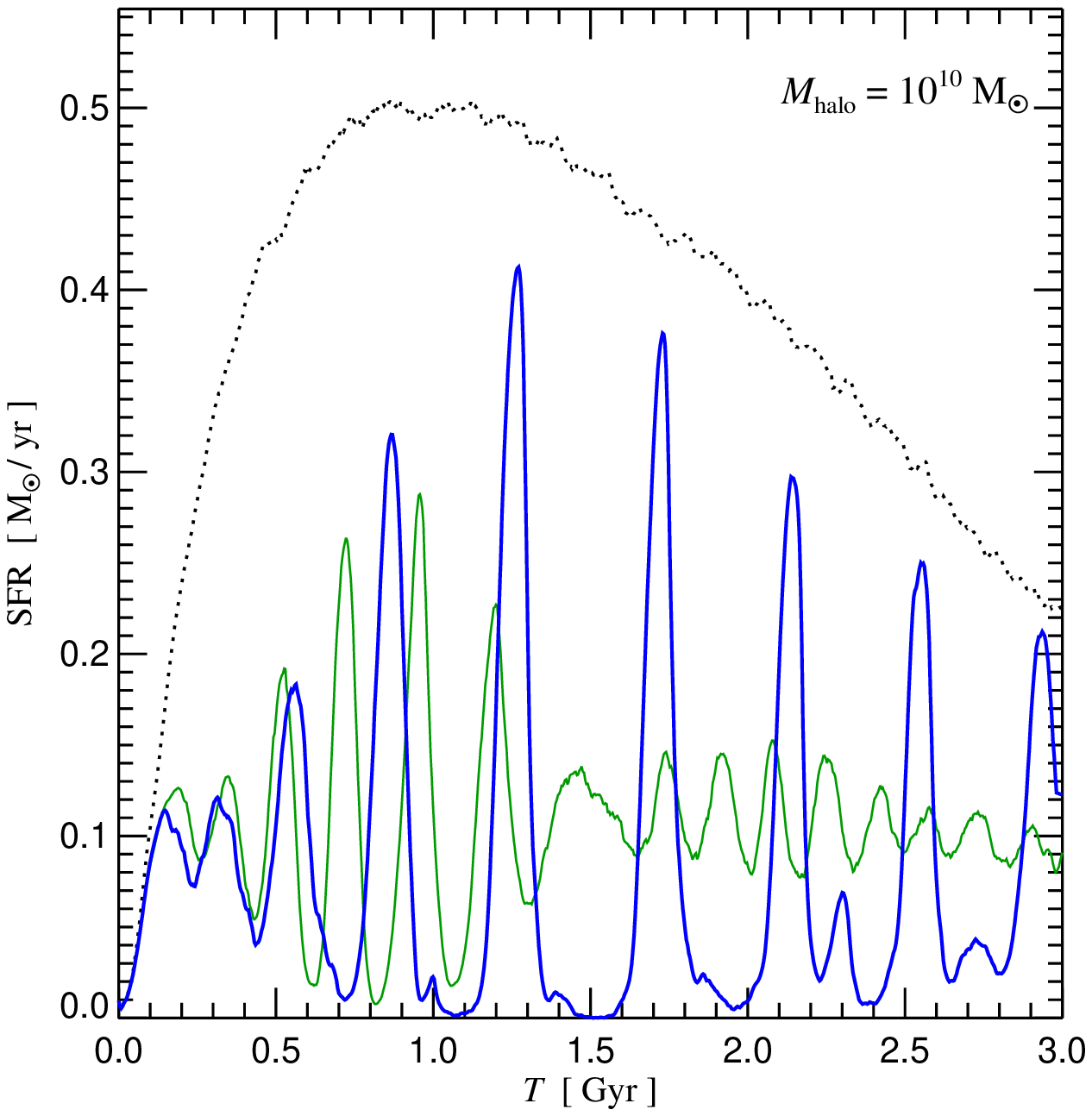}}\\%
\resizebox{8.0cm}{!}{\includegraphics{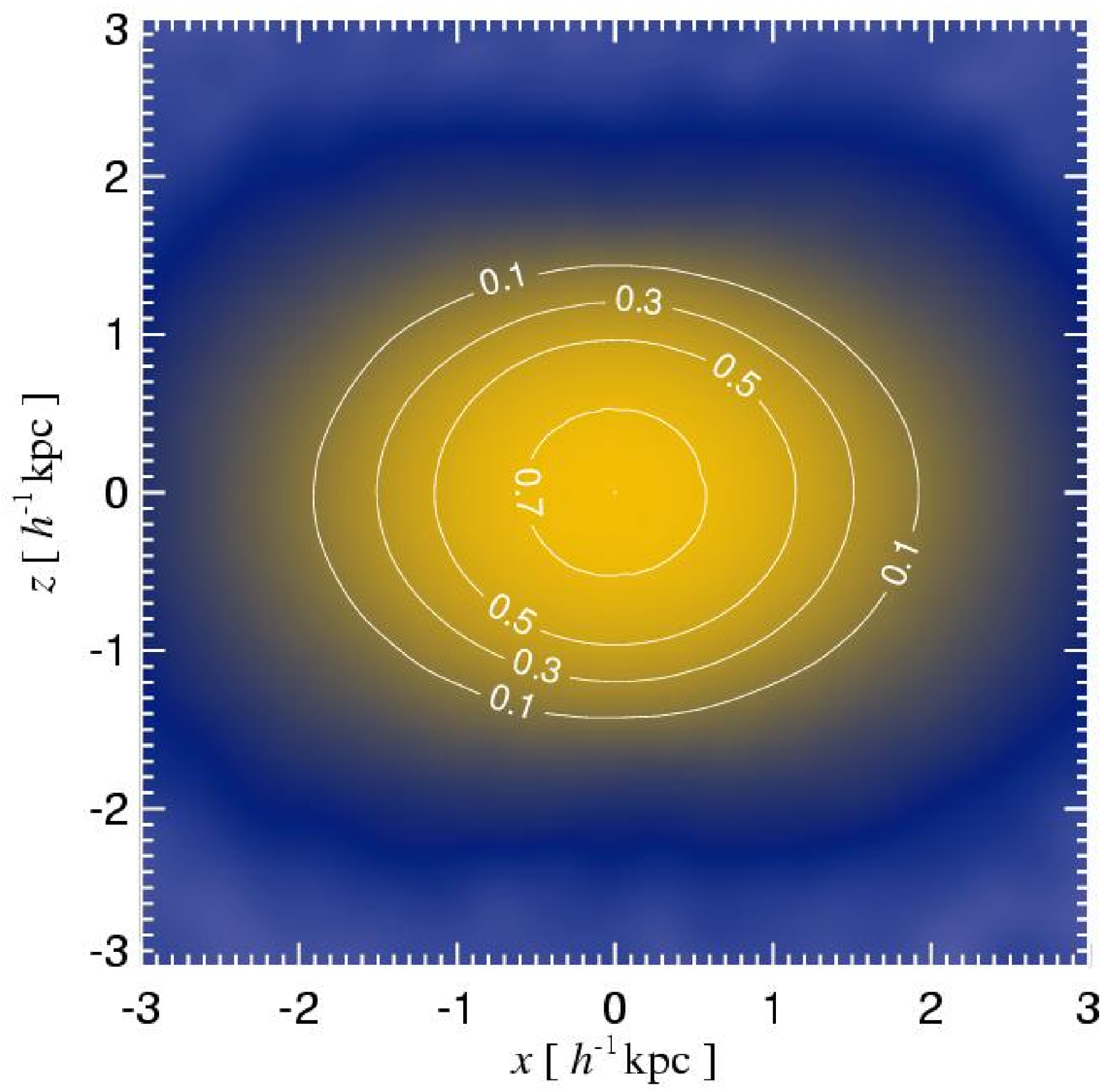}}%
\resizebox{8.0cm}{!}{\includegraphics{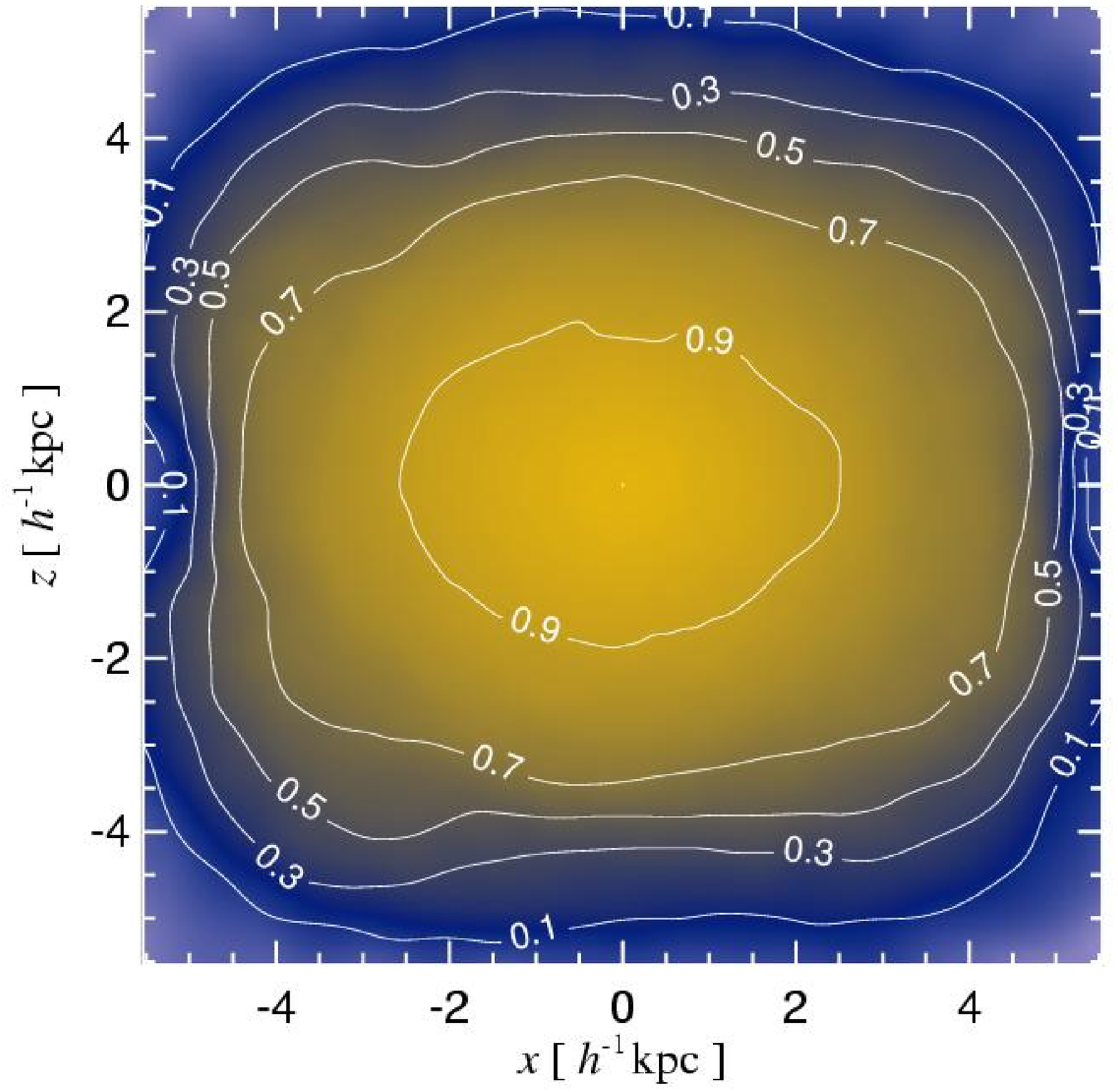}}\vspace*{-.5cm}\\%
\end{center}
  \caption{Effects of cosmic ray diffusion on the star formation and the
    pressure distribution in isolated halos of mass \mhalo{9} and \mhalo{10}.
    The panels on top compare the star formation rate when CR diffusion is
    included (thick blue line) to the case where it is neglected (thin green
    line). The dotted lines show the result when CR-production by
    supernovae is not included. In the bottom panels, we show projected gas
    density fields through the halos at time $t=2.0\,{\rm Gyr}$, with contours
    overlaid that give the fractional contribution of the projected CR energy
    to the total projected energy. These panels correspond directly to
    equivalent maps shown in Figure~\ref{fig:Galaxy:Slices} for the case
    without CR diffusion.
            \label{FigDiffusion}}
\end{figure*}

In Figure~\ref{FigClusterRad}, we show the equivalent results for the
radial profiles for the 3 simulations that include radiative cooling
and star formation. Compared to the non-radiative calculations, the
ICM shows a markedly different structure. Due to the presence of a
strong cooling flow, the temperature profile rises towards the centre,
until it eventually drops sharply at around $20\, {\rm kpc}$ due to
the onset of efficient cooling. The gas density has become
significantly lower in the bulk of the cluster volume due to the large
amount of gas that has cooled out, and correspondingly, the total
pressure has fallen in much of the cluster volume. But there are also
interesting differences in the simulations with and without cosmic
rays.  Recall that both simulations included cosmic ray production by
supernovae feedback, while only one of them accounted for cosmic rays
by structure formation shocks. The CR pressure contribution in both
simulations is quite similar through most of the cluster, at the level
of a few percent of the thermal pressure. Also note that in the very
inner parts, where the gas drops out through cooling, the cosmic ray
pressure rises sharply, even reaching and exceeding the thermal
pressure. In this small region, the thermal pressure is dissipated
more rapidly than the cosmic ray pressure.

Finally, in Figure~\ref{FigClusterStars} we compare the cumulative stellar
profile of the cluster in the three simulations with radiative cooling and
star formation. While the total mass of stars formed within the virial radius
has become smaller by the inclusion of cosmic ray feedback, the stellar mass
in the central cluster galaxy has actually increased. The cluster cooling flow
has therefore slightly increased in strength, consistent with the results we
obtained for isolated halos of this mass range. On the other hand, the
luminosity of the smaller galaxies orbiting within the cluster has become
smaller, in line with our finding that small galaxies experience a reduction
of their star formation activity. It is clear however that our results do not
suggest that cosmic rays could provide a solution to the cooling flow problem
in clusters of galaxies, at least not with the CR sources we have considered
here.  This conclusion could potentially change in interesting ways when CR
production by AGN in clusters of galaxies is included as well
\citep{Churazov2001,EnsslinVogt2005}.

\subsection{The influence of cosmic ray diffusion}

In all of our previous results, we have ignored the effects of cosmic ray
diffusion, largely because of the uncertainty involved in constraining an
appropriate diffusivity. However, diffusion could potentially be important in
several environments, depending of course on the details of the magnetic field
structure and the strength of the resulting diffusivity. While our present
formalism implemented in the simulation code is capable of dealing with
isotropic diffusion, in reality the diffusion is likely to be anisotropic,
governed by the local magnetic field configuration. In principle, cosmological
structure formation calculations with SPH are capable of following
magneto-hydrodynamics \citep{Dolag1999,Dolag2005,Price2004,Price2005},
although this is presently still fraught with numerical and physical
difficulties. We therefore postpone a detailed analysis of the influence of
cosmic ray diffusion to future work.  Here, we investigate instead a simple
example, that gives a first illustration of the effects that can be expected.

To this end, we repeat our simulations of isolated disk galaxy formation with
CR injection by supernovae, but this time with diffusion included.  We use a
parameterized diffusivity as described in section \ref{section:diffusion},
setting the values of the density and temperature scaling exponents to
$n_T=1/6$ and $n_\rho=-1/2$, respectively, with a baseline diffusivity of
$\sim 10\,{\rm kpc^2 Gyr^{-1}}$ at the threshold for star formation, i.e.~our
diffusivity model is given by equation~(\ref{eqndiffusivity}). The simulations
we repeat are the ones considered in Section~\ref{SecIsolatedDisks} with an
injection efficiency of $\zeta_{\rm SN}=0.3$ for the production of CRs by
supernovae.

In Figure~\ref{FigDiffusion}, we compare the resulting star formation rates
for halos of mass \mhalo{9} and \mhalo{10} as a function of time with the
corresponding results without diffusion. Interestingly, the oscillations due
to the unstable dynamics of a cosmic ray dominated ISM are substantially
suppressed when diffusion is included. This effect is quite strong in the
results for the \mhalo{9} halo, where we now observe a nearly constant,
quiescent star formation rate. For the \mhalo{10} halo, the oscillations are
only partially washed out and happen less frequently, but if they occur, they
are stronger.  Here the star formation rate of the galaxy develops a `bursty'
character.  Interestingly, diffusion actually reduces the integrated star
formation still further; it drops by about 30\% for the \mhalo{9} halo, and by
21\% for the \mhalo{10} halo compared to the case without diffusion.
Apparently, the cosmic rays that escape from the star-forming ISM into the
less-dense gas in the halo are  able to supply a partial pressure support
that effectively reduces the rate at which gas cools.

The more extended and smoother spatial distribution of cosmic rays due to
diffusion can also be appreciated in the bottom panels of
Figure~\ref{FigDiffusion}, where we show projections of the gas density field
with contours for the cosmic ray to thermal energy content overlaid. These
panels directly correspond to the ones shown in Figure~\ref{fig:Galaxy:Slices}
for the case without diffusion. However, for halos of mass \mhalo{11} and
more, diffusion with the strength considered here has no significant impact on
the dynamics. The progressively larger size of more massive systems makes it
ever more important for diffusion to efficiently transport CR energy across
the system.

\section{Conclusions}

In this paper, we have presented the details of the first practical
implementation of a simulation code capable of carrying out high-resolution
simulations of cosmological structure formation with a self-consistent
treatment of cosmic ray physics. In particular, our method takes dynamical
effects of pressure forces due to cosmic rays into account and therefore
allows us to carry out studies of CR feedback in the context of galaxy
formation. The underlying formalism for the treatment of cosmic rays is
discussed in a companion paper \citep{Ensslin2005} and forms a compromise
between the complexity of cosmic ray physics and the requirements of
computational efficiency. In particular, we use a simplified spectral
representation in terms of power laws for the momentum distribution with a low
momentum cut-off. This allows for a rather significant simplification at the
prize of a moderate loss of accuracy. As \citet{Ensslin2005} have shown, the
cosmic ray pressure is expected to be accurate to about 10 per cent in our
model under steady state conditions. We argue that this is sufficiently
accurate for our purposes given the other uncertainties and approximations
involved.

Our formalism also makes used of an on-the-fly shock detection scheme for SPH
developed in a second companion paper \citep{Pfrommer2005}. This method allows
us to estimate Mach numbers in shocks captured during SPH simulations, such
that we can use an appropriate efficiency for the CR injection at large-scale
structure shock waves.

We have given an initial analysis of the principal effects of two sources of
cosmic rays, namely CRs produced by supernova explosions, and by diffusive
shock acceleration during structure formation.  The loss processes we
considered were Coloumb cooling and hadronic losses, which should be the most
important ones. If desired, the modelling of these CR sink terms can be
refined in the future within our methodology, and additional sources like
cosmic rays from AGN can in principle be added as well.

There are several noteworthy results we obtained with our cosmic ray treatment
in this study. First of all, our simulations of galaxy formation with cosmic
ray production by supernovae indicate that cosmic ray pressure can play an
important role in regulating star formation in small galaxies. Here we find a
significant reduction of the star formation rate compared to the one without
CR physics, provided cosmic ray production efficiencies of several tens of
percent are assumed. In small galaxies, the mean densities reached in the ISM
stay sufficiently low such that the CR pressure can exceed the effective
pressure produced by the thermal supernova feedback. Once this occurs, the gas
of the ISM is puffed up, quenching the star formation rate. Due to the
comparatively long cosmic ray dissipation timescale, the CR-pressure survives
for a sufficiently long time in these systems and develops a sizable impact on
the star formation rate. In massive galaxies on the other hand, the ISM
becomes so dense that the CR-pressure is unable to exceed the effective pressure
predicted by the multi-phase model of \citet{Springel2003}, such that the star
formation rates are not altered. 

This effect on star formation also manifests itself in a reduction of the
cosmic star formation rate density in cosmological simulations of galaxy
formation. Here the SFR history is reduced at high redshift, where the bulk of
star formation is dominated by small dwarf galaxies. As the star formation
shifts to the scale of more massive halos towards low redshift, the reduction
becomes progressively smaller. An interesting implication of the strong effect
of CR feedback on small galaxies is that this reduces the faint-end slope of
the resulting galaxy luminosity function, an area that continues to be a
problematic issue for hydrodynamical simulations of galaxy formation within
the $\Lambda$CDM scenario. We have indeed detected this flattening, although
with a weak strength overall. Another tantalizing effect of cosmic rays is
that they help to keep gas in small galaxies more diffuse. This should in
principle help to alleviate the ``angular momentum problem'', which describes
the problem of an efficient angular momentum loss of gas to the dark matter
caused by the early collapse of large amounts of gas in small halos. It is
believed that this is a primary reason why present simulations
generally fail to produce large spiral galaxies at low redshift. Cosmic rays
physics might help to resolve this problem.

In simulations where we included cosmic ray injection by structure formation
shocks, we find that they are produced efficiently at high redshift when
structure formation ensues, driven by the high Mach-number shocks found at low
to moderate overdensities. At low redshift on the other hand, most of the
energy is thermalized in weaker shocks where the injection efficiency is much
smaller. As a result, the mean energy content in cosmic rays can reach above
40\% at redshifts $z\simeq 5$, but drops to $\sim 10\%$ at low redshift.
However, the relative energy content also shows a strong density dependence.
It is highest at low to moderate overdensities and declines continously with
density, such that deep inside halos, only comparatively little cosmic ray
energy produced by shock waves survive. An important factor in this trend is
the strong density dependence of cosmic ray loss processes, and the softer
adiabatic index of CRs.

When full cosmological simulations of the formation of galaxy clusters are
considered, it is therefore not very surprising that we find that structure
formation shocks can build up only a comparatively small cosmic ray pressure
contribution inside clusters. Even at the virial radius this contribution
reaches only about 10\%, but lies much lower in the inner parts of the
cluster. When radiative cooling and cosmic ray production by supernovae are
included, we find that supernovae can boost the mean CR energy density in the
cluster, but the averaged contribution still stays at the percent level
throughout the cluster volume, except at places where the gas rapidly cools.
Here the CR pressure can temporarily dominate the pressure and delay the
collapse briefly.  Nevertheless, we find that CR production by supernovae and
structure formation shocks is unable to reduce central cluster cooling flows.
Instead, we in fact detect a slight increase of the cooling in the \mhalo{14}
cluster we have simulated. This can be understood as a result of the higher
compressibility of the cluster gas in the cosmic ray simulations, leading to
an increased central concentration of the gas and an elevated baryon fraction
in the cluster, and thereby to higher cooling of gas in the centre
overall. Note however that the currently discussed AGN feedback for
cooling-flow quenching was not included in our simulations.
On the other hand, the bulk of the cluster galaxies experience a reduction of
their star formation rate when CR feedback is included, such that the cluster
galaxy luminosity function is expected to develop a shallower faint-end slope.

Overall, our results suggest that cosmic ray physics is unlikely to
drastically modify the physics of galaxy formation in the $\Lambda$CDM model.
However, cosmic rays help in areas where current model-building faces
important problems, like for the faint-end slope of the galaxy luminosity
function and the angular momentum problem. Our formalism for treating CRs in
cosmological simulations should therefore be very valuable for future studies
on the role of cosmic rays in cosmological structure formation.  In
particular, it would be highly interesting to examine the effects of CRs on
the metal distribution of the universe, or on the dynamics of buoyant bubbles
inflated by AGN in clusters of galaxies. It will also be important to provide
an in-depth analysis of the role of cosmic ray diffusion in future work.

\section*{Acknowledgements}

The simulations of this study were computed at the `Rechenzentrum der
Max-Planck-Gesellschaft' (RZG), Garching.

\bibliographystyle{mnras}
\bibliography{paper}

\end{document}